\documentclass[a4paper,11pt]{article}
\pdfoutput=1
\usepackage[english]{babel}
\usepackage{jheppub}
\usepackage{amsmath,amssymb,graphicx,textcomp,enumerate,xspace}
\usepackage{soul} %Use \st from soul instead of \sout from ulem; ulem breaks the biblio

\usepackage[utf8]{inputenc}
\newcommand\sss{\mathchoice%
{\displaystyle}%
{\scriptstyle}%
{\scriptscriptstyle}%
{\scriptscriptstyle}%
}

\def\beq{\begin{equation}}
\def\beqn{\begin{eqnarray}}
\def\eeq{\end{equation}}
\def\eeqn{\end{eqnarray}}

\def\lq{\left[} 
\def\rq{\right]} 
\def\rg{\right\}} 
\def\lg{\left\{} 
\def\({\left(} 
\def\){\right)} 
 
\newcommand\nf{n_{\rm f}}

\newcommand\as{\alpha_{\sss\rm S}}

\newcommand\CF{C_{\sss\rm F}}
\newcommand\CA{C_{\sss\rm A}}

\newcommand\TR{T_{\sss\rm R}}
\newcommand\NC{N_{\rm c}}

\newcount\minutes 
\newcount\scratch 
\def\timestamp{% 
\scratch=\time 
\divide\scratch by 60 
\edef\hours{\the\scratch} 
\multiply\scratch by 60 
\minutes=\time 
\advance\minutes by -\scratch 
---$\,$\hours:\null 
\ifnum\minutes< 10 0\fi 
\the\minutes}

\definecolor{mygray}{gray}{0.5}

\newcommand\qt{q_{\sss\rm T}}
\newcommand\pt{p_{\sss\rm T}}
\newcommand{\qtcut}{q_{\sss \rm T}^{\sss\rm cut}}
\newcommand{\qtmax}{q_{\sss \rm T}^{\sss\rm max}}
\newcommand{\zmax}{z^{\sss\rm max}}

\newcommand{\qtoQsq}{{\displaystyle\frac{\qt^2}{Q^2}}}
\def\beeq{\begin{eqnarray}} 
\def\eeeq{\end{eqnarray}} 
\def\to{\rightarrow}
\def\ito{\leftarrow}

\newcommand\HCCF{\lq \HCCF H^{F} C_1 C_2 \rq_{c\bar{c};\,ab}}

\newcommand\Lum{{\cal L}}
\newcommand\ord[1]{\mathcal{O}\!\(#1\)}
\newcommand\lum{\Lum(\tau)}
\newcommand\deronelum{\Lum^{(1)}(\tau)}
\newcommand\dertwolum{\Lum^{(2)}(\tau)}
\newcommand\derthreelum{\Lum^{(3)}(\tau)}
\newcommand\derfourlum{\Lum^{(4)}(\tau)}

\newcommand\APreg{p}
\newcommand\APnoreg{\hat{p}}

\newcommand\dsigdqtz{\frac{d\hat\sigma_{ab}\!\(\qt,z\)}{d\qt^2}}
\newcommand\piT{\pi_{\sss\rm T}}
\newcommand\gu{\hat{g}^{\sss\rm U (1)}}
\newcommand\gd{\hat{g}^{\sss \rm R (1)}}
\newcommand\Iu{I^{\sss\rm U}}
\newcommand\Id{I^{\sss \rm R}}

\newcommand\sigLTab{\sigma_{ab}^{\sss \rm LT }}
\newcommand\sigNLTab{ \sigma_{ab}^{\sss \rm NLT }}
\newcommand\sigNNLTab{\sigma_{ab}^{\sss \rm N^2LT }}

\newcommand\sigLT{ \sigma^{\sss \rm LT }}
\newcommand\sigNLT{ \sigma^{\sss \rm NLT }}
\newcommand\sigNNLT{\sigma^{\sss \rm N^2LT }}

\newcommand\MLS{\left| {\cal M} \right|^2_{\rm\sss LS} }

\def\MNLS#1{\left| {\cal M} \right|^2_{\rm\sss N^{#1}LS} }

\newcommand\dontshow[1]{}

\title{Higher-order power corrections in a
  transverse-momentum cut for colour-singlet production at NLO}

\author[a]{Leandro Cieri,}
\author[a,b]{Carlo Oleari,}
\author[a,b]{Marco Rocco}

\emailAdd{leandro.cieri@mib.infn.it}
\emailAdd{carlo.oleari@mib.infn.it}
\emailAdd{m.rocco10@campus.unimib.it}

\affiliation[a] {INFN, Sezione di Milano-Bicocca,
Piazza della Scienza 3, I-20126 Milano, Italy}

\affiliation[b] {Universit\`a di Milano\,-\,Bicocca and INFN, Sezione di
  Milano\,-\,Bicocca, Piazza della Scienza 3, 20126 Milano, Italy}

\abstract{We consider the production of a colourless system at
  next-to-leading order in the strong coupling constant $\as$.
  We impose a transverse-momentum cutoff, $\qtcut$, on the colourless final
  state and we compute the power corrections for the inclusive cross section
  in the cutoff, up to the fourth power.
  
  The study of the dependence of the cross section on $\qtcut$ allows for an
  understanding of its behaviour at the boundaries of the phase space, giving
  hints on the structure at all orders in $\as$ and on the identification of
  universal patterns.
  The knowledge of such power corrections is also a required ingredient in
  order to reduce the dependence on the transverse-momentum cutoff of the QCD
  cross sections at higher orders, when the $\qt$-subtraction
  method is applied.

  We present analytic results for both Drell--Yan vector boson and Higgs
  boson production in gluon fusion and we illustrate a process-independent
  procedure for the calculation of the all-order power corrections in the
  cutoff.
  In order to show the impact of the power-correction terms, we present
  selected numerical results and discuss how the residual dependence on
  $\qtcut$ affects the total cross section for Drell--Yan $Z$ production and
  Higgs boson production via gluon fusion at the LHC.

}


\begin{document}

\maketitle

%%%%%%%%%%%%%%%%%%%%%%%%%%%%%%%%%%%%%%%%%%%%%%%%%%%%%%%%%%%%%%
%%%%%%%%%%%%%%%%%%%%%%%%%%%%%%%%%%%%%%%%%%%%%%%%%%%%%%%%%%%%%%
\section{Introduction}
\label{sec:intro}
%%%%%%%%%%%%%%%%%%%%%%%%%%%%%%%%%%%%%%%%%%%%%%%%%%%%%%%%%%%%%%
%%%%%%%%%%%%%%%%%%%%%%%%%%%%%%%%%%%%%%%%%%%%%%%%%%%%%%%%%%%%%%

%% \comment{ LT = leading terms = LL, NLL + cost}
%% \comment{ NLT = $a, a \log(a)$}
%% \comment{ N2LT = $a^2, a^2 \log(a)$}

The current precision-physics program at the Large Hadron Collider~(LHC)
requires Standard Model~(SM) theoretical predictions at the highest
accuracy. Data belonging to ``benchmark'' processes, which are measured with
the utmost precision at the LHC, need to be tested against theoretical
results at the same level of accuracy. This is not only important for the
extraction of SM parameters per se, but also for searches of signals of new
physics, that can appear as small deviations in kinematic distributions with
respect to the SM predictions. Reaching the highest possible level of
precision is then the main goal and the calculation of perturbative QCD
corrections plays a dominant role in this context.

Until a few years ago, the standard for such calculations was next-to-leading
order~(NLO) accuracy. In recent years, a continuously-growing number of
next-to-next-to-leading order~(NNLO) results for many important processes has
appeared in the literature, giving birth to the so called ``NNLO
revolution''.
For several ``standard candles'' processes, the first steps towards the
calculation of differential cross sections at N$^3$LO have also been taken
(see e.g.~\cite{Cieri:2018oms, Dulat:2018bfe}).

The computation of higher-order terms in the perturbative series becomes more
involved due to the technical difficulties arising in the evaluation of
virtual contributions and to the increasing complexity of the infrared~(IR)
structure of the real contributions. In order to expose the cancellation of
the IR divergences between real and virtual contributions, the knowledge of
the behaviour of the scattering amplitudes at the boundaries of the phase
space is then a crucial ingredient and it is indeed what is used by the
subtraction methods in order to work.
These methods
%(or regularisation prescriptions)
can be roughly divided into local and slicing. Among the first, the most
extensively used at NLO were proposed in refs.~\cite{Catani:1996vz,
  Frixione:1995ms}. As far as the NNLO subtraction methods are concerned, the
past few years have witnessed a great activity in their development: the
transverse-momentum~($\qt$) subtraction method~\cite{Catani:2007vq,
  Bozzi:2005wk, Bonciani:2015sha, Catani:2019iny}, the $N$-jettiness
subtraction~\cite{Boughezal:2015eha, Gaunt:2015pea}, the
projection-to-Born~\cite{Cacciari:2015jma}, the residue
subtraction~\cite{Czakon:2011ve, Boughezal:2011jf} and the antenna
subtraction method~\cite{GehrmannDeRidder:2005cm, Daleo:2006xa,
  Currie:2013vh} have all been successfully applied to LHC phenomenology.
The first application of the $\qt$-subtraction method to differential cross
sections at N$^3$LO was recently proposed in ref.~\cite{Cieri:2018oms}, in
the calculation of the rapidity distribution of the Higgs boson.

While a local subtraction is independent of any regularising parameter,
slicing methods require the use of a cutoff to separate the different IR
regions. Such separation of the phase space introduces instabilities in the
numerical evaluation of cross sections and differential
distributions~\cite{Catani:2011qz, Catani:2018krb, Grazzini:2017mhc,
  Boughezal:2016wmq}, and some care has to be taken in order to obtain stable
and reliable results.

The knowledge of logarithmic and power-correction terms in the cutoff plays a
relevant role in the identification of universal structures, in the
development of regularisation prescriptions and in resummation
programs~\cite{Dokshitzer:1978yd, Dokshitzer:1978hw, Parisi:1979se,
  Curci:1979bg, Collins:1981uk, Kodaira:1981nh, Kodaira:1982az,
  Collins:1984kg, Catani:1988vd, deFlorian:2000pr, Alioli:2015toa}.
According to their behaviour in the zero limit, the cutoff-dependent terms
can be classified into logarithmically-divergent or finite contributions, and
power-correction terms, that vanish in that limit.  In particular, the terms
that are singular in the small-cutoff limit are universal and are cancelled
by the application of the subtraction methods, while finite and vanishing
terms are, in general, process dependent. However, after the subtraction
procedure, a residual dependence on the cutoff remains as power corrections.
While these terms formally vanish in the null cutoff limit, they give a
non-zero numerical contribution for any finite choice of the cutoff.

From a theoretical point of view, the knowledge of the power corrections
greatly increases our understanding of the perturbative behaviour of the QCD
cross sections, since more non-trivial (universal and non-universal) terms
appear. The origin of these terms can be traced back both to the scattering
amplitudes, evaluated at phase-space boundaries, and to the phase space
itself.
Thus, several papers have tackled the study of power corrections in the soft and
collinear limits~\cite{Bern:2014oka, Larkoski:2014bxa,
%  Dixon:2011pw, Dixon:2014iba, Dixon:2015iva, Caron-Huot:2016owq,
%  Dixon:2016nkn,
  Luo:2014wea}, while studies in the general framework of fixed-order and
threshold-resummed computations have also been
performed~\cite{vanBeekveld:2019cks, vanBeekveld:2019prq, DelDuca:2017twk,
  Bonocore:2015esa, Bonocore:2014wua, Laenen:2010uz, Laenen:2008ux,
  Beneke:2018gvs}.

From a practical point of view, the knowledge of the power corrections makes
the numerical implementation of a subtraction method more robust, since the
power terms weaken the dependence of the final result on the arbitrary
cutoff.  This is not only valid when the subtraction method is applied to NLO
computations, but it is numerically more relevant when applied to
higher-order calculation, as pointed out, for example, in the evaluation of
NNLO cross sections in refs.~\cite{Grazzini:2017mhc, Boughezal:2016wmq}.

Power corrections at NLO have been extensively studied in
refs.~\cite{Moult:2016fqy, Boughezal:2016zws, Boughezal:2018mvf,
  Moult:2017jsg, Ebert:2018lzn, Ebert:2018gsn, Bhattacharya:2018vph,
  Campbell:2019gmd, Moult:2018jjd, Boughezal:2019ggi} in the context of the $N$-jettiness
subtraction method, and in refs.~\cite{Bauer:2000ew, Bauer:2000yr,
  Bauer:2001ct, Bauer:2001yt, Bauer:2002aj, Moult:2019mog} within SCET-based
subtraction methods. A numerical extraction of power corrections in the
context of NNLL'+NNLO calculations was done in
$N$-jettiness~\cite{Alioli:2015toa}, and a general discussion in the context
of the fixed-order implementation of the $N$-jettiness subtraction can be
found in ref.~\cite{Gaunt:2015pea}.

In this paper we consider the production of a colourless system at
next-to-leading order in the strong coupling constant $\as$.  In particular,
we discuss Drell--Yan~(DY) $V$ production and Higgs boson production in gluon
fusion at NLO, in the infinite top-mass limit.
We impose a transverse-momentum cutoff, $\qtcut$, on the colourless system
and we compute the power corrections in the cutoff, up to order $(\qtcut)^4$,
for the inclusive cross sections. The knowledge of these terms will shed
light upon the non-trivial behaviour of cross sections at the boundaries of
the phase space, and upon the resummation structure at subleading orders.  In
addition, it allows to have a better control on the cutoff-dependent terms,
when the $\qt$-subtraction method of ref.~\cite{Catani:2007vq} is applied to
the numerical calculation of cross sections, allowing for a use of larger
values of the cutoff. We also describe a process-independent procedure that
can be used to compute the all-order power corrections in the cutoff.

The outline of this paper is as follows. In sec.~\ref{sec:kinematics} we
introduce our notation, and we briefly summarize the expressions of the
partonic and hadronic cross sections, in a form that is suitable for what
follows. In sec.~\ref{sec:calculation} we outline the calculation we have
done and in sec.~\ref{sec:results} we present and discuss our analytic
results for $V$ and $H$ production, along with a study of their numerical
impact. We draw our conclusions in sec.~\ref{sec:conclusions}.  We leave to
the appendixes all the technical details of our calculation.

\section{Kinematics and notation}
\label{sec:kinematics}
We briefly introduce the notation used in our theoretical framework and we
recall some kinematic details of the calculations presented in this paper.

\subsection{Hadronic cross sections}
We consider the production of a colourless system $F$ of squared invariant
mass $Q^2$ plus a coloured system $X$ at a hadron collider
\begin{equation}
\label{eq:proc}
h_1+h_2 \to F+X \,.
\end{equation}
We call $S$ the hadronic squared center-of-mass energy and we write the hadronic
differential cross section for this process as
\begin{equation}
\label{eq:had_tot_XS}
d\sigma = \sum_{a,b} \int_\tau^1 dx_1 \int_\frac{\tau}{x_1}^1 dx_2\,
f_a\!\(x_1\) f_b\!\(x_2\) d\hat\sigma_{ab}\,,
\end{equation}
where 
\begin{equation}
\label{eq:tau_def}
\tau = \frac{Q^2}{S}\,,
\end{equation}
$f_{a/b}$ are the parton densities of the partons $a$ and $b$, in the hadron
$h_1$ and $h_2$ respectively, and $d\hat\sigma_{ab}$ is the partonic cross
section for the process $a+b \to F + X$. The dependence on the
renormalisation and factorisation scales and on the other kinematic
invariants of the process are implicitly assumed. 

In appendix~\ref{app:partonic_PS} we have collected all the formulae for the
calculation of the partonic cross section. Using eqs.~(\ref{eq:disgma_hat})
and~(\ref{eq:diffXS_M}), we write the hadronic cross section as
\begin{equation}
\sigma = \sum_{a,b} \int_\tau^1 dx_1 \int_\frac{\tau}{x_1}^1 dx_2\,
f_a\!\(x_1\) f_b\!\(x_2\) \int  d\qt^2\, dz \,\dsigdqtz\,
\delta\!\(\! z- \frac{Q^2}{s}\),
\end{equation}
where $s$ is the partonic center-of-mass energy, equal to
\begin{equation}
\label{eq:s_S}
s = S \, x_1\, x_2\,.
\end{equation}
We have also made explicit the dependence on $z$, the ratio between the
squared invariant mass of the system $F$ and the partonic center-of-mass
energy, and on $\qt$, the transverse momentum of the system $F$ with respect to
the hadronic beams.  Using eqs.~(\ref{eq:s_S}) and~(\ref{eq:tau_def})
%% we can write
%% \begin{equation}
%% %\label{eq:had_tot_XS}
%% \sigma = \sum_{a,b} \int_0^1 dz \, \delta\!\(\! z - \frac{\tau}{x_1 x_2}\)
%% \int_\tau^1 dx_1 \int_\frac{\tau}{x_1}^1 dx_2\, 
%% f_a\!\(x_1\) f_b\!\(x_2\) \!\int\! d\qt^2 \, \dsigdqtz    \,,
%% \end{equation}
%% and we use the $\delta$ function to integrate over $x_2$, obtaining
and integrating over $x_2$ we obtain
\begin{equation}
\sigma = \sum_{a,b} \tau \int_\tau^1 \frac{dz}{z}  \int_\frac{\tau}{z}^1 \frac{dx_1}{x_1}\,
f_a\!\(x_1\) f_b\!\(\frac{\tau}{z\,x_1}\) \frac{1}{z} \int\! d\qt^2 \, \dsigdqtz \,.
\end{equation}
We then introduce the parton luminosity ${\cal L}_{ab}(y)$ defined by
\begin{equation}
  \label{eq:lum_def}
{\cal L}_{ab}(y) \equiv \int_y^1 \frac{dx}{x}\, f_a\!\(x\) f_b\! \(\frac{y}{x}\),
\end{equation}
so that we can finally write
\begin{equation}
\label{eq:sigma_had_z_one}
\sigma = \sum_{a,b} \tau \int_\tau^1 \frac{dz}{z}  \, {\cal L}_{ab}\!\(\frac{\tau}{z}\)
 \frac{1}{z} \int\! d\qt^2 \,  \dsigdqtz \,.
\end{equation}

\subsection{Partonic differential cross sections}
\label{sec:part_diff_XS}
In this section we recall the formulae for the first-order real corrections
to the Drell--Yan production of a weak boson $V$ ($W$ or $Z$) and to the
Higgs boson production in gluon fusion, in the infinite top-mass limit.

The partonic cross sections $d\hat\sigma_{ab}(\qt,z)/d\qt^2$ in
eq.~(\ref{eq:sigma_had_z_one}) are computable in perturbative QCD as power
series in $\as$
\begin{equation}
  \dsigdqtz = \frac{d\hat\sigma^{\sss (0)}(\qt,z)}{d\qt^2}
  + \frac{\as}{2\pi} \, \frac{d\hat\sigma_{ab}^{\sss (1)}(\qt,z)}{d\qt^2} + \ldots
\end{equation}
The Born contribution ${d\hat\sigma^{\sss (0)}(\qt,z)}/{d\qt^2}$ and the
virtual contributions to ${d\hat\sigma_{ab}^{\sss (1)}(\qt,z)}/{d\qt^2}$ are
proportional to $\delta(\qt)$.

Applying the formulae detailed in appendix~\ref{app:partonic_PS}, with a
little abuse of notation,\footnote{In the rest of the paper we deal only with
  the real corrections to $V$ and $H$ production. We then use
  ${d\hat\sigma_{ab}^{\sss (1)}(\qt,z)}/{d\qt^2}$ to indicate them.} we can
write the partonic differential cross sections for the real corrections to
$V$ and $H$ production as

\subsubsection*{$\boldsymbol{V}$ production}

\begin{itemize}
  
\item  $q(\bar{q}) +  g \to V + q(\bar{q})$

\begin{equation}
  \label{eq:dsigmahat_V_qg}
%  \frac{d^2\hat\sigma_{qg}(\qt,z)}{d\qt^2\,dz}
  \frac{d\hat\sigma_{qg}^{\sss (1)}(\qt,z)}{d\qt^2}
  = \sigma_{qq}^{\sss (0)} %\asotpi
  \,\TR\,
z\,\frac{z\,(1+3z)\,\qtoQsq +(1-z)\,
  \APreg_{qg}(z)}{\sqrt{(1-z)^2-4z\,\qtoQsq }}\,\frac{1}{\qt^2}\,, 
\end{equation}

\item  $q + \bar{q} \to V + g$ 
  
\begin{equation}
%\\[2mm]
\label{eq:dsigmahat_V_qq}
%\frac{d^2\hat\sigma_{qq}(\qt,z)}{d\qt^2\,dz}
\frac{d\hat\sigma_{q\bar q}^{\sss (1)}(\qt,z)}{d\qt^2}
=  \sigma_{qq}^{\sss (0)} %\asotpi
\, \CF \,
z\, \frac{-4z\,\qtoQsq
  +2\,(1-z)\,\APnoreg_{qq}(z)}{\sqrt{(1-z)^2-4z\,\qtoQsq}}\,\frac{1}{\qt^2}\,, 
%\end{eqnarray}
\end{equation}
\end{itemize}
where
\begin{equation}
\label{eq:sig0_V_qq} 
\sigma_{q q}^{\sss (0)} = \frac{\pi}{\NC}\,\frac{g^2 \(g_v^2 + g_a^2\)}
      {c_{\rm \sss W}^2}  \frac{1}{Q^2}
\end{equation}
is the Born-level cross section for the process $q \bar{q} \to V$, with
$c_{\rm\sss W}$ the cosine of the weak angle and $g$, $g_v$, $g_a$ the weak,
the vector and the axial coupling, respectively.  With $\NC$ we denote the
number of colours, $\CF=\(\NC^2-1\)/2\NC=4/3$ and $\TR=1/2$. For $W$
production, the flavours of the quarks in
eqs.~(\ref{eq:dsigmahat_V_qg})--(\ref{eq:sig0_V_qq}) are different, and the
corresponding Cabibbo--Kobayashi--Maskawa matrix element has to be included
in eq.~(\ref{eq:sig0_V_qq}).  The expression of the Altarelli--Parisi
splitting functions $\APreg_{qg}(z)$ and $\APnoreg_{qq}(z)$ are given in
appendix~\ref{app:AP}.

\subsubsection*{$\boldsymbol{H}$ production}

\begin{itemize}

\item $g + q(\bar q) \to H + q(\bar q)$ 

  \begin{equation}
  \label{eq:dsigmahat_H_gq}
%  \frac{d^2\hat\sigma_{gq}(\qt,z)}{d\qt^2\,dz}
  \frac{d\hat\sigma_{gq}^{\sss (1)}(\qt,z)}{d\qt^2} 
  = \sigma_{gg}^{\sss (0)} %\asotpi
  \,\CF \,
z\,\frac{-3\,(1-z)\,\qtoQsq+(1-z)\,\APreg_{gq}(z)}{\sqrt{(1-z)^2-4z\,\qtoQsq}}
\,\frac{1}{\qt^2}\,, 
\end{equation}

\item  $g + g \to H + g$
  
%\\[2mm]
\begin{equation}
  \label{eq:dsigmahat_H_gg}
%  \frac{d^2\hat\sigma_{gg}(\qt,z)}{d\qt^2\,dz}
  \frac{d\hat\sigma_{gg}^{\sss (1)}(\qt,z)}{d\qt^2}
  =\sigma_{gg}^{\sss (0)} % \asotpi
  \,\CA \,
z\,\frac{4z \(\qtoQsq\)^{\!\!2} - 8\,(1-z)^2\,\qtoQsq+2\,(1-z)\,\APnoreg_{gg}(z)}
{\sqrt{(1-z)^2 - 4 z\,\qtoQsq}}\,\frac{1}{\qt^2}\,,\phantom{aaa}
\end{equation}

\end{itemize}
where
\begin{equation}
\sigma_{gg}^{\sss (0)} = \frac{\as^2}{72\pi} \,\frac{1}{ \NC^2-1}\, \frac{1}{v^2}
\end{equation}
is the Born-level cross section for the process $gg \to H$, with $v$ the
Higgs vacuum expectation value and  $\CA=\NC=3$.
The expression of the Altarelli--Parisi splitting functions $\APreg_{gq}(z)$ and
$\APnoreg_{gg}(z)$ are given in appendix~\ref{app:AP}.

We notice that the terms proportional to the Altarelli--Parisi splitting
functions in eqs.~(\ref{eq:dsigmahat_V_qg})--(\ref{eq:dsigmahat_H_gg}) embody
in a single expression the whole infrared behaviour of the amplitudes,
i.e.~their soft and collinear limits. The structure of these terms was
derived in a completely general form, from the universal behaviour of the
scattering amplitudes in those limits, in ref.~\cite{deFlorian:2001zd}.

\section{Description of the calculation}
\label{sec:calculation}
In the small-$\qt$ region, i.e.~$\qt\ll Q$, the real contribution to the
perturbative cross sections of
eqs.~(\ref{eq:dsigmahat_V_qg})--(\ref{eq:dsigmahat_H_gg}) contains well-known
logarithmically-enhanced terms that are singular in the $\qt\to 0$
limit~\cite{Dokshitzer:1978yd, Dokshitzer:1978hw, Parisi:1979se,
  Curci:1979bg, Collins:1981uk, Kodaira:1981nh, Kodaira:1982az,
  Collins:1984kg, Catani:1988vd, deFlorian:2000pr}. In the context of
inclusive NLO fixed-order calculations, the logarithmic terms are cancelled
when using the subtraction prescriptions. For more exclusive quantities, such
as the transverse-momentum distribution of the colourless system, the same
logarithmic terms need to be resummed at all orders in the strong coupling
constant to produce reliable results. Although our studies are of value in
the context of the transverse-momentum resummation, here we limit ourselves
to the case of inclusive fixed-order predictions at NLO, leaving the
resummation program to future investigations.  In this paper we
compute power-correction terms to the cross section that, although vanishing
in the small-$\qt$ limit, may give a sizable numerical contribution when
using a slicing subtraction method.

To explicitly present the perturbative structure of these terms at small
$\qt$, it is customary in the literature~\cite{deFlorian:2001zd,
  Ebert:2018lzn} to compute the following cumulative partonic cross section,
integrating the differential cross section in the range $0 \le \qt \le
\qtcut$,
\begin{equation}
  \label{eq:XS_cumul}
  %  \ref{eq:int_0^qtcut}
 \hat\sigma_{ab}^{\sss  <}(z) \equiv
\int_0^{\(\qtcut\)^2}\!\! d\qt^2 \,\frac{d\hat\sigma_{ab}(\qt,z)}{d\qt^2}.
\end{equation}
The cross section in eq.~(\ref{eq:XS_cumul}) receives contributions from the
Born and the virtual terms, both proportional to $\delta(\qt)$, and from the
part of the real amplitude that describes the production of the $F$ system
with transverse momentum less than $\qtcut$. The virtual and real
contributions are separately divergent and are typically regularised in
dimensional regularisation.  Since the total partonic cross section is finite
and analytically known for the processes under study, following what was done
in refs.~{\cite{Catani:2011kr, Catani:2012qa}, we compute the above integral
  as
\begin{equation}
  \label{eq:int_0^qtcut-sub}
 \hat\sigma_{ab}^{\sss <}(z)  = \hat\sigma^{\rm\sss tot}_{ab}(z) -
 \hat\sigma_{ab}^{\sss >}(z)\,, 
\end{equation}
with
\begin{eqnarray}
 \hat\sigma^{\rm\sss tot}_{ab}(z)  &=&
 \int_0^{\(\qtmax\)^2}\!\! d\qt^2 \,\frac{d\hat\sigma_{ab}(\qt,z)}{d\qt^2}\,,
 \\
   \label{eq:int_qtcut^qmax}
\hat\sigma_{ab}^{\sss >}(z) &=& \int_{\(\qtcut\)^2}^{\(\qtmax\)^2}\!\!
d\qt^2\,\frac{d\hat\sigma_{ab}(\qt,z)}{d\qt^2} \,,
\end{eqnarray}
where $\qtmax$ is the maximum transverse momentum allowed by the kinematics,
$\hat\sigma^{\rm\sss tot }_{ab}(z)$ is the total partonic cross section
%(up to the order included in $d\hat\sigma_{ab}$)
and $\hat\sigma_{ab}^{\sss >}(z)$ is the partonic cross section integrated
above $\qtcut$. The advantage of using eq.~(\ref{eq:int_0^qtcut-sub}) is that
the partonic cross section integrated in the range $0\le \qt \le \qtcut$ is
obtained as difference of the total cross section (formally free from any
dependence on $\qtcut$) and the partonic cross section integrated in the
range above $\qtcut$ of eq.~(\ref{eq:int_qtcut^qmax}).  Since $\qt>\qtcut >
0$, the last integration can be performed in four space-time dimensions, with
no further use of dimensional regularisation. In refs.~\cite{Catani:2011kr,
  Catani:2012qa} the computation of the cumulative cross section was
performed in the limit $\qtcut \ll Q$, neglecting terms of $\ord{(\qtcut)^2}$
on the right-hand side of eq.~(\ref{eq:int_0^qtcut-sub}). In this paper, we
compute these terms up to $\ord{(\qtcut)^4}$ included.

\subsection[${\qt}$-integrated partonic cross
  sections]{$\boldsymbol{\qt}$-integrated partonic cross sections} 
\label{sec:qt_integration}
In this section we present the results for the partonic cross section in
eq.~(\ref{eq:int_qtcut^qmax}), integrated in $\qt$, from an arbitrary value
$\qtcut$ up to the maximum transverse momentum $\qtmax$ allowed by the
kinematics of the event, given by
\begin{equation}
\(\qtmax\)^2 = Q^2\, \frac{(1-z)^2}{4\,z}\,,
\end{equation}
at a fixed value of $z$.  The integrations are straightforward and do not
need any dedicated comment.  To lighten up the notation, we introduce the
dimensionless quantity\footnote{In the literature, the parameter $a$ is also
  referred to as $r^2_{\rm \sss cut}$ (see e.g.~\cite{Grazzini:2017mhc}).}
\begin{equation}
  \label{eq:a_def}
a \equiv  \frac{\(\qtcut\)^2}{Q^2}\,,
\end{equation}
that will be our expansion parameter in the rest of the paper, and we define
\begin{equation}
  \piT^2 \equiv \frac{4az}{(1-z)^{2}}\,,
\end{equation}
that will allow us to write the upcoming differential cross sections in a
more compact form.

\subsubsection*{$\boldsymbol{V}$ production}
\begin{itemize}
\item $q(\bar{q}) +  g \to V + q(\bar{q})$
  
\begin{eqnarray}
\label{eq:dXS0dqt_V_gq}
%\frac{d\hat\sigma_{qg}}{dz}
\hat\sigma_{qg}^{\sss > (1)}(z)
&=&  \int_{\(\qtcut\)^2}^{\(\qtmax\)^2}\!\!d\qt^2 \,
%\frac{d^2\hat\sigma_{qg}(\qt,z)}{d\qt^2\,dz}
\frac{d\hat\sigma_{qg}^{\sss (1)}(\qt,z)}{d\qt^2}
\nonumber\\
&=& \sigma_{qq}^{\sss (0)} %\asotpi
\,\TR\, z\lg\frac{1}{2}(1+3z)(1-z)\,\sqrt{1-\frac{4az}{(1-z)^{2}}} \right.
\nonumber\\
&&\hspace{0.7cm} \left. +\,\APreg_{qg}(z)\!
\lq -\log{\frac{az}{(1-z)^{2}}}+2\log \frac{1}{2} \(
\sqrt{1-\frac{4az}{(1-z)^2}}+1 \) \rq\rg 
\nonumber \\
&=& \sigma_{qq}^{\sss (0)} %\asotpi
\,\TR\, z\lg\frac{1}{2}(1+3z)(1-z)\,\sqrt{1-\piT^2} + \APreg_{qg}(z)
\log\frac{1+\sqrt{1-\piT^2}}{1-\sqrt{1-\piT^2}} \rg\! ,\phantom{aaaa}
\end{eqnarray}

\item $q + \bar{q}  \to V + g$
  
  \begin{eqnarray}
    \label{eq:dXS0dqt_V_qq}
\hat\sigma_{q\bar q}^{\sss >(1)}(z)
&=& \int_{\(\qtcut\)^2}^{\(\qtmax\)^2}\!\!
%\frac{d^2\hat\sigma_{q\bar q}(\qt,z)}{d\qt^2\,dz}
\frac{d\hat\sigma_{q\bar q}^{\sss (1)}(\qt,z)}{d\qt^2}
\nonumber\\
&=&  \sigma_{qq}^{\sss (0)} %\asotpi
\, \CF \, z
\lg - 2\, (1-z)\,\sqrt{1-\frac{4az}{(1-z)^{2}}} \right.
\nonumber \\
&& \hspace{0.7cm}\left.{} + 2 \,\APnoreg_{qq}(z)\! \lq- \log\frac{az}{(1-z)^{2}} + 2
\log \frac{1}{2} \(  \sqrt{1-\frac{4az}{(1-z)^2}}+1 \) \rq \!\rg \phantom{aaaaaa}
\nonumber \\
&=&  \sigma_{qq}^{\sss (0)} %\asotpi
\, \CF \, z
\lg - 2\, (1-z)\,\sqrt{1-\piT^2}  +
2 \,\APnoreg_{qq}(z) \log\frac{1+\sqrt{1-\piT^2}}{1-\sqrt{1-\piT^2}} \rg \! .
\end{eqnarray}

\end{itemize}

\subsubsection*{$\boldsymbol{H}$ production}

\begin{itemize}
 \item $g + q(\bar q) \to H + q(\bar q)$ 
   \begin{eqnarray}
\label{eq:dXS0dqt_H_qg}
%  \frac{d\hat\sigma_{gq}}{dz}
  \hat\sigma_{gq}^{\sss >(1)}(z)
&=&\int_{\(\qtcut\)^2}^{\(\qtmax\)^2}\!\!
%  \frac{d^2\hat\sigma_{gq}(\qt,z)}{d\qt^2\,dz}
  \frac{d\hat\sigma_{gq}^{\sss (1)}(\qt,z)}{d\qt^2}
\nonumber\\
&=& \sigma_{gg}^{\sss (0)} %\asotpi
\,\CF\,  z
\lg  -\frac{3(1-z)^2}{2z}\,\sqrt{1-\frac{4az}{(1-z)^2}} \right. 
  \nonumber \\
&& \hspace{0.7cm} \left. {} + \APreg_{gq}(z) \!
  \lq- \log\frac{az}{(1-z)^2} + 2 \log \frac{1}{2}
  \(\sqrt{1-\frac{4az}{(1-z)^2}}+1 \) \rq \rg \phantom{aaaaa}
\nonumber \\
&=&
\sigma_{gg}^{\sss (0)} %\asotpi
\,\CF\,  z
\lg  -\frac{3(1-z)^2}{2z}\,\sqrt{1- \piT^2}  + \APreg_{gq}(z) 
 \log\frac{1+\sqrt{1-\piT^2}}{1-\sqrt{1-\piT^2}}  \rg\!,
  \end{eqnarray}

  \item  $g + g \to H + g$

\begin{eqnarray}
\label{eq:dXS0dqt_H_gg}
%\frac{d\hat\sigma_{gg}}{dz}
\hat\sigma_{gg}^{\sss >(1)}(z)
&=& \int_{\(\qtcut\)^2}^{\(\qtmax\)^2}\!\!
%\frac{d^2\hat\sigma_{gg}(\qt,z)}{d\qt^2\,dz}
\frac{d\hat\sigma_{gg}^{\sss (1)}(\qt,z)}{d\qt^2} 
\nonumber\\
&=& 
\sigma_{gg}^{\sss (0)} %\asotpi
\,\CA \, z
\lg -\frac{4(1-z)^{3}}{z}\,\sqrt{1-\frac{4az}{(1-z)^{2}}} \right.
  \nonumber\\
  &&\hspace{0.7cm} {} + 4 \,z \lq\frac{1-z}{2z}\,\frac{(1-z)^{2}}{6z}\,
  \sqrt{1-\frac{4az}{(1-z)^{2}}}\,\(1+\frac{2az}{(1-z)^{2}}\) \rq
  \nonumber\\
  &&\hspace{0.7cm}{} \left.{} + 2 \, \APnoreg_{gg}(z) 
  \lq -\log{\frac{az}{(1-z)^{2}}}+2\log{\frac{1}{2} \(
    \sqrt{1-\frac{4az}{(1-z)^{2}}}+1 \)} \rq \rg \phantom{aaaaa}
\nonumber  \\
  &=&
  \sigma_{gg}^{\sss (0)} %\asotpi
\,\CA \, z\!
\lg\! -\frac{11}{3} \frac{(1-z)^3}{z}\( 1 - \frac{\piT^2}{22}\) \sqrt{1-\piT^2} 
  + 2 \, \APnoreg_{gg}(z)
  \log\frac{1+\sqrt{1-\piT^2}}{1-\sqrt{1-\piT^2}} \rg\! .
  \nonumber\\
   \end{eqnarray}
   
\end{itemize}
We do not consider the process $q\bar{q} \to H g$ since it is not singular in
the limit $\qt \to 0$ and the corresponding analytic/numeric integration in
the transverse momentum can be performed setting explicitly $\qtcut = 0$.

A couple of further comments about the above expressions are also in
order. In first place, the part of the cross sections proportional to the
Altarelli--Parisi splitting functions in
eqs.~(\ref{eq:dXS0dqt_V_gq})--(\ref{eq:dXS0dqt_H_gg}) has a universal origin,
due to the factorisation of the collinear singularities on the underlying
Born.  The rest of the above cross sections is, in general, not universal.
In addition, for Higgs boson production, the NLO cumulative cross sections
that we have computed coincide exactly with the jet-vetoed cross sections
$\sigma^{\rm veto}(\pt^{\rm veto})$ of ref.~\cite{Catani:2001cr}, provided we
identify $\qtcut=\pt^{\rm veto}$.

\subsection[Extending the integration in ${z}$]{Extending the integration in
  $\boldsymbol{z}$} 
\label{sec:z_integration}

According to eq.~(\ref{eq:sigma_had_z_one}), in order to compute the hadronic
cross section we need to integrate the partonic cross sections convoluted
with the corresponding luminosities.  In the calculation of the total cross
sections, the upper limit in the $z$ integration is unrestricted and is equal
to 1. When a cut on the transverse momentum $\qt$ is applied, the reality of
eqs.~(\ref{eq:dXS0dqt_V_gq})--(\ref{eq:dXS0dqt_H_gg}) imposes the non
negativity of the argument of the square roots, i.e.
\begin{equation}
1- \piT^2\ge 0\,,
\end{equation}
that in turn gives
\begin{equation}
z \le \zmax \equiv 1-f(a)\,, \qquad \qquad f(a) \equiv 2 \sqrt{a}\(\sqrt{1 +
  a} - \sqrt{a}\). 
\end{equation}
Since our aim is to make contact with the transverse-momentum subtraction
formulae, that describe the behaviour of the cross sections in the soft and
collinear limits, we need to extend the integration range of the $z$ variable
up to 1, i.e.~the upper integration limit of $z$ in a Born-like
kinematics. In fact, only in the $z \to 1$ limit we recover the
logarithmic structure from the soft region of the emission.
In order to obtain explicitly all the logarithmic-enhanced terms in the
small-$\qtcut$ limit, we have then to expand our results in powers of $a$.
Since both the integrand and the upper limit of the integral depend on $a$,
the na\"ive approach of expanding only the integrand does not work, due to
the appearance of divergent terms in the $z \to 1$ limit, that have to be
handled with the introduction of plus distributions.

Using the notation of ref.~\cite{Catani:2011kr}, we first introduce the
function $\hat{R}_{ab}(z)$, defined by
\begin{equation}
  \label{eq:Rhat_def}
 \sigma_{ab}^{\sss <} =
 \tau \int_\tau^{1-f(a)} \frac{dz}{z}  \, {\cal L}_{ab}\!\(\frac{\tau}{z}\) 
 \frac{1}{z} \,\hat\sigma_{ab}^{\sss <}(z)
 \equiv
  \tau \int_\tau^1 \frac{dz}{z}  \, {\cal L}_{ab}\!\(\frac{\tau}{z}\) 
  \hat{\sigma}^{\sss (0)} \hat{R}_{ab}^{\sss}(z)\,,
\end{equation}
where the upper integration limit in $z$ in the last integral is exactly 1
and $\hat{\sigma}^{\sss (0)}$ is the partonic Born-level cross section for
the production of the colourless system $F$.  The function $\hat{R}_{ab}(z)$
can be written as a perturbative expansion in~$\as$
\begin{equation}
  \label{eq:Rhat_expans}
\hat{R}_{ab}(z) = \delta_{\rm\sss B}\,\delta(1-z) + \sum_{n=1}^\infty
\(\frac{\as}{2\pi}\)^n \hat R_{ab}^{\sss(n)}(z)\,, 
\end{equation}
where the $\delta(1-z)$ term is the Born-level contribution, and
$\delta_{\sss\rm B}=1$ when partons $a$ and $b$ are such that $a+b \to F$ is
a possible Born-like process, otherwise its value is 0.

The coefficient functions $\hat R_{ab}^{\sss(n)}(z)$ can be computed as power
series in $a$. It is in fact well known in the
literature~\cite{Catani:2011kr} that the NLO coefficient $\hat
R_{ab}^{\sss(1)}(z)$ has the following form\footnote{The notation for the
  expansion of $R_{ab}^{(1)}(z)$ follows from the number of powers of $\as$,
  $\log(a)$ and $a^\frac{1}{2}$, i.e.
  \begin{equation}
  \label{Rhat_coeff}
  \hat{R}_{ab}^{\sss (1)}(z) =  \sum_{m,r} \,\log^m(a) \, a^\frac{r}{2}
  \hat{R}_{ab}^{\sss(1,m,r)}(z) \,. 
  \nonumber
\end{equation}
In refs.~\cite{Catani:2011kr, Bozzi:2005wk}, the leading-logarithmic
$R_{ab}^{(1,2,0)}(z)$ and next-to-leading-logarithmic $R_{ab}^{(1,1,0)}(z)$
coefficient functions are directly associated to $\Sigma_{c \bar{c} \ito
  ab}^{F(1;2)}(z)$ and $\Sigma_{c \bar{c} \ito ab}^{F(1;1)}(z)$,
respectively. The hard-virtual coefficient function ${\cal H}_{c \bar{c} \ito
  ab}^{F(1)}$ corresponds to $R_{ab}^{(1,0,0)}(z)$. }
\begin{equation}
\label{eq:Rhat_one_expans}
\hat{R}_{ab}^{\sss(1)}(z) = \log^2(a) \, \hat R_{ab}^{\sss(1,2,0)}(z) +
\log(a)  \, \hat  R_{ab}^{\sss(1,1,0)}(z) + 
\hat R_{ab}^{\sss(1,0,0)}(z) + \ord{a^\frac{1}{2} \log{a}} ,
\end{equation}
and the aim of this paper is to compute the first unknown terms in
eq.~(\ref{eq:Rhat_one_expans}) that were neglected in
refs.~\cite{Catani:2011kr} and~\cite{Catani:2012qa}, namely
$R_{ab}^{(1,m,r)}(z)$, for $r$ up to 4 and for any $m$.

In a way similar to what was done in eq.~(\ref{eq:Rhat_def}) for
$\hat{R}_{ab}^{\sss}(z)$, we introduce the function $\hat{G}_{ab}^{\sss}(z)$ 
defined by
\begin{equation}
  \label{eq:Ghat_def}
  \sigma_{ab}^{\sss >} =
 \tau \int_\tau^{1-f(a)} \frac{dz}{z}  \, {\cal L}_{ab}\!\(\frac{\tau}{z}\) 
 \frac{1}{z} \,\hat\sigma_{ab}^{\sss >}(z)
 \equiv
  \tau \int_\tau^1 \frac{dz}{z}  \, {\cal L}_{ab}\!\(\frac{\tau}{z}\) 
  \hat{\sigma}^{\sss (0)} \, \hat{G}_{ab}(z)\,.
\end{equation}
Since
\begin{equation}
  \sigma_{ab}^{\sss <}  + \sigma_{ab}^{\sss >} =
 \tau \int_\tau^1 \frac{dz}{z}  \, {\cal L}_{ab}\!\(\frac{\tau}{z}\)
 \hat{\sigma}^{\rm\sss tot}_{ab}(z)
 \equiv \sigma^{\rm\sss tot}_{ab}\,,
\end{equation}
and $\sigma^{\rm\sss tot}_{ab}$ is independent of $a$, the coefficients of
the terms that vanish in the small-$\qt$ limit in the series expansion in $a$
of $\hat{R}_{ab}(z)$ and $\hat{G}_{ab}(z)$ are equal but with opposite sign,
at any order in $\as$. We recall that $\hat{R}_{ab}(z)$ contains terms of the
form $\delta(1-z)$, coming from the Born and the virtual contributions, that
are independent of $a$ and are obviously absent in $\hat{G}_{ab}(z)$.

In the rest of the paper we compute the first terms of the expansion in $a$
of $\hat{G}_{ab}^{\sss(1)}(z)$, that will be obtained from the following
identity
\begin{equation}
 \label{eq:Ghat_one_def} 
 \sigma_{ab}^{\sss > (1)} = \tau \int_\tau^{1-f(a)} \frac{dz}{z}  \, {\cal
   L}_{ab}\!\(\frac{\tau}{z}\)  
 \frac{1}{z} \,\hat\sigma_{ab}^{\sss >(1)}(z)=
 \tau \int_\tau^1 \frac{dz}{z}  \, {\cal L}_{ab}\!\(\frac{\tau}{z}\) 
  \hat{\sigma}^{\sss (0)} \, \hat{G}_{ab}^{\sss(1)}(z)  \,.
\end{equation}
We have elaborated a process-independent formula to transform an integral of
the form of the first one in eq.~(\ref{eq:Ghat_one_def}) into the form of the
second one, producing the series expansion of $\hat{G}_{ab}^{\sss(1)}(z)$ in
$a$.
The application of our formula reorganizes the divergent terms in the $z \to
1$ limit into terms that are integrable up to $z=1$ and logarithmic terms in
$a$.
Since this is a very technical procedure, we have collected all the details
in appendix~\ref{app:integration_recipe}, and we refer the interested reader
to that appendix for the description of the method.

\section{Results}
\label{sec:results}

In this section we summarize our findings.  We present in
sec.~\ref{sec:res_Ghat} the analytic results for the $\hat{G}_{ab}^{\sss
  (1)}(z)$ functions we have computed.  In the calculation of these
functions, we kept trace of all the terms originating from the manipulation
of the contributions proportional to the Altarelli--Parisi splitting
functions, in the partonic cross sections of
eqs.~(\ref{eq:dsigmahat_V_qg})--(\ref{eq:dsigmahat_H_gg}). These terms
constitute what we call the ``universal part'' of our results, as detailed in
secs.~\ref{sec:part_diff_XS} and~\ref{sec:qt_integration}.  We will indicate
these terms with the superscript ``U'', while the remaining terms will have a
superscript ``R''. 
We stress here that the distinction between universal and non-universal part
is purely formal, and it does not have a physical implication.  The reason of
this separation is to have hints on the general structure of the $\qtcut$
dependence of inclusive cross sections for the production of arbitrary
colorless systems.
We comment on the results that we have obtained in sec.~\ref{sec:comments}.

In sec.~\ref{sec:LA_numerica} we study the numerical significance of the
power-correction terms we have computed, discussing first their impact on the
different production channels for Drell--Yan $Z$ boson and Higgs boson
production in gluon fusion. Then we present their overall effect, normalising
the results with respect to the total NLO cross section, in order to have a
better grasp on the size of these contributions.

\subsection[Results for the ${\hat{G}_{ab}^{\sss (1)}(z)}$
  functions]{Results for the $\boldsymbol{\hat{G}_{ab}^{\sss (1)}(z)}$
  functions}
\label{sec:res_Ghat}

We indicate with $\gu_{ab}(z)$ the universal part of the $\hat{G}_{ab}^{\sss
  (1)}(z)$ functions, and with $\gd_{ab}(z)$ the remaining part, stripped off
of a common colour factor.  Our expressions for $\hat{G}^{\sss (1)}_{ab}(z)$
contain derivatives of the Dirac $\delta$ function, $\delta^{(n)}(z)$, up to
$n=5$, and plus distributions up to order~5. We report here the definition of
a plus distribution of order~$n$
\begin{equation}
\int_0^1 dz\, l(z) \lq g(z) \rq_{n+} \equiv \int_0^1 dz \lg
l(z) - \sum_{i=0}^{n-1}  \frac{1}{i!} \, l^{(i)}(1)\, (z-1)^i \rg  g(z) \,,
\end{equation}
where $g(z)$ has a pole of order $n$ for $z=1$, and $l(z)$ is a continuous
function in $z=1$, together with all its derivatives up to order $(n-1)$.
For completeness, we collect in appendix~\ref{app:plus_distribs} more details
on the plus distributions, and the identities we have used to simplify our
results.

\subsection*{$\boldsymbol{V}$ production}

%\subsubsection*{$\boldsymbol{qg}$ channel}

\begin{itemize}

\item $q(\bar{q}) +  g \to V + q(\bar{q})$

\begin{equation}
  \label{eq:Ghat_V_qg}
\hat{G}^{ \sss (1)}_{qg}(z)= \TR \, \hat{g}^{\sss (1)}_{qg}(z)\,, \hspace{2cm}
\hat{g}^{\sss (1)}_{qg}(z) =   \gu_{qg}(z) + \gd_{qg}(z) \, ,
\end{equation}
where
%% \begin{eqnarray}
%% \label{eq:ghat_V_qg}
%% \hat{g}^{\sss (1)}_{qg}(z) &=& {}  -\APreg_{qg}(z)  \log(a)   -  \APreg_{qg}(z)\,
%% \log\frac{z}{(1-z)^{2}} + \frac{1}{2}\, (1+3z) (1-z)\,
%% \nonumber \\
%% &&{} + \lg \delta^{(1)}(1 - z) - \delta(1 - z) \rg a \log(a)
%% \nonumber \\
%% &&{} - \lg   \delta(1 - z) + 2 \, z\, \APreg_{qg}(z) \lq \frac{1}{(1-z)^2}\rq_{2+}
%% +  z (1+3z) \lq \frac{1}{1-z} \rq_+ \rg a
%% \nonumber \\
%% &&{} + \lg   - 4 \,\delta(1 - z) + 5 \,\delta^{(1)}(1 - z) - 2 \,\delta^{(2)}(1 - z)
%% + \frac{1}{4}\, \delta^{(3)}(1 - z) \rg a^2 \log(a)
%% \nonumber \\
%% &&{} + \lg \delta^{(1)}(1 - z) - \frac{3}{4}\, \delta^{(2)}(1 - z) +
%% \frac{1}{6}\, \delta^{(3)}(1-z) \right .
%% \nonumber\\
%% &&{} \left . \hspace{7mm} -z^2   (1+3z)  \lq \frac{1}{(1-z)^3}\rq_{3+} 
%%  - 3 \, z^2  \APreg_{qg}(z) \lq \frac{1}{(1-z)^4} \rq_{4+} \rg a^2
%% \nonumber \\
%% &&{} + \comment{\ord{a^\frac{5}{2}\log(a)}} \,,\phantom{aaa}
%% \end{eqnarray}

\begin{eqnarray}
   \label{eq:ghat_un_V_qg}  
\gu_{qg}(z)   &=& {} - \APreg_{qg}(z) \log(a)  -  \APreg_{qg}(z)\, \log\frac{z}{(1-z)^{2}}
\nonumber \\
&&{} + \lg \delta^{(1)}(1 - z) - 3\,\delta(1 - z) \rg a \log(a)
\nonumber \\
&&{} + \lg   \delta(1 - z) - 2 \, z\, \APreg_{qg}(z) \lq \frac{1}{(1-z)^2}\rq_{2+} \rg a
\nonumber \\
&&{} + \lg  - 9\, \delta(1 - z) + \frac{21}{2}\, \delta^{(1)}(1 - z)  - 3
\,\delta^{(2)}(1 - z) + \frac{1}{4} \,\delta^{(3)}(1 - z) \!  \rg \! a^2 \log(a) 
\nonumber \\
&&{} + \lg + 2 \,\delta(1 - z)   + \frac{7}{4} \,\delta^{(1)}(1 - z) -
\frac{5}{4} \,\delta^{(2)}(1 - z) + \frac{1}{6}\, \delta^{(3)}(1 - z)
\right.  
\nonumber\\
&&{} \left . \hspace{7mm}
 - 3 \, z^2  \APreg_{qg}(z) \lq \frac{1}{(1-z)^4} \rq_{4+} \rg a^2
 + \ord{a^\frac{5}{2}\log(a)}, \phantom{aaa}
\end{eqnarray}

\begin{eqnarray}
 \label{eq:ghat_nun_V_qg}  
  \gd_{qg}(z)  &=&
  \frac{1}{2}\, (1+3z) (1-z)
   - z (1+3z) \lq \frac{1}{1-z} \rq_+  \!\! a -z^2   (1+3z)  \lq
 \frac{1}{(1-z)^3}\rq_{3+} \!\! a^2
\nonumber \\
&&{} + 2\, \delta(1 - z) \, a \log(a) - 2 \, \delta(1 - z) \, a
\nonumber \\
&&{} + \lg 5 \,\delta(1 - z)  - \frac{11}{2}\, \delta^{(1)}(1 - z) +
\delta^{(2)}(1 - z) \rg a^2 \log(a) 
\nonumber \\
&&{} + \lg - 2\, \delta(1 - z)  - \frac{3}{4}\, \delta^{(1)}(1 - z) +
\frac{1}{2}\, \delta^{(2)}(1 - z) \rg a^2  +
\ord{a^\frac{5}{2}\log(a)}, \phantom{aaaa}
\end{eqnarray}
and 
\begin{equation}
 \APreg_{qg}(z) = 2z^2-2z+1 \,.
\end{equation}

%\subsubsection*{$\boldsymbol{q\bar{q}}$ channel}
\item $q + \bar{q}  \to V + g$

\begin{equation}
  \label{eq:Ghat_V_qq}  
  \hat{G}^{\sss (1)}_{q\bar{q}}(z)  = \CF \, \hat{g}^{\sss (1)}_{q\bar{q}}(z)\,,
  \hspace{2cm} 
 \hat{g}^{\sss (1)}_{q\bar{q}}(z) =  \gu_{q\bar{q}}(z)+ \gd_{q\bar{q}}(z) \,,
\end{equation}
where
%% \begin{eqnarray}
%% \label{eq:ghat_V_qq}  
%% \hat{g}^{\sss (1)}_{q\bar{q}}(z) &=&   \delta(1 - z)  \log^2(a) + \big\{3\,
%% \delta(1 - z) - 2 \, \APreg_{qq}(z)\big\} \log(a) 
%% \nonumber \\
%% &&{}   - \frac{\pi^2}{3} \delta(1 - z)  -  2 \,  (1-z)\,
%% - 2  \, \APnoreg_{qq}(z) \log(z)\, 
%% + 4  \,  (1-z) \, \hat{p}_{qq}(z)  \lq \frac{\log(1-z)}{1-z} \rq_+
%% \nonumber \\
%% &&{} + \lg 4\, \delta(1 - z)   - 8\, \delta^{(1)}(1 - z) + 2 \,\delta^{(2)}(1 - z)
%% \rg a \log(a)
%% \nonumber \\
%% &&{} + \lg   - 4\, \delta(1 - z) +  4 \,\delta^{(1)}(1 - z)
%% + 4\,z  \lq \frac{1}{1-z} \rq_+
%% \right.
%% \nonumber \\
%% && {} \left.\phantom{1cm} -4  \,z \, (1-z) \, \hat{p}_{qq}(z)  \lq
%% \frac{1}{(1-z)^3}\rq_{3+} \rg a 
%% \nonumber \\
%% &&{} + \lg  \delta(1 - z)   - 8\, \delta^{(1)}(1 - z) +
%% \frac{19}{2} \,\delta^{(2)}(1 - z) - 3 \,\delta^{(3)}(1 - z)
%% \right.
%% \nonumber \\
%% && {} \left.\phantom{1cm}
%% + \frac{1}{4} \,\delta^{(4)}(1 - z) \!\rg  a^2 \log(a)
%% \nonumber \\
%% &&{} + \lg - 2\, \delta(1 - z) + 6 \,\delta^{(1)}(1 - z) - \delta^{(2)}(1 -
%% z)  - \delta^{(3)}(1 - z)
%% + \frac{1}{6} \,\delta^{(4)}(1 - z)   \right.
%% \nonumber\\
%% && {} \left. \phantom{1cm} + 4   z^2 \, \lq\frac{1}{(1-z)^3}\rq_{3+}
%% - 6  z^2 (1-z)\, \APnoreg_{qq}(z) \lq \frac{1}{(1-z)^5}\rq_{5+}
%% \rg a^2
%% \nonumber \\
%% && {} +\comment{\ord{a^{\frac{5}{2}}\log(a)}} \,,
%% \end{eqnarray}
\begin{eqnarray}
\label{eq:ghat_un_V_qq} 
\gu_{q\bar{q}}(z) &=&   \delta(1 - z)  \log^2(a)
% + \big\{3\, \delta(1 - z) - 2 \, \APreg_{qq}(z)\big\} \log(a)
 - 2 \, \APreg_{qq}(z) \log(a) 
\nonumber \\
&&{}   - \frac{\pi^2}{3} \delta(1 - z) 
- 2  \, \APnoreg_{qq}(z) \log(z)\, 
+ 4  \,  (1-z) \, \hat{p}_{qq}(z)  \lq \frac{\log(1-z)}{1-z} \rq_+
\nonumber \\
&&{} + \lg 6\, \delta(1 - z)   - 8\, \delta^{(1)}(1 - z) + 2 \,\delta^{(2)}(1 - z)
\rg a \log(a)
\nonumber \\
&&{} + \lg   - 6\, \delta(1 - z) +  4 \,\delta^{(1)}(1 - z)
-4  \,z \, (1-z) \, \hat{p}_{qq}(z)  \lq \frac{1}{(1-z)^3}\rq_{3+} \rg a
\nonumber \\
&&{} + \lg 3 \, \delta(1 - z)   - 12 \, \delta^{(1)}(1 - z)
+ \frac{21}{2}\, \delta^{(2)}(1 - z) - 3 \, \delta^{(3)}(1 - z) \right.
\nonumber \\
&&{} \hspace{1cm}
\left. + \frac{1}{4}\, \delta^{(4)}(1 - z)   \!\rg  a^2 \log(a)
\nonumber \\
&&{} + \lg - 4\, \delta(1 - z)  + 6 \,\delta^{(1)}(1 - z) - \frac{1}{2}\,\delta^{(2)}(1 - z)
- \delta^{(3)}(1 - z)   \right. 
\nonumber\\
&& {}  \hspace{1cm} \left. + \frac{1}{6}\, \delta^{(4)}(1 - z)  - 6  z^2 (1-z)\, \APnoreg_{qq}(z) \lq
\frac{1}{(1-z)^5}\rq_{5+} \rg a^2
\nonumber \\
&& {} + \ord{a^{\frac{5}{2}}\log(a)} ,
\end{eqnarray}
\begin{eqnarray}
  \label{eq:ghat_nun_V_qq} 
\gd_{q\bar{q}}(z) &=& {}  -  2 \,  (1-z) + 4\,z  \lq \frac{1}{1-z} \rq_+ \!\!  a +
4  \, z^2 \, \lq\frac{1}{(1-z)^3}\rq_{3+}\!\!  a^2
\nonumber \\
&&{} - 2 \,\delta(1 - z) \, a \log(a)  +  2 \,\delta(1 - z) \,  a
\nonumber \\
&&{} + \lg - 2\, \delta(1 - z) + 4 \, \delta^{(1)}(1 - z) - \delta^{(2)}(1 - z)   \rg a^2 \log(a)
\nonumber \\
&&{} + \lg  2 \,\delta(1 - z) - \frac{1}{2}\, \delta^{(2)}(1 - z)  \rg a^2
+ \ord{a^{\frac{5}{2}}\log(a)},
\end{eqnarray}
and
\begin{equation}
  \APnoreg_{qq}(z) = \frac{1+z^2}{1-z}\,, \qquad \qquad
  \APreg_{qq}(z)= \frac{1+z^2}{(1-z)_+}  \,.
\end{equation}
In eq.~(\ref{eq:ghat_un_V_qq}) we have written the $(1+z^2)$ terms coming
from the numerator of the $ \APnoreg_{qq}(z)$ splitting function as
\begin{equation}
1+z^2 = (1-z) \, \APnoreg_{qq}(z)\,,
\end{equation}
in order to keep track of the universal origin of those terms.

\end{itemize}

\subsection*{$\boldsymbol{H}$ production}

\begin{itemize}

\item $g + q(\bar q) \to H + q(\bar q)$ 

\begin{equation}
  \label{eq:Ghat_H_gq}
  \hat{G}^{\sss (1)}_{gq}(z) = \CF \, \hat{g}^{\sss (1)}_{gq}(z)\,,
  \hspace{2cm} 
  \hat{g}^{\sss (1)}_{gq}(z) =  \gu_{gq}(z) + \gd_{gq}(z)\,,
\end{equation}
where
%% \begin{eqnarray}
%% \label{eq:ghat_H_gq}
%% \hat{g}^{\sss (1)}_{gq}(z) &=&
%% - \APreg_{gq}(z)  \log(a) - \frac{3}{2z}  (1-z)^2
%% - \APreg_{gq}(z)\log\frac{z}{(1-z)^{2}} 
%% \nonumber \\
%% &&{} + \delta^{(1)}(1 - z)\,  a \log(a)
%%  + \lg \delta(1 - z)   + 3    - 2\, z \, \APreg_{gq}(z) \lq
%% \frac{1}{(1-z)^2}\rq_{2+}\rg a
%% \nonumber \\
%% &&{} + \lg  - \frac{3}{4}\, \delta^{(2)}(1 - z) + \frac{1}{4}\, \delta^{(3)}(1 - z) \rg a^2 \log(a)
%% \nonumber \\
%% &&{} + \lg - \frac{1}{2}\, \delta(1 - z)  - \frac{1}{2}\, \delta^{(1)}(1 - z) 
%% + \frac{1}{4}\, \delta^{(2)}(1 - z) + \frac{1}{6}\, \delta^{(3)}(1 - z)    \right.
%% \nonumber \\
%% &&{} \left. \phantom{1cm}  +  3 \,  z   \lq \frac{1}{(1-z)^2}\rq_{2+}
%% - 3 \,  z^2 \, \APreg_{gq}(z) \lq \frac{1}{(1-z)^4} \rq_{4+} \rg a^2
%% \nonumber \\
%% &&{} + \comment{\ord{a^\frac{5}{2}\log(a)}} \,, \phantom{aaaaa}
%% \end{eqnarray}
%
\begin{eqnarray}
  \label{eq:ghat_un_H_gq} 
\gu_{gq}(z) &=&
- \APreg_{gq}(z)  \log(a) 
- \APreg_{gq}(z)\log\frac{z}{(1-z)^{2}} 
\nonumber \\
&&{} + \delta^{(1)}(1 - z)\,  a \log(a)
 + \lg \delta(1 - z)      - 2\, z \, \APreg_{gq}(z) \lq
\frac{1}{(1-z)^2}\rq_{2+}\rg a
\nonumber \\
&&{} + \lg - \frac{3}{2} \,\delta(1 - z) + \frac{3}{2}\, \delta^{(1)}(1 - z)  - \frac{3}{4}\, \delta^{(2)}(1 -
z) + \frac{1}{4} \,\delta^{(3)}(1 - z) \rg a^2 \log(a) 
\nonumber \\
&&{} + \lg   \frac{1}{4} \,\delta(1 - z)  +
\frac{1}{4} \,\delta^{(1)}(1 - z) + \frac{1}{4} \,\delta^{(2)}(1 - z)+ \frac{1}{6} \,\delta^{(3)}(1 - z)
\right. 
\nonumber \\
&&{} \left. \phantom{1cm}  
- 3 \,  z^2 \, \APreg_{gq}(z) \lq \frac{1}{(1-z)^4} \rq_{4+} \rg a^2
\nonumber \\
&&{} + \ord{a^\frac{5}{2}\log(a)} , \phantom{aaaaa}
\end{eqnarray}

\begin{eqnarray}
  \label{eq:ghat_nun_H_gq}
\gd_{gq}(z) &=& - \frac{3}{2z}  (1-z)^2 + 3\, a +  3 \,  z   \lq
\frac{1}{(1-z)^2}\rq_{2+} \!\!  a^2
\nonumber \\
&&{} + \frac{3}{2} \lg  \delta(1 - z)   -  \delta^{(1)}(1 - z) \rg a^2 \log(a)
- \frac{3}{4} \lg \delta(1 - z)   + \delta^{(1)}(1 - z) \rg a^2
\nonumber \\
&&{} + \ord{a^\frac{5}{2}\log(a)},
\end{eqnarray}
and   
\begin{equation}
\APreg_{gq}(z) = \frac{z^2-2z+2}{z} \,.
\end{equation}

\item  $g + g \to H + g$

\begin{equation}
  \label{eq:Ghat_H_gg}
  \hat{G}^{\sss (1)}_{gg}(z) = \CA \, \hat{g}^{\sss (1)}_{gg}(z)\,,
  \hspace{2cm}
 \hat{g}^{\sss (1)}_{gg}(z) =  \gu_{gg}(z) + \gd_{gg}(z)\,,
\end{equation}
where
%% \begin{eqnarray}
%% \label{eq:ghat_H_gg}
%%  \hat{g}^{\sss (1)}_{gg}(z) &=& \delta(1 - z)  \log^2(a)
%% -2  \, (1-z) \, \APnoreg_{gg}(z) \lq \frac{1}{1-z} \rq_+ \log(a)
%%   - \frac{\pi^2}{3} \delta(1 - z) 
%% %% \nonumber\\
%% %% &&{} - \frac{11}{3 z} \, (1-z)^3
%% %% + 2 \, (1-z) \, \APnoreg_{gg}(z)  \lg  2 \lq \frac{\log(1-z)}{1-z} \rq_+ -
%% %%  \log(z)  \lq\frac{1}{1-z} \rq_+ \rg
%% \nonumber\\
%% &&{} - \frac{11}{3 z} \, (1-z)^3
%%  + 4 \, (1-z) \, \APnoreg_{gg}(z)  \lq \frac{\log(1-z)}{1-z} \rq_+ - 2 \, \APnoreg_{gg}(z)
%%  \log(z)  
%% \nonumber \\
%% &&{} + \Big\{ 12 \,\delta(1 - z)  - 8 \, \delta^{(1)}(1 - z)  + 2  \,\delta^{(2)}(1 - z) \Big\}\, a \log(a)
%% \nonumber\\
%% &&{} + \lg - 6\, \delta(1 - z) + 4\, \delta^{(1)}(1 - z)  + 8\,(1-z)
%% \phantom{\frac{1}{1}} \right.
%% \nonumber\\
%% &&{} \left. \phantom{1cm} - 4\, z \,(1-z) \,\APnoreg_{gg}(z)   \lq \frac{1}{(1-z)^3}\rq_{3+} \rg a
%% \nonumber\\
%% &&{}
%% + \lg 15 \, \delta(1 - z) - 30 \, \delta^{(1)}(1- z) + 15 \, \delta^{(2)}(1 - z)
%% - 3\, \delta^{(3)}(1 - z) \phantom{\frac{1}{1}} \right.
%% \nonumber\\
%% &&{} \left. \phantom{1cm} + \frac{1}{4} \,\delta^{(4)}(1 - z) \rg a^2 \log(a)
%% \nonumber \\
%% &&{} + \lg - 10 \,\delta(1 - z) + 3 \,\delta^{(1)}(1 - z)  + \frac{5}{2}\, \delta^{(2)}(1 - z)
%% - \delta^{(3)}(1 - z)    \right.
%% \nonumber\\
%% && {}\left.  \phantom{1cm} + \frac{1}{6} \,\delta^{(4)}(1 - z) + 6 \, z \! \lq \frac{1}{1-z}
%% \rq_+ \! \!  - 6\, z^2 \, (1-z)  \,\APnoreg_{gg}(z)  \!  \lq 
%% \frac{1}{(1-z)^5}\rq_{5+} \rg a^2 
%% \nonumber\\
%% && {} + \comment{\ord{a^\frac{5}{2}\log(a)}} \,,
%% \end{eqnarray}

\begin{eqnarray}
\label{eq:ghat_un_H_gg}
\gu_{gg}(z) &=& \delta(1 - z)  \log^2(a)
-2   \, \APreg_{gg}(z)  \log(a) 
\nonumber\\
&&{}
- \frac{\pi^2}{3} \delta(1 - z) - 2 \, \APnoreg_{gg}(z) \log(z)
+ 4 \, (1-z) \, \APnoreg_{gg}(z)  \lq \frac{\log(1-z)}{1-z} \rq_+  
\nonumber \\
&&{} + \Big\{ 12 \,\delta(1 - z)  - 8 \, \delta^{(1)}(1 - z)  + 2  \,\delta^{(2)}(1 - z) \Big\}\, a \log(a)
\nonumber\\
&&{} + \lg - 6\, \delta(1 - z) + 4\, \delta^{(1)}(1 - z) 
 - 4\, z \,(1-z) \,\APnoreg_{gg}(z)   \lq \frac{1}{(1-z)^3}\rq_{3+} \rg a
\nonumber\\
&&{} + \lg 18 \,\delta(1 - z) - 30\,\delta^{(1)}(1 - z) + 15\, \delta^{(2)}(1
- z) - 3 \,\delta^{(3)}(1 - z) \phantom{\frac{1}{1}}\right.
\nonumber \\
&&{} \hspace{1cm} \left. + \frac{1}{4}\, \delta^{(4)}(1 - z) \rg a^2 \log(a)
\nonumber \\
&&{} + \lg - \frac{15}{2}\, \delta(1 - z) + 3 \,\delta^{(1)}(1 - z) + \frac{5}{2} \,\delta^{(2)}(1 - z) -
\delta^{(3)}(1 - z) \right.
\nonumber \\
&&{} \hspace{1cm} \left. + \frac{1}{6} \,\delta^{(4)}(1 - z)   - 6\, z^2 \, (1-z)  \,\APnoreg_{gg}(z)   \lq
\frac{1}{(1-z)^5}\rq_{5+} \rg a^2
\nonumber\\
&& {} + \ord{a^\frac{5}{2}\log(a)} ,
\end{eqnarray}

\begin{eqnarray}
\label{eq:ghat_nun_H_gg}
\gd_{gg}(z) &=&  - \frac{11}{3 z} \, (1-z)^3 + 8\,(1-z) \, a + 6 \, z
\lq \frac{1}{1-z} \rq_+ \!\!  a^2
\nonumber \\
&&{}  - 3\, \delta(1 - z) \, a^2 \log(a)
 - \frac{5}{2} \, \delta(1 - z) \,  a^2 + \ord{a^\frac{5}{2}\log(a)},
\end{eqnarray}
and
\begin{equation}
\APnoreg_{gg}(z)= \frac{2 (z^2-z+1)^2}{z(1-z)}\,.
\end{equation}
In eq.~(\ref{eq:ghat_un_H_gg}) we have written the $2 (z^2-z+1)^2/z$ terms
coming from the numerator of the $\APnoreg_{gg}(z)$ splitting function as
\begin{equation}
\frac{2 (z^2-z+1)^2}{z} = (1-z) \, \APnoreg_{gg}(z)\,,
\end{equation}
in order to keep track of the universal origin of those terms.

\end{itemize}

\subsection[Comments on ${\hat{G}_{ab}^{\sss (1)}(z)}$]{Comments on
  $\boldsymbol{\hat{G}_{ab}^{\sss (1)}(z)}$} 
\label{sec:comments}
The leading-logarithmic~(LL) and next-to-leading-logarithmic~(NLL)
coefficients of the $\hat{G}_{ab}^{\sss (1)}(z)$ functions that we have
computed agree with the ones in the literature, along with the finite term.
Their values have been known for a while~\cite{Kodaira:1981nh, Catani:1988vd}
and are related to the perturbative coefficients of the transverse-momentum
subtraction/resummation formulae for $V$~\cite{Collins:1984kg} and Higgs
boson production~\cite{Catani:2010pd}, as pointed out in
sec.~\ref{sec:z_integration}.
The coefficients of the terms of order $a\log(a)$ and $a$, and of order
$a^2\log(a)$ and $a^2$ are instead the new results computed in this paper.

The general form of the $\hat{G}_{ab}^{\sss(1)}(z)$ functions we have
computed reads\footnote{The notation for the
  expansion of $G_{ab}^{(1)}(z)$ follows from the number of powers of $\as$,
  $\log(a)$ and $a^\frac{1}{2}$ (in the same way as for $\hat{R}_{ab}^{\sss
  (1)}(z)$), i.e.
\begin{equation}
  \label{Ghat_coeff}
  \hat{G}_{ab}^{\sss (1)}(z) =  \sum_{m,r} \,\log^m(a) \,
  \(a^\frac{1}{2}\)^r \hat{G}_{ab}^{\sss(1,m,r)}(z) \,. 
  \nonumber
\end{equation}
}
\begin{eqnarray}
  \label{Ghat_one_expans}
\hat{G}_{ab}^{\sss(1)}(z) &=& {} \log^2(a) \, \hat G_{ab}^{\sss(1,2,0)}(z) +
\log(a)  \, \hat  G_{ab}^{\sss(1,1,0)}(z) + \hat G_{ab}^{\sss(1,0,0)}(z)
\nonumber\\[2mm]
&& {} +   a\log(a) \,\hat G_{ab}^{\sss(1,1,2)}(z)  +  a \, \hat G_{ab}^{\sss(1,0,2)}(z)
\nonumber\\[2mm]
&& {} + a^2 \log(a) \,\hat G_{ab}^{\sss(1,1,4)}(z)  + a^2 \, \hat G_{ab}^{\sss(1,0,4)}(z)  +
\ord{a^\frac{5}{2}\log(a)} ,
\end{eqnarray}
all the other coefficients being zero.

We will refer to the terms in the first line of eq.~(\ref{Ghat_one_expans})
as leading terms~(LT). These terms are either logarithmically divergent or
finite in the $a\to 0$ limit.  We name the terms in the sum in the second
line of eq.~(\ref{Ghat_one_expans}) as next-to-leading terms~(NLT), and the
first two terms in the third line as next-to-next-to-leading terms~(N$^2$LT),
and so forth.

We notice that the NLT and N$^2$LT terms are at most linearly dependent on
$\log(a)$, consistently with the fact that the LL contribution is a squared
logarithm.
In addition, no odd-power corrections of $\sqrt{a} = \qtcut/Q$ appear in the
NLT and N$^2$LT terms. This behaviour is in agreement with what found, for
example, in ref.~\cite{Ebert:2018gsn}, i.e.~that, at NLO, the power expansion
of the differential cross section for colour-singlet production is in
$(\qtcut)^2$. We do not expect this to be true in general when cuts are
applied to the final state.

\subsubsection{Soft behaviour of the universal part}
\label{sec:soft_universal}
The origin of some of the terms in the diagonal channels, i.e.~the $q\bar{q}$
channel for $V$ production and the $gg$ channel for $H$ production, can be
traced back to the behaviour of the Altarelli--Parisi splitting functions in
the soft limit, i.e.~$z\to 1$.  In fact, in this limit,
\begin{equation}
  \hat{P}_{qq}(z) \approx \frac{2\CF}{1-z}\,, \qquad \qquad \hat{P}_{gg}(z)
  \approx \frac{2\CA}{1-z}\,,  
\end{equation}
so that
\begin{equation}
  \hat{p}_{qq}(z)  \approx \hat{p}_{gg}(z) \approx \frac{2}{1-z}\equiv\hat{p}(z)\,.
\end{equation}
Inserting $\hat{p}_{qq}(z)$ and $\hat{p}_{gg}(z)$ in
eqs.~(\ref{eq:dXS0dqt_V_qq}) and~(\ref{eq:dXS0dqt_H_gg}), respectively, they
give rise to a contribution of the form
\begin{eqnarray}
  \label{eq:soft_contr}
 && \int_\tau^{1-f(a)} \frac{dz}{z} \, \mathcal{L}\(\frac{\tau}{z}\)
 \, 2\, \hat{p}(z)\,
 \log\frac{1+\sqrt{1-\piT^2}}{1-\sqrt{1-\piT^2}}
 \nonumber\\
&& {}\hspace{1cm}
 = \int_0^1  \frac{dz}{z} \,\Lum\(\frac{\tau}{z}\)\lg
  \delta(1 - z)  \log^2(a) - 2\, p(z) \log(a)
 - \frac{\pi^2}{3} \, \delta(1 - z)  + \ldots \rg\phantom{aaaaa}
\end{eqnarray}
where, following the notation of appendix~\ref{app:AP}, we have defined
\begin{equation}
p(z) = \lq \frac{2}{1-z}\rq_+ .
\end{equation}
The details for the derivation of eq.~(\ref{eq:soft_contr}) are collected in
appendix~\ref{app:soft_coll_univ}.  Inspecting the first three terms of the
universal function $\gu_{q\bar{q}}(z)$ in eq.~(\ref{eq:ghat_un_V_qq}) and
$\gu_{gg}(z)$ in eq.~(\ref{eq:ghat_un_H_gg}), we recognize exactly the three
terms on the right-hand side of eq.~(\ref{eq:soft_contr}).

\subsubsection{The non-universal part}
\label{sec:log_non_un}

It is also interesting to notice that the non-universal part of the
$\hat{G}_{ab}^{\sss (1)}(z)$ functions contains terms proportional to
$\log(a)$, multiplied by powers of $a$.  These powers are controlled by the
form of the non-universal parts in
eqs.~(\ref{eq:dXS0dqt_V_gq})--(\ref{eq:dXS0dqt_H_gg}), and to the way they
enter in our generating procedure described in
appendix~\ref{app:integration_recipe}. In fact, by inspecting
eq.~(\ref{eq:Itilde2}), we see that they contribute to $\gd_{ab}$ with terms
of the form
\begin{equation}
  \label{eq:logterms_n}
  \int_\tau^{1-f(a)} dz \, (1-z)^n \sqrt{1-\piT^2} = \lg
\begin{array}{ll}
 {} + a^2 \log(a) + a\log(a) + \ldots \qquad  & n=1\\[1.5mm]
 {} -2\, a^2 \log(a) + \ldots \qquad  &  n=2\\[1.5mm]
  {} + a^2 \log(a) + \ldots \qquad  &  n=3\\[1.5mm]
  {} -6\,a^3 \log(a) + \ldots \qquad  &  n=4\\[1.5mm]
  {} +2\,a^3 \log(a) + \ldots \qquad  &  n=5
  \end{array}
\right.
\end{equation}
where the dots stand for power terms in $a$ with no logarithms attached.
%% The origin of the logarithmic terms in eq.~(\ref{eq:logterms_n}) is then
%% connected to the boundaries of the phase-space integration.
This also
explains why, for $V$ production, $\gd_{qg}(z)$ in
eq.~(\ref{eq:ghat_nun_V_qg}) and $\gd_{q\bar{q}}(z)$ in
eq.~(\ref{eq:ghat_nun_V_qq}) contain both terms $a\log(a)$ and $a^2\log(a)$:
they receive contributions from all the terms in eq.~(\ref{eq:logterms_n})
starting from $n=1$, since eqs.~(\ref{eq:dXS0dqt_V_gq})
and~(\ref{eq:dXS0dqt_V_qq}) contain a term proportional to
$(1-z)\sqrt{1-\piT^2}$.  Instead, $\gd_{gq}(z)$ in
eq.~(\ref{eq:ghat_nun_H_gq}) and $\gd_{gg}(z)$ in
eq.~(\ref{eq:ghat_nun_H_gg}) contain only the term $a^2\log(a)$, since they
receive contributions from the terms in eq.~(\ref{eq:logterms_n}) starting
from $n=2$, due to the fact that eqs.~(\ref{eq:dXS0dqt_H_qg})
and~(\ref{eq:dXS0dqt_H_gg}) contain a term proportional to
$(1-z)^2\sqrt{1-\piT^2}$ and $(1-z)^3\sqrt{1-\piT^2}$, respectively.

As far as the finite term in the diagonal channels is concerned, we notice
that, in the $q\bar{q}$ channel of DY production, the first term in
eq.~(\ref{eq:ghat_nun_V_qq}) happens to correspond to the first-order
collinear coefficient function defined in the ``hard-resummation scheme'',
introduced in ref.~\cite{Catani:2013tia, Catani:2000vq} within the
$\qt$-subtraction formalism. Instead, the first term in the $gg$ channel of
$H$ production in eq.~(\ref{eq:ghat_nun_H_gg}) has no connection with the
first-order collinear coefficient function, that is zero for this production
channel.  In conclusion, the structure of the terms in the non-universal part
depends on the peculiar form of the differential cross sections.

\subsubsection{Higher-order soft behaviour of the squared amplitudes}
\label{sec:NLS}
In this section, we extend the study performed in
Sec.~\ref{sec:soft_universal} in order to investigate the origin of the
power-suppressed terms $a \log(a)$ and $a^2 \log(a)$, present both in the
universal and in the non-universal parts.  We will show that their origin can
be connected to the higher-order soft behaviour of the squared amplitudes.
To this aim, we have performed a Laurent expansion in the energy $k^0$ of the
final-state parton of the exact squared amplitudes of
eqs.~(\ref{eq:M_qg_Vq}), (\ref{eq:M_qq_Vg}), (\ref{eq:M_gq_Hq})
and~(\ref{eq:M_gg_Hg}).  In the following, we call leading soft~(LS) the term
proportional to the highest negative power of $k^0$, next-to-leading
soft~(N$^1$LS) the subsequent term, and so on.  All the technical details and
expressions of the expansion terms are collected in
appendix~\ref{app:squared_ampl}.

We have then applied the algorithm described in this paper to each of the
terms of the expansions that we have calculated, in order to compute their
behaviour as a function of $a$.
This has allowed us to trace the origin of the $a\log(a)$ and $a^2 \log(a)$ terms.
Our findings are collected in the following:
\begin{itemize}

\item $q(\bar{q}) +  g \to V + q(\bar{q})$
    
  We reproduce the $a\log(a)$ behaviour of
  $\hat{g}^{\sss (1)}_{qg}(z)$
  in eq.~(\ref{eq:Ghat_V_qg})
  if we consider the soft-expansion of the exact amplitude up to the
  N$^1$LS level, i.e.~if we sum
  eqs.~(\ref{eq:M_qg_Vq_LS}) and~(\ref{eq:M_qg_Vq_NLS}),
  and the $a^2\log(a)$ behaviour if we consider the soft-expansion up to the
  N$^3$LS level, i.e.~if we sum
  eqs.~(\ref{eq:M_qg_Vq_LS})--(\ref{eq:M_qg_Vq_N3LS}).

\item $q + \bar{q}  \to V + g$

  We reproduce the $a\log(a)$ behaviour of
  $\hat{g}^{\sss (1)}_{q\bar{q} }(z)$
  in eq.~(\ref{eq:Ghat_V_qq})
  if we consider the soft-expansion of the exact amplitude up to the
  N$^1$LS level, i.e.~if we sum
  eqs.~(\ref{eq:M_qq_Vg_LS}) and~(\ref{eq:M_qq_Vg_NLS}),
  and the $a^2\log(a)$ behaviour if we consider the soft-expansion up to the
  N$^2$LS level, i.e.~if we sum
  eqs.~(\ref{eq:M_qq_Vg_LS})--(\ref{eq:M_qq_Vg_N2LS}).

\item $g + q(\bar q) \to H + q(\bar q)$ 

  We reproduce the $a\log(a)$ behaviour of
  $\hat{g}^{\sss (1)}_{gq}(z)$
  in eq.~(\ref{eq:Ghat_H_gq})
  if we consider the soft-expansion of the exact amplitude up to the
  N$^1$LS level, i.e.~if we sum
  eqs.~(\ref{eq:M_gq_Hq_LS}) and~(\ref{eq:M_gq_Hq_NLS}),
  and the $a^2\log(a)$ behaviour if we consider the soft-expansion up to the
  N$^2$LS level, i.e.~if we sum
  eqs.~(\ref{eq:M_gq_Hq_LS})--(\ref{eq:M_gq_Hq_N2LS}).

\item $g + g \to H + g$
  
  We reproduce the $a\log(a)$ behaviour of
  $\hat{g}^{\sss (1)}_{gg}(z)$
  in eq.~(\ref{eq:Ghat_H_gg})
  if we consider the soft-expansion of the exact amplitude up to the
  N$^2$LS level, i.e.~if we sum
  eqs.~(\ref{eq:M_gg_Hg_LS})--(\ref{eq:M_gg_Hg_N2LS}),
  and the $a^2\log(a)$ behaviour if we consider the soft-expansion up to the
  N$^4$LS level, i.e.~if we sum
  eqs.~(\ref{eq:M_gg_Hg_LS})--(\ref{eq:M_gg_Hg_N4LS}).

\end{itemize}
Collecting our result in a table, we have:
  \begin{center}
 \begin{tabular}{l|c|c|c|c|}
 \cline{2-5}
 &  \multicolumn{2}{ |c|}{$V$}
 &  \multicolumn{2}{ |c|}{ \phantom{\Big|} $H$}\\
 \cline{2-5}
 & \phantom{\Big|} $\hat{g}^{\sss (1)}_{qg}(z)$ & $\hat{g}^{\sss
   (1)}_{q\bar{q} }(z)$ & \phantom{\Big|} $\hat{g}^{\sss (1)}_{gq}(z)$ & $\hat{g}^{\sss (1)}_{gg}(z)$ \\
 \cline{1-5}
 \multicolumn{1}{ |c|  }{ \phantom{\Big|}  $a\log(a)$ }
 & N$^1$LS
 & N$^1$LS
 & N$^1$LS
 & N$^2$LS
 \\ \cline{1-5}
 \multicolumn{1}{ |c|  }{ \phantom{\Big|} $a^2 \log(a)$}
  & N$^3$LS
  & N$^2$LS
  & N$^2$LS
  & N$^4$LS
 \\ \cline{1-5}
 \end{tabular}
  \end{center}
In summary, the next-to-leading-soft approximation of the exact amplitudes
reproduces the $a\log(a)$ term only for $V$ production and for the
non-diagonal channel of $H$ production. For the diagonal channel of $H$
production, only the expansion up to next-to-next-to-leading-soft order
reproduces the $a\log(a)$ term. We would like to point out that, of the three
terms contributing to the N$^2$LS of eq.~(\ref{eq:M_gg_Hg_N2LS}), only the
constant one, i.e.~the number 16, is needed to reproduce the
$a\log(a)$ coefficient. The $u/t$ and $t/u$ terms do not give rise to
any $a\log(a)$ contribution.

Moreover, only the expansion up to next-to-next-to-leading-soft order in the
diagonal channel for $V$ production and in the non-diagonal channel for $H$
production reproduces the $a^2\log(a)$ coefficient. Higher orders in the
expansion in the softness of the final-state parton are needed for the
non-diagonal channel of $V$ production and for the diagonal channel of $H$
production.

\subsubsection[${\qt}$-subtraction method]{$\boldsymbol{\qt}$-subtraction method}
In the original paper on the $\qt$-subtraction method~\cite{
%  Bozzi:2005wk,
  Catani:2007vq}, the expansion in $\as$ of the transverse-momentum
resummation formula generates exactly the three terms in
eq.~(\ref{Rhat_coeff}), plus extra power-correction terms.

In the formula for $\hat{R}_{ab}^{\sss (1)}(z)$ that we can build from our
expression of $\hat{G}_{ab}^{\sss (1)}(z)$, by changing the overall sign and
adding the $\delta(1-z)$ contribution from the virtual correction, the
power-correction terms are exactly those produced by the expansion of the
real amplitudes.
If one is interested in using our formula for $\hat{R}_{ab}^{\sss (1)}(z)$ to
reduce the dependence on the transverse-momentum cutoff, within the
${\qt}$-subtraction method, the aforementioned extra terms need then to be
subtracted from our expression of $\hat{R}_{ab}^{\sss (1)}(z)$.

\subsection{Numerical results}
\label{sec:LA_numerica}

As previously pointed out, NLO (and NNLO) cross sections computed with the
$\qt$-subtracted formalism exhibit a residual dependence on $\qtcut$,
i.e.~the parameter $a$ we have introduced in eq.~(\ref{eq:a_def}). This
residual dependence is due to power terms which remain after the subtraction
of the IR singular contributions, and vanish only in the limit $a\to 0$
(limit which is unattainable in a numerical computation).  In this section we
discuss the residual systematic dependence on $\qtcut$ due to terms beyond
LT, NLT and N$^2$LT accuracy.

We present our results for $Z$ and $H$ production at the LHC, at a
center-of-mass energy of $\sqrt{S}=13$~TeV.  In our NLO calculations we have
set the renormalisation and factorisation scales equal to the mass of the
corresponding produced boson, and we have used the {\tt MSTW2008nlo}
parton-distribution function set~\cite{Martin:2009iq}.  The mass of the $Z$
boson $m_{\rm \sss Z}$ and of the Higgs boson $m_{\rm \sss H}$ have been set
to the values 91.1876~GeV and 125~GeV, respectively.

As an overall check of our calculation, we compared the results obtained with
the analytically $\qt$-integrated cross sections in
eqs.~(\ref{eq:dXS0dqt_V_gq})--(\ref{eq:dXS0dqt_H_gg}) with the
numerically-integrated results computed with both the {\tt
  DYqT-v1.0}~\cite{Bozzi:2008bb, Bozzi:2010xn} and {\tt
  HqT2.0}~\cite{
%  Catani:2000vq,
  Bozzi:2005wk, deFlorian:2011xf} codes, and found an excellent
agreement.

Then, in order to study the residual $\qtcut$ dependence of the NLO cross
sections for all the partonic subprocesses, we insert the expansion in
eq.~(\ref{Ghat_one_expans}) into eq.~(\ref{eq:Ghat_one_def}), and we
introduce the following definitions
\begin{eqnarray}
 \label{eq:sigma_LT} 
\sigLTab &\equiv& 
 \tau \int_\tau^1 \frac{dz}{z}  \, {\cal L}_{ab}\!\(\frac{\tau}{z}\) 
 \hat{\sigma}^{\sss (0)}
 \lq  \log^2(a) \, \hat G_{ab}^{\sss(1,2,0)}(z) + \log(a)  \, \hat
 G_{ab}^{\sss(1,1,0)}(z) + \hat G_{ab}^{\sss(1,0,0)}(z) \rq ,\phantom{aa}
\\
 \label{eq:sigma_NLT} 
\sigNLTab  &\equiv& 
 \tau \int_\tau^1 \frac{dz}{z}  \, {\cal L}_{ab}\!\(\frac{\tau}{z}\) 
 \hat{\sigma}^{\sss (0)}
 \lq  a\log(a) \,\hat G_{ab}^{\sss(1,1,2)}(z)  +  a \, \hat G_{ab}^{\sss(1,0,2)}(z) \rq,
\\
 \label{eq:sigma_N2LT} 
 \sigNNLTab &\equiv& 
 \tau \int_\tau^1 \frac{dz}{z}  \, {\cal L}_{ab}\!\(\frac{\tau}{z}\) 
 \hat{\sigma}^{\sss (0)}
 \lq  a^2 \log(a) \,\hat G_{ab}^{\sss(1,1,4)}(z)  + a^2 \, \hat G_{ab}^{\sss(1,0,4)}(z)  \rq,
\end{eqnarray}
where we have dropped the $^{\sss >}$ and $^{\sss (1)}$ superscripts for ease
of notation, since there is no possibility of misunderstanding in this
section, because we present only the NLO results we have computed for the
$\hat G_{ab}^{\sss (1)}(z)$ functions.

The $\hat{G}_{ab}^{\sss(1,n,m)}(z)$ functions in
eqs.~(\ref{eq:sigma_LT})--(\ref{eq:sigma_N2LT}) contain plus distributions up
to order~5 and to compute these cross sections we have first built
interpolations of the luminosity functions ${\cal L}_{ab}(y)$, defined in
eq.~(\ref{eq:lum_def}), for the channels that contribute to $Z$ and $H$
production at NLO.  We have expanded the luminosity functions on the basis of
the
%\v{C}eby\v{s}\"{e}v
Chebyshev polynomials up to order 30. In this way, the computation of the
derivatives of the luminosity functions can be performed in a fast and sound
way.

In the forthcoming figures, we plot the following quantities as a function of
$\qtcut$ (the corresponding value of $a$ is given on top of each figure):
\begin{enumerate}
\item  $(\sigma_{ab}^{\sss > (1)} - \sigLTab)$, %in blue;
\item $(\sigma_{ab}^{\sss > (1)} - \sigLTab - \sigNLTab)$, %in black;
\item $(\sigma_{ab}^{\sss > (1)} - \sigLTab - \sigNLTab-\sigNNLTab )$,% in  red.
\end{enumerate}
where $\sigma_{ab}^{\sss > (1)}$ is the cumulative cross section defined on
the left-hand side of eq.~(\ref{eq:Ghat_one_def}), obtained by integrating
the exact differential cross sections of
eqs.~(\ref{eq:dXS0dqt_V_gq})--(\ref{eq:dXS0dqt_H_gg}).
We expect that, by adding higher-power terms in $a$, these differences tend
to zero more and more quickly when $\qtcut \to 0$.  And in fact, the results
shown in the following figures confirm this behaviour.

\begin{figure}[htb!]
  \begin{center}
    \includegraphics[width=0.49\textwidth]{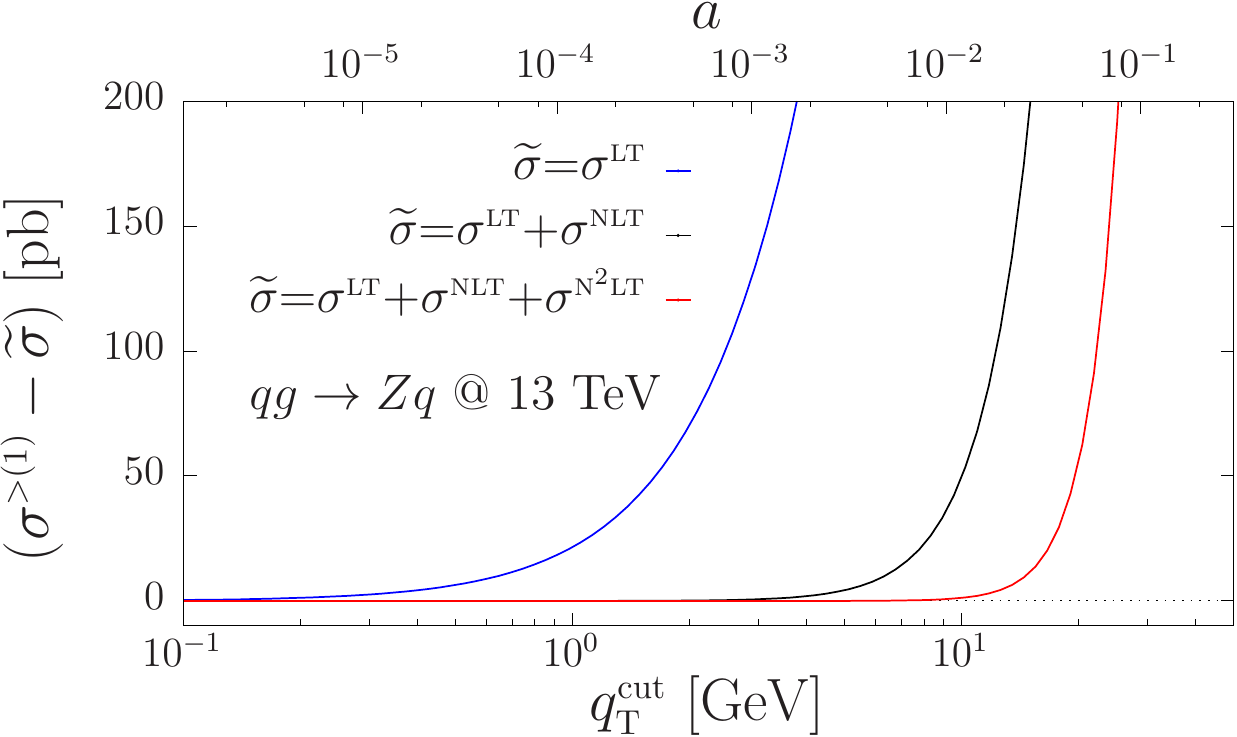}
    \includegraphics[width=0.49\textwidth]{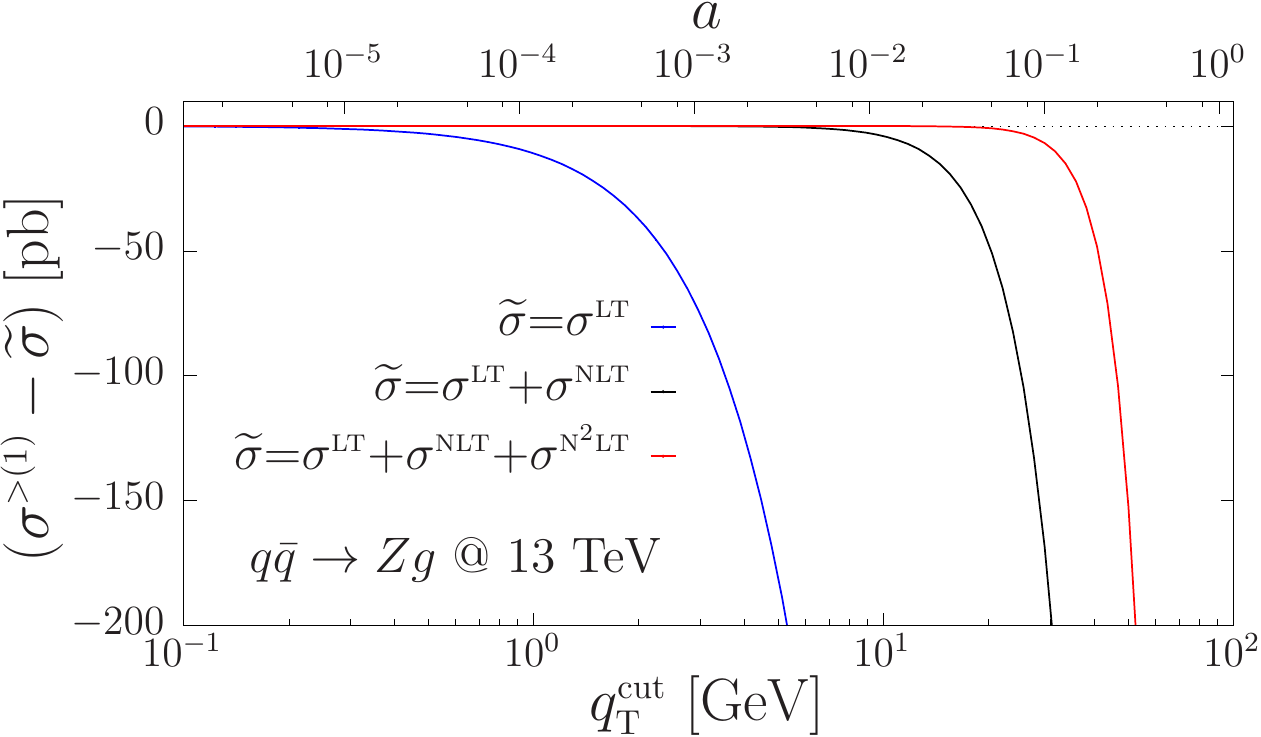}
  \end{center}
  \caption{Difference of the total cross sections $\(\sigma^{\sss > (1)} -
    \widetilde{\sigma}\)$ as a function of $\qtcut$, for $Z$ boson
    production, in the $qg \to Z q$~(left pane) and in the $q\bar{q} \to Z g$
    channel~(right pane).  The three curves correspond to the three possible
    choices of $\widetilde{\sigma}$: results for $\widetilde{\sigma} =
    \sigLT$ are displayed in blue, for $\widetilde{\sigma} = \sigLT+ \sigNLT$
    are displayed in black and for $\widetilde{\sigma} = \sigLT+ \sigNLT +
    \sigNNLT$ are displayed in red. The corresponding values of
    $a=\(\qtcut/m_{\rm\sss Z}\)^2$ are displayed on the top of the figure.
    The statistical errors of the integration are also shown, but they are
    totally negligible on the scale of the figure.}
  \label{fig:Z_qg_qq}
\end{figure}

\begin{figure}[htb!]
  \begin{center}
    \includegraphics[width=0.49\textwidth]{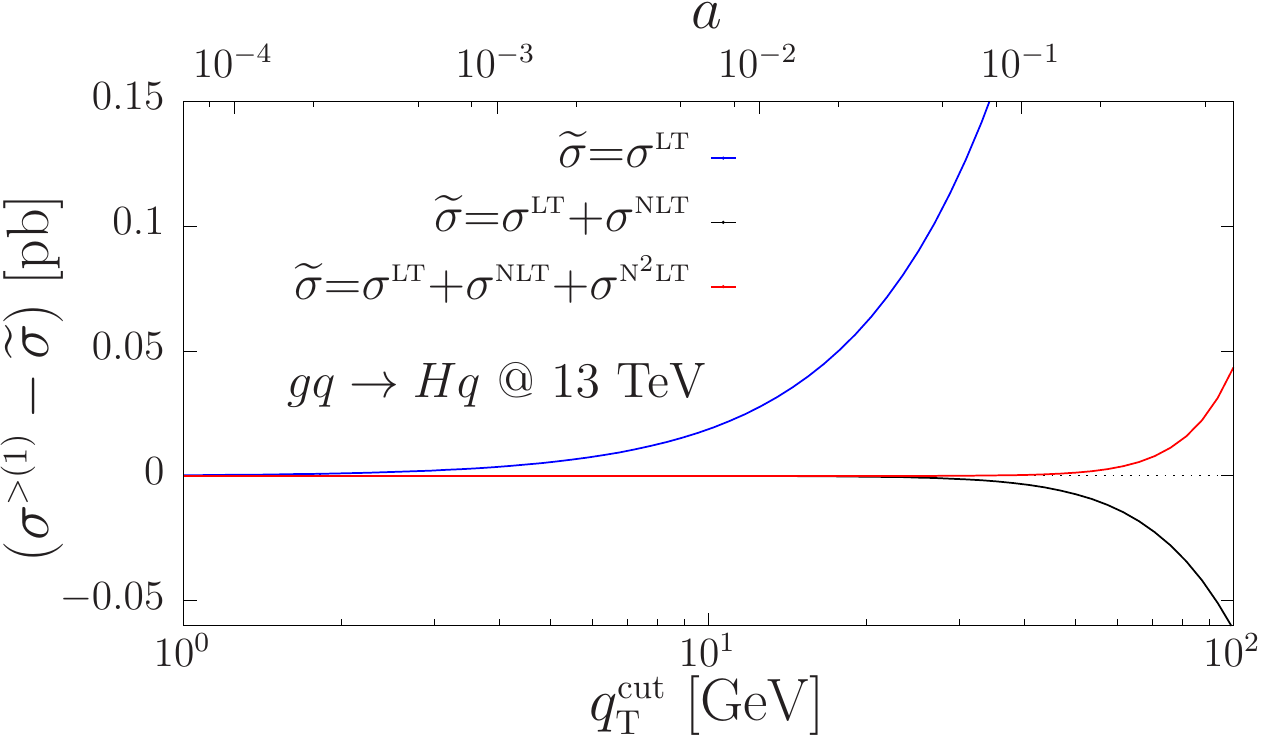}
    \includegraphics[width=0.49\textwidth]{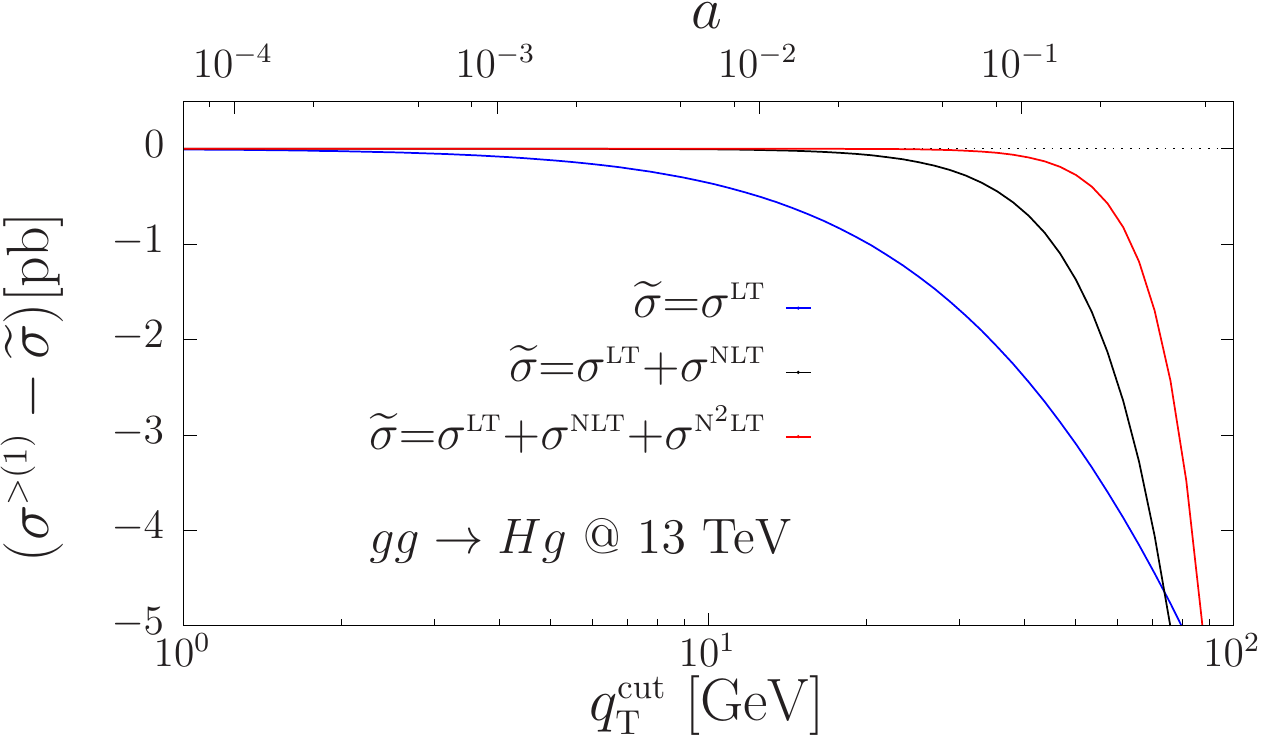}
  \end{center}
  \caption{Difference of the total cross sections $\(\sigma^{\sss > (1)} -
    \widetilde{\sigma}\)$ as a function of $\qtcut$, for $H$ boson
    production, in the $qg \to H q$~(left pane) and in the $gg \to H g$
    channel~(right pane). Same legend as in fig.~\ref{fig:Z_qg_qq}.  The
    corresponding values of $a=\(\qtcut/m_{\rm\sss H}\)^2$ are displayed on
    the top of the figure.}
  \label{fig:H_qg_gg}
\end{figure}

We first present our findings separated according to the partonic production
channels. In all the figures presented in this paper, the statistical errors
of the integration procedure are also displayed, but they are always totally
negligible on the scales of the figures.

In fig.~\ref{fig:Z_qg_qq} we collect the results for the aforementioned
cross-section differences, as a function of $\qtcut$, for the $qg \to
Zq$~(left) and $q \bar{q} \to Z g$~(right) channels, and in
fig.~\ref{fig:H_qg_gg} we collect similar results for the $gq \to Hq$~(left)
and $gg\to Hg$~(right) channels.  As expected, NLT and N$^2$LT contributions
increase the accuracy of the expanded cross section, with respect to the
exact one.

\begin{figure}[htb!]
  \begin{center}
    \includegraphics[width=0.49\textwidth, height=6.5cm]{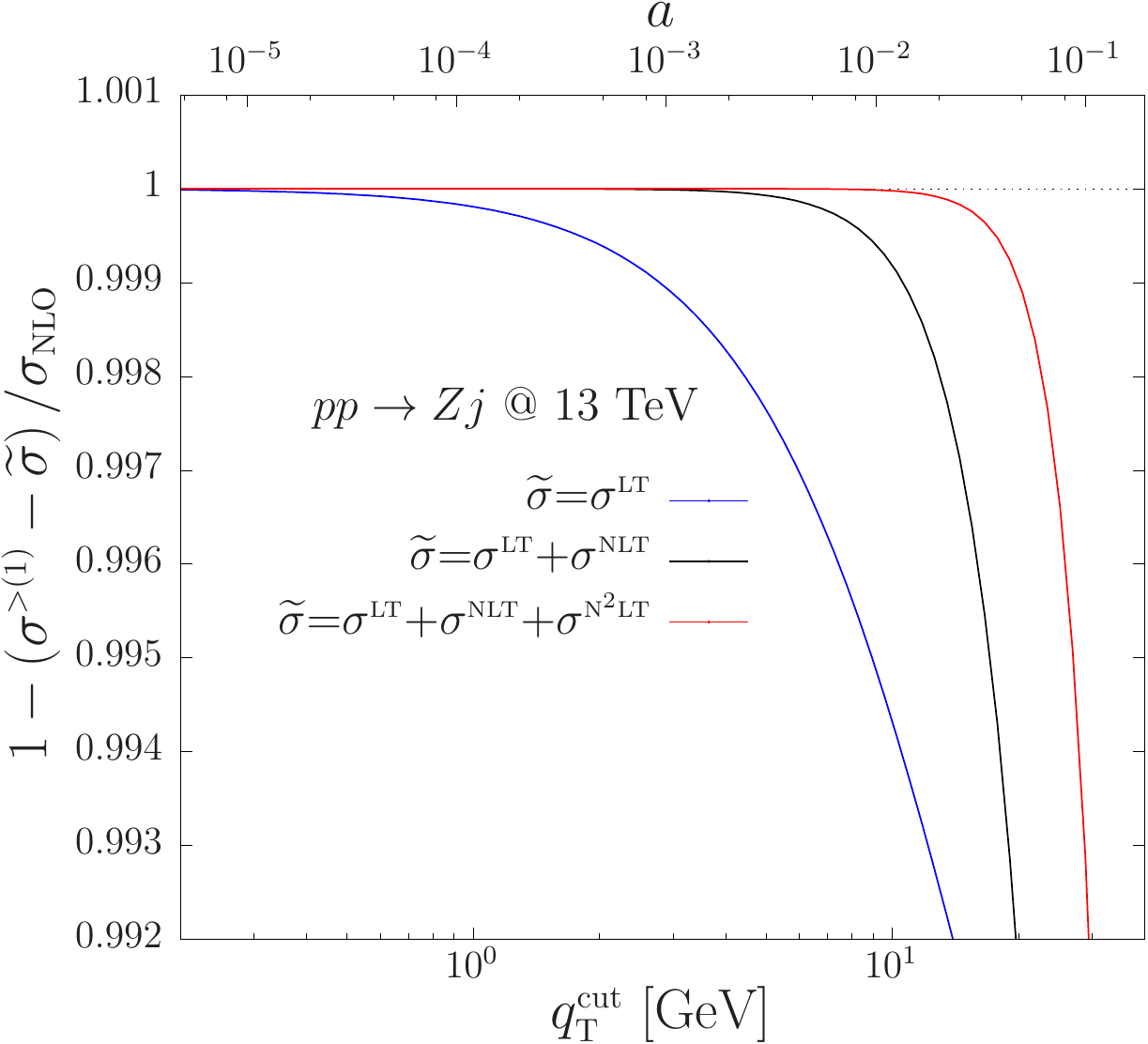}
    \includegraphics[width=0.49\textwidth, height=6.5cm]{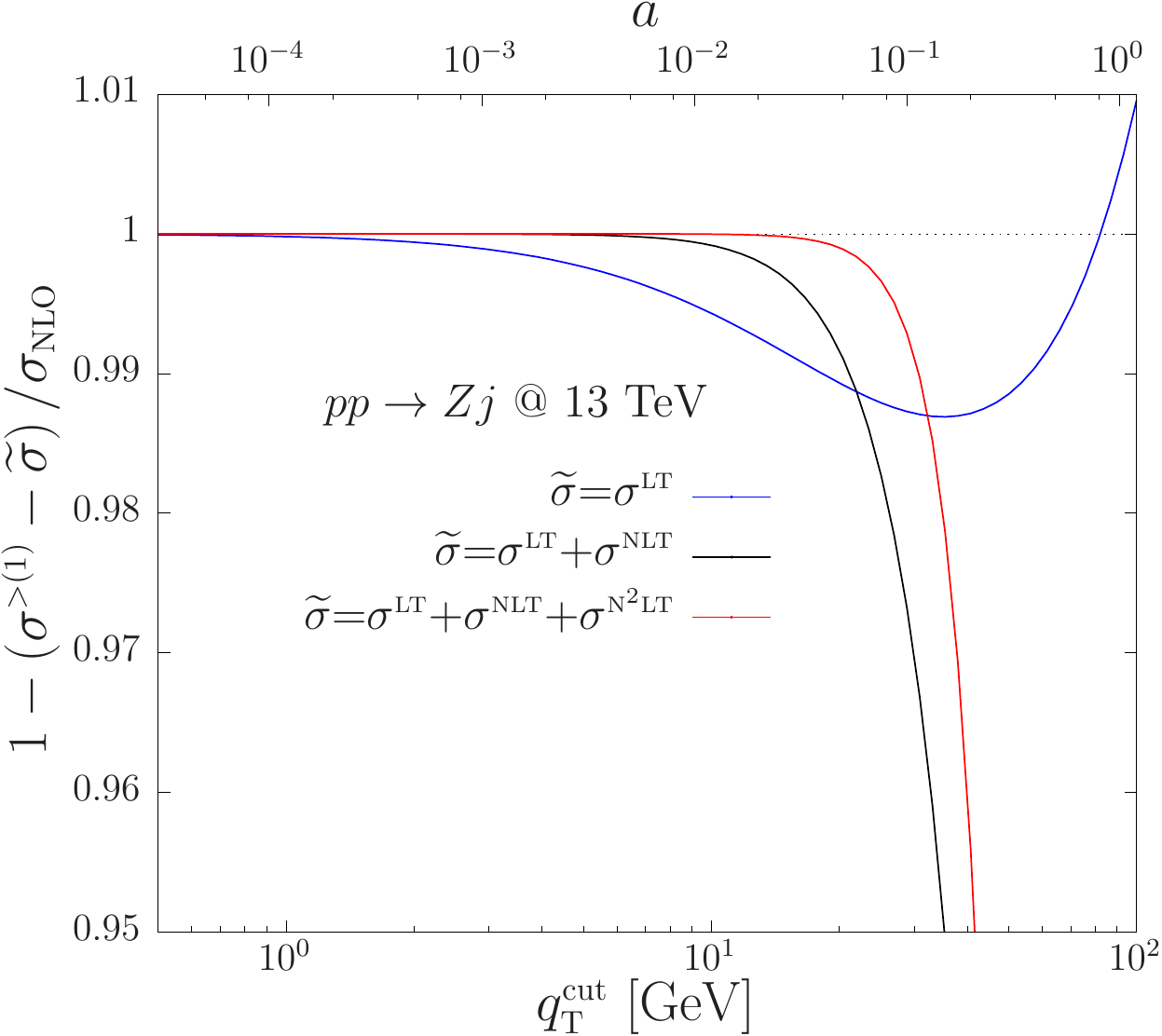}
  \end{center}
  \caption{Results for $1 - \(\sigma^{\sss > (1)} -
    \widetilde{\sigma}\)/\sigma_{\rm\sss NLO}$ as a function of $\qtcut$, for
    $Z$ boson production, in $pp\to Zj$. Same legend as in
    fig.~\ref{fig:Z_qg_qq}. In the left pane, the low-$\qtcut$ region is
    displayed, while, in the right pane, a larger region in $\qtcut$ is
    shown. The total cross section at NLO for $Z$ production,
    $\sigma_{\rm\sss NLO}$, has been taken equal to 55668.1~pb.}
  \label{fig:Z_norm}
\end{figure}

\begin{figure}[htb!]
  \begin{center}
    \includegraphics[width=0.49\textwidth]{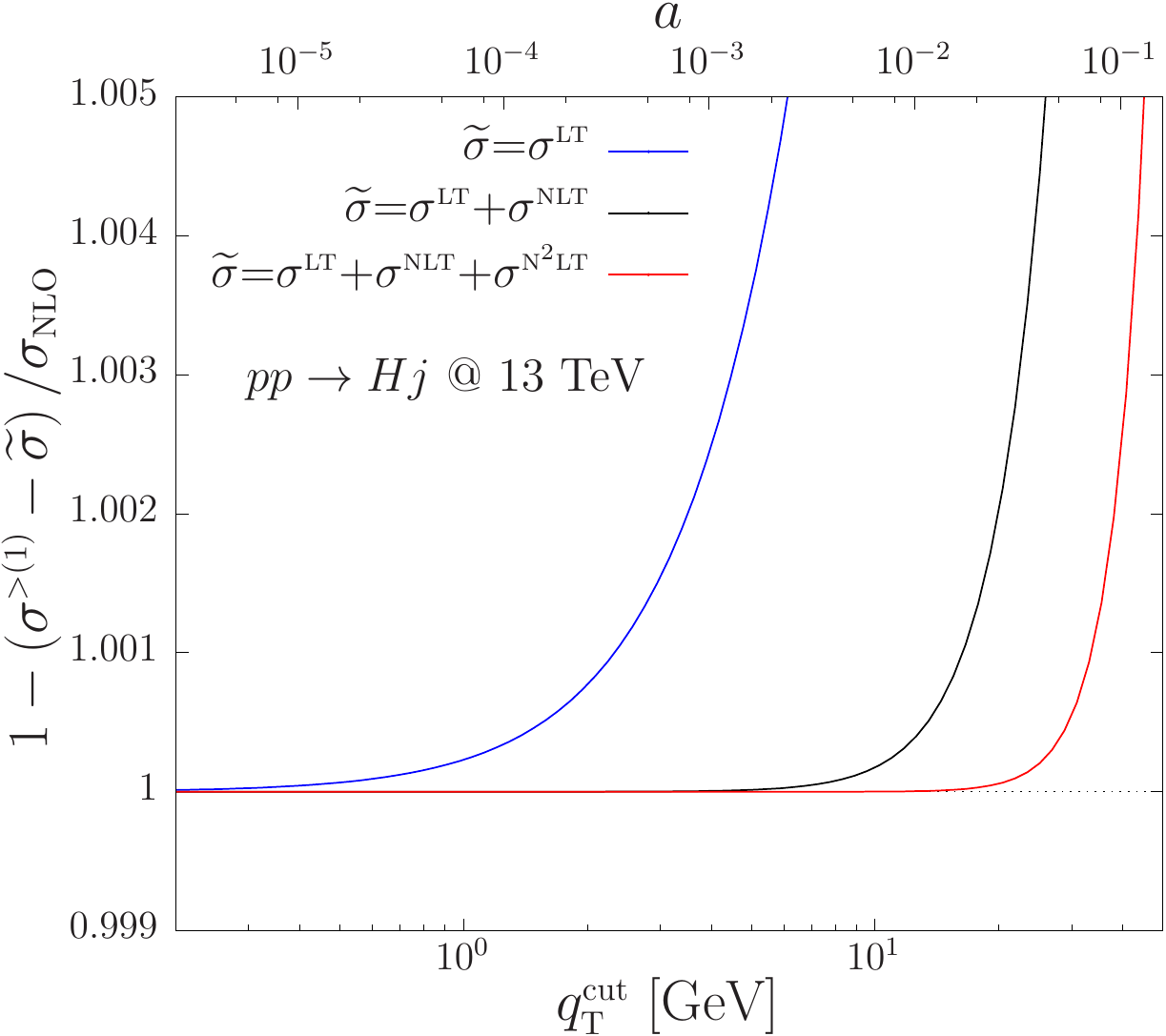}
    \includegraphics[width=0.49\textwidth]{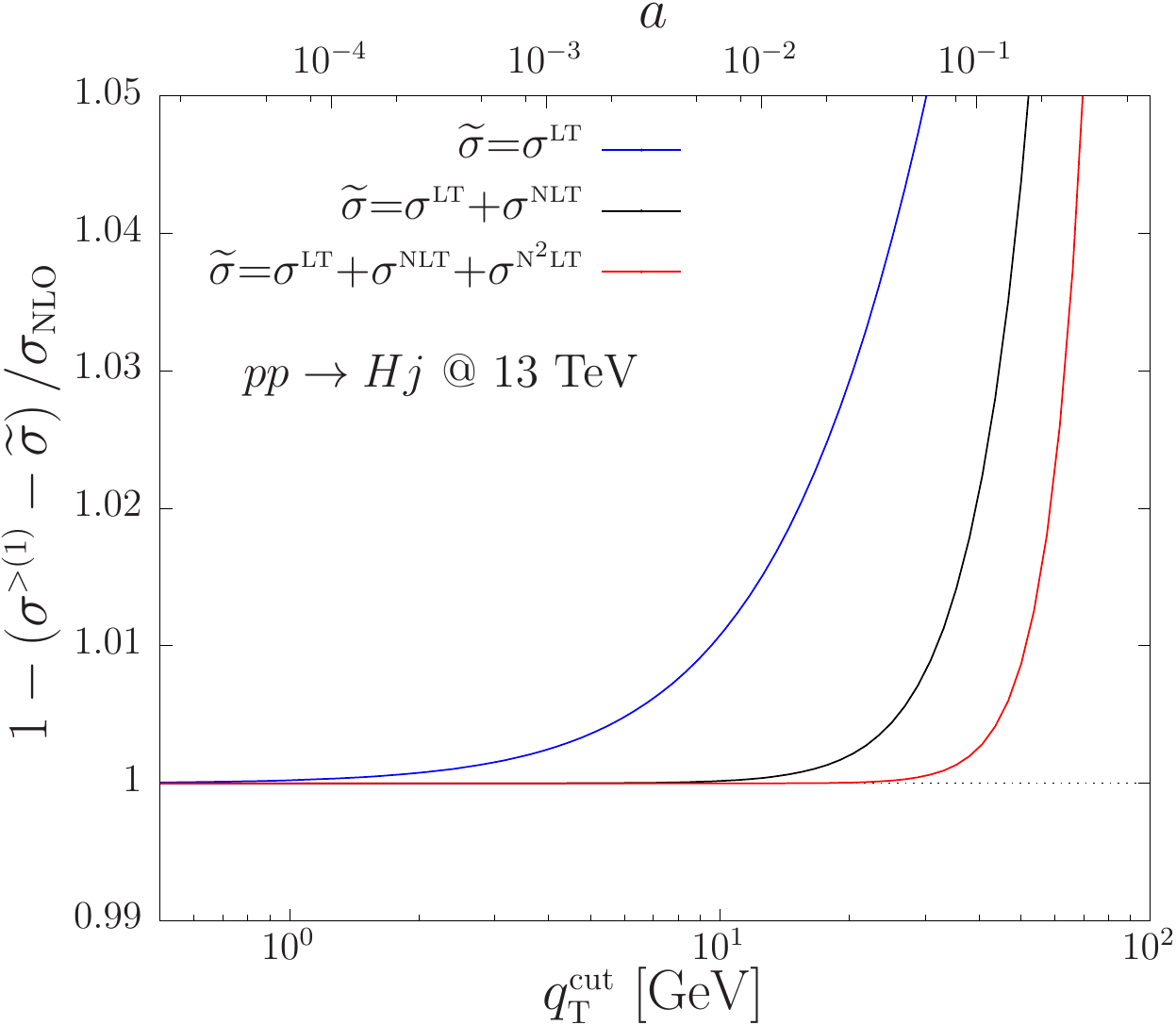}
  \end{center}
 \caption{Results for $1 - \(\sigma^{\sss > (1)} -
   \widetilde{\sigma}\)/\sigma_{\rm\sss NLO}$ as a function of $\qtcut$, for
   $H$ boson production, in $pp\to Hj$. Same legend as in
   fig.~\ref{fig:Z_qg_qq}. In the left pane, the low-$\qtcut$ region is
   displayed, while, in the right pane, a larger region in $\qtcut$ is
   shown. The total cross section at NLO for $H$ production, $\sigma_{\rm\sss
     NLO}$, has been taken equal to 31.52~pb.}
  \label{fig:H_norm}
\end{figure}

To give a more quantitative estimation of the power-suppressed corrections,
we present results for the total hadronic cross section, normalised with
respect to the corresponding exact NLO cross section $\sigma_{\rm\sss NLO}$
(i.e.~including also the virtual contributions). The results are shown in
figs.~\ref{fig:Z_norm} and~\ref{fig:H_norm}, where we have used
$\sigma_{\rm\sss NLO}=55668.1$~pb for $Z$ production and 31.52~pb for $H$
production. On the left panes we plot results in a smaller $\qtcut$ region,
while, on the right panes, we extend the $\qtcut$ interval to higher values.

These plots show exactly how the residual cutoff dependence of the cross
sections changes when the $\qt$-subtraction counterterm is corrected by the
NLT and N$^2$LT power terms.  For example, for $Z$ production and for
$\qtcut=10$~GeV, corresponding to $a=0.012$, the LT cross section gives an
estimate of the exact cross section within the 5\textperthousand, that
reduces to below the 1\textperthousand{} when the NLT contribution is added
and becomes less than 0.01\textperthousand{} when also the N$^2$LT is
present.  For Higgs boson production, the residual cutoff dependence is even
more pronounced: in fact, at $\qtcut=10$~GeV, corresponding to $a=0.0064$,
the LT is precise within the 1\%{} level. When the NLT is added, the
precision reaches the 0.2\textperthousand{}, and is below
0.001\textperthousand{} with the addition of the N$^2$LT.

\begin{figure}[htb!]
  \begin{center}
    \includegraphics[width=0.49\textwidth]{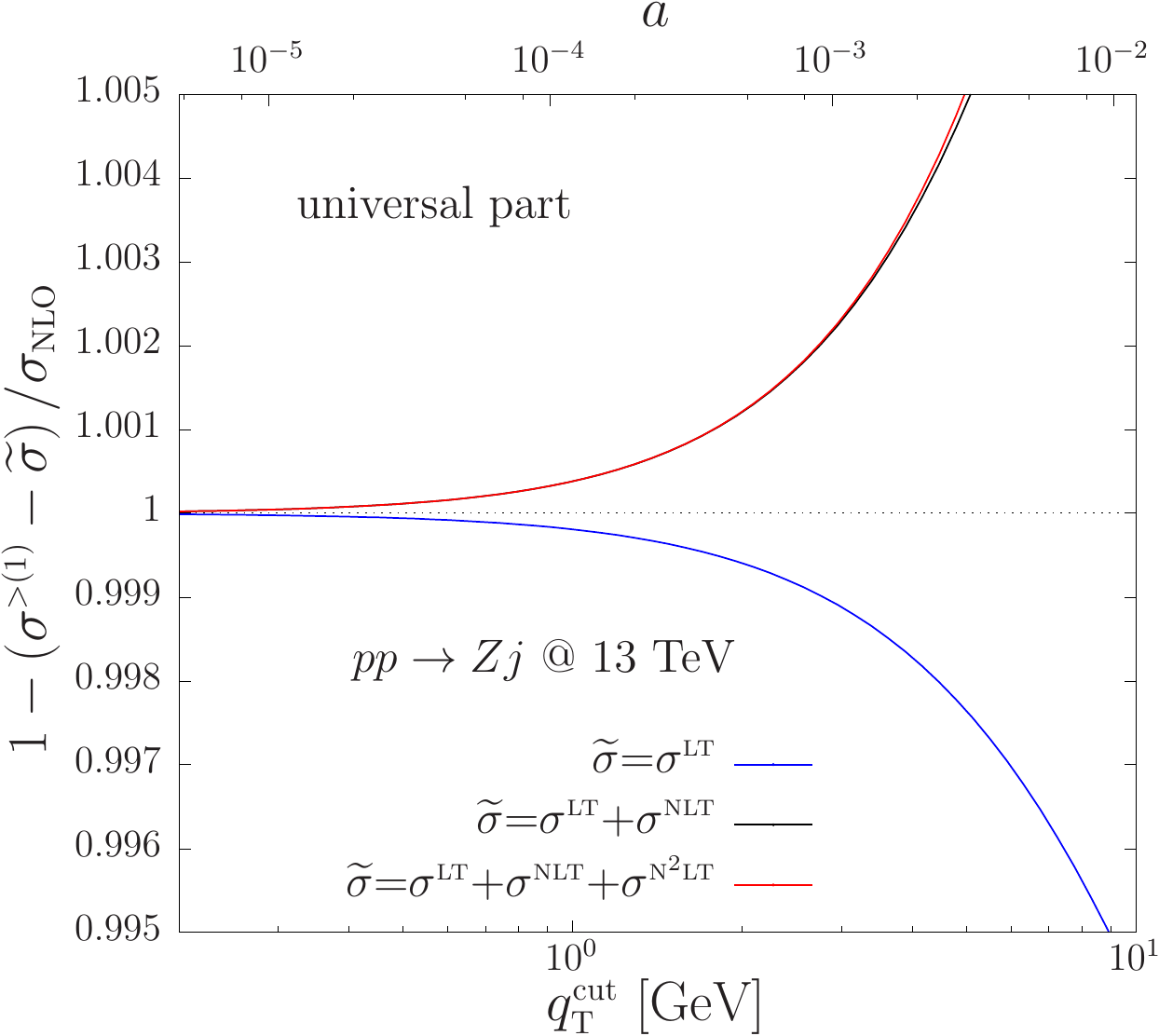}
    \includegraphics[width=0.49\textwidth]{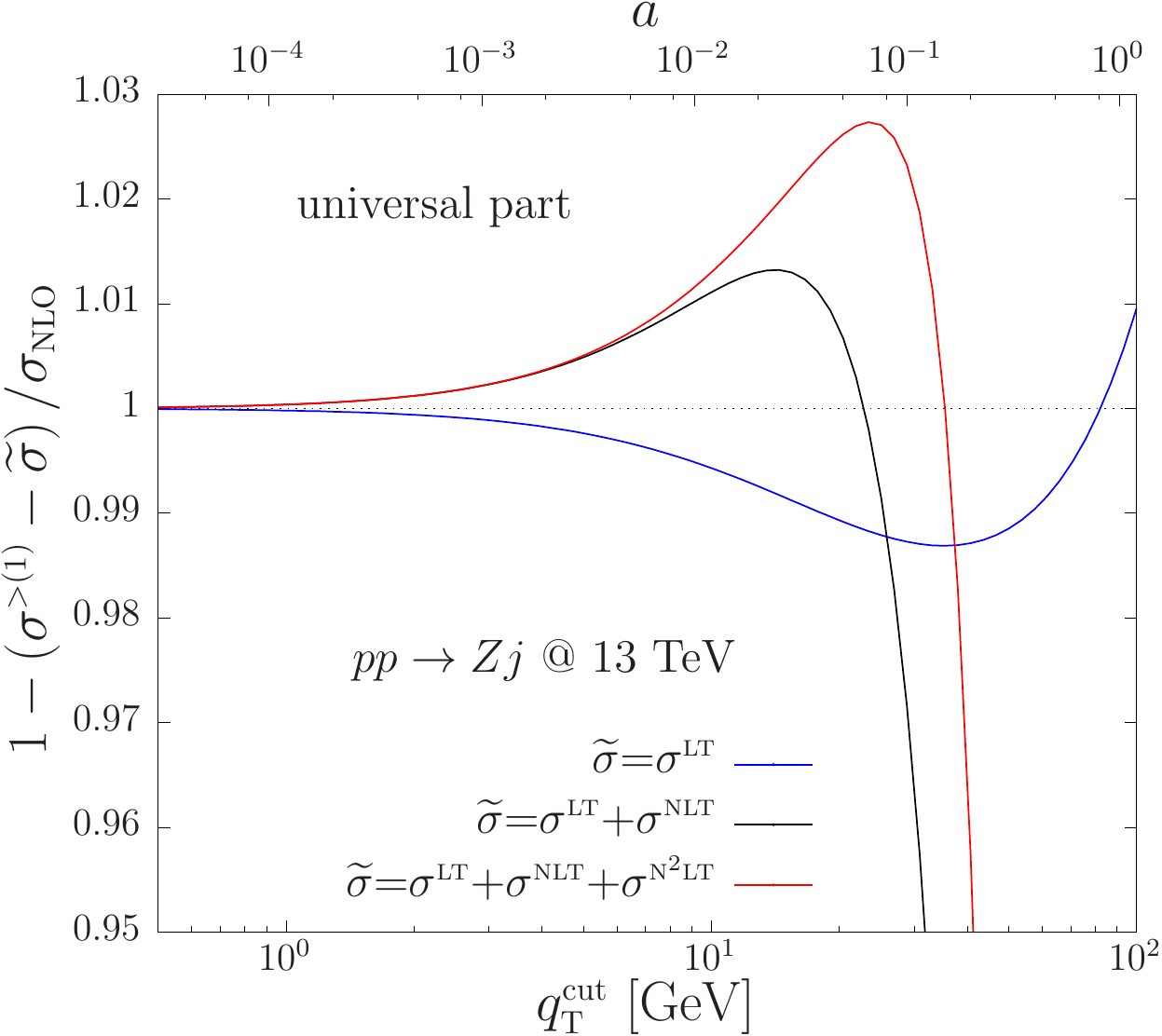}
  \end{center}
  \caption{Same as fig.~\ref{fig:Z_norm}, but using only the  universal part
    of $\hat{G}_{ab}^{\sss(1,n,m)}(z)$ in computing the cross sections of
    eqs.~(\ref{eq:sigma_LT})--(\ref{eq:sigma_N2LT}).}
  \label{fig:Z_un_norm}
\end{figure}

\begin{figure}[htb!]
  \begin{center}
    \includegraphics[width=0.49\textwidth]{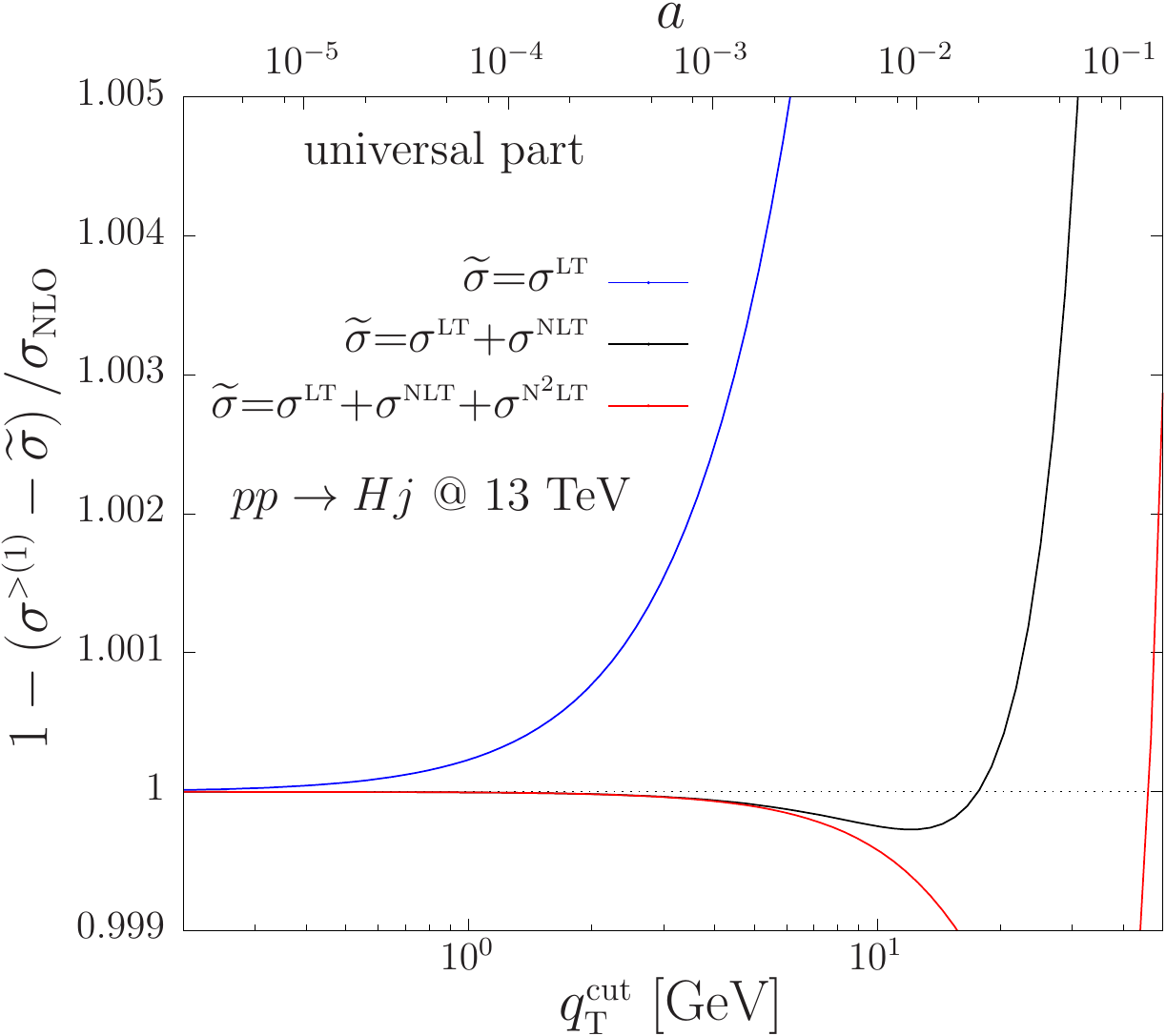}
    \includegraphics[width=0.49\textwidth]{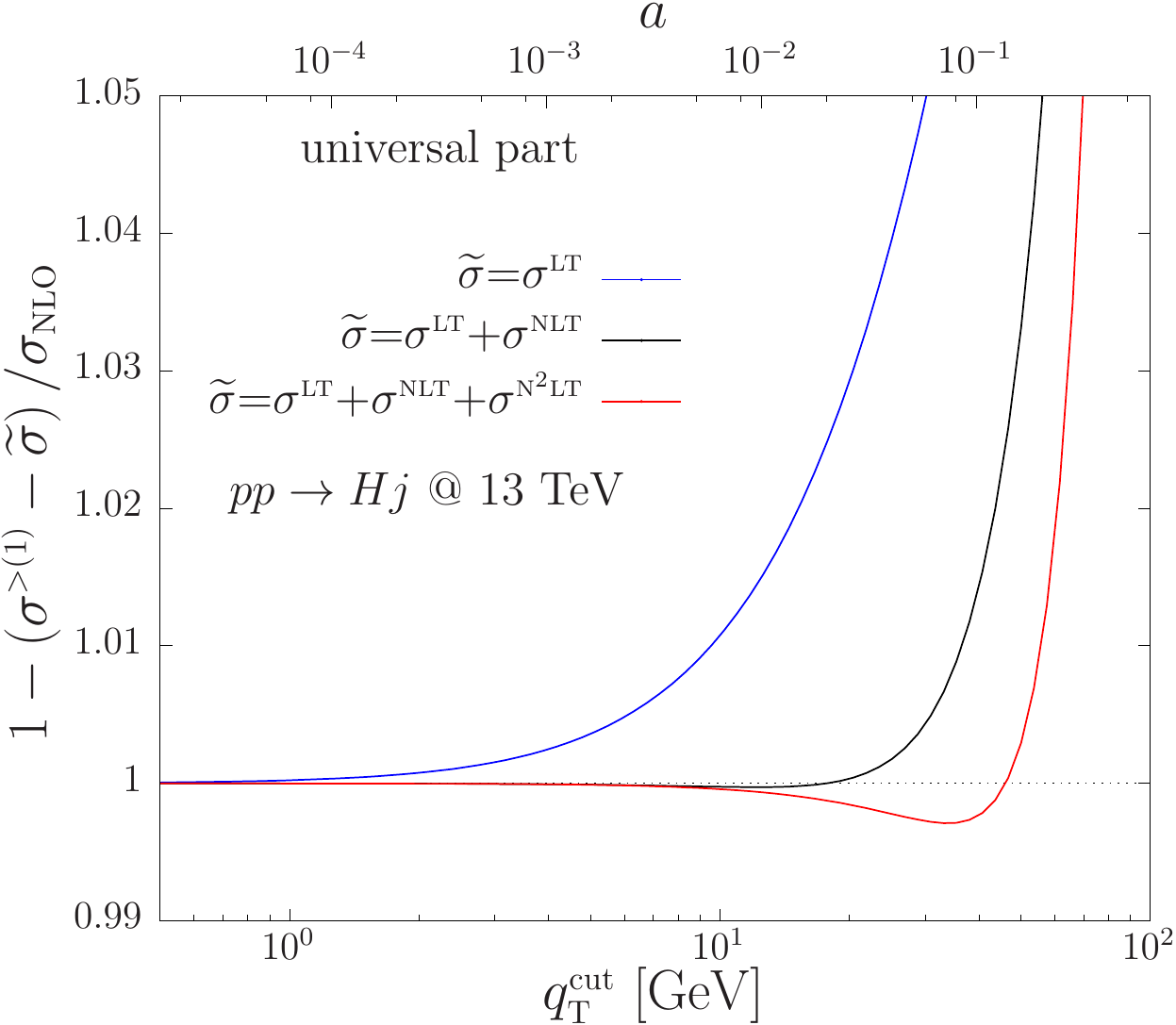}
  \end{center}
  \caption{Same as fig.~\ref{fig:Z_un_norm} but for Higgs boson production.}  
  \label{fig:H_un_norm}
\end{figure}

An interesting question is to estimate the impact of the universal parts of
the $\hat{G}_{ab}^{\sss(1,n,m)}(z)$ functions, with respect to the
non-universal ones.  We have then computed the cross sections in
eqs.~(\ref{eq:sigma_LT})--(\ref{eq:sigma_N2LT}), taking into account only the
universal parts of the $\hat{G}_{ab}^{\sss(1,n,m)}(z)$ functions.  Our
results are displayed in fig.~\ref{fig:Z_un_norm}, for $Z$ production, and in
fig.~\ref{fig:H_un_norm}, for $H$ production.

Comparing these figures with the corresponding ones with the full
$\hat{G}_{ab}^{\sss(1,n,m)}(z)$ functions, i.e.~figs.~\ref{fig:Z_norm}
and~\ref{fig:H_norm}, we see that the non-universal contributions play a
crucial role for $Z$ production, while their role is minor in Higgs boson
production.  This is due to the fact that the non-universal part in
eqs.~(\ref{eq:dXS0dqt_H_qg}) and~(\ref{eq:dXS0dqt_H_gg}) is suppressed by
higher powers of $(1-z)$, with respect to the corresponding expression for
$Z$ production, in eqs.~(\ref{eq:dXS0dqt_V_gq}) and~(\ref{eq:dXS0dqt_V_qq}),
confirming the conclusions drawn in sec.~\ref{sec:log_non_un}.

\section{Conclusions}
\label{sec:conclusions}

In this paper we considered the production of a colourless system at
next-to-leading order in the strong coupling constant $\as$.
We imposed a transverse-momentum cutoff, $\qtcut$, on the colour-singlet
final state and we computed the power corrections for the inclusive cross
section in the cutoff, up to the fourth power.
Although we studied Drell--Yan vector boson production and Higgs boson
production in gluon fusion, the procedure we followed is general and can be
applied to other similar cases, up to any order in the powers of $\qtcut$.

We presented analytic results, reproducing the known logarithmic terms from
collinear and soft regions of the phase space, along with the finite
contribution, and adding new terms as power corrections in $\qtcut$.
We found that the logarithmic terms in $\qtcut$ show up at most linearly in
the power-correction contributions, consistently with the fact that the LL
contribution is a squared logarithm.
In addition, no odd-power corrections in $\qtcut$ appeared in our
calculation, in agreement with known results in the literature for the NLO
differential cross section in colour-singlet production.  We do not expect
this to be true in general when cuts are applied to the final state.

Along the calculation we kept track of the origin of the newly-computed
terms, so that we were able to separate them into a universal part, and a
part that depends on the process at stake. In particular we derived and
identified the contribution to the universal part coming from soft radiation,
present in the diagonal partonic channels for $Z$ and $H$ production.  We
could also explain some features about the presence of power-suppressed
logarithmic terms, appearing in the non-universal part of the power
corrections. Furthermore we showed that the knowledge of the squared
amplitudes at the next-to-leading-soft approximation is not enough to predict
the $(\qtcut)^2\log\qtcut$ behaviour of the power corrections. The same
conclusion can be drawn for the knowledge of the next-to-next-to-leading-soft
approximation in predicting the $(\qtcut)^4\log\qtcut$ power correction.

We also studied the numerical impact of the power terms in the hadronic cross
sections for $Z$ and $H$ production at the LHC at 13~TeV, both by keeping
track of the different partonic production channels and by summing over all
of them.  We plotted the behaviours of the cross sections while adding more
and more orders of the power-correction terms, as a function of $\qtcut$, and
comparing them with the exact cross sections. For example, in $Z$ production
and for a value of $\qtcut=10$~GeV, the sensitivity on the cutoff can be
reduced from 1\textperthousand{} to 0.01\textperthousand{}, when adding the
$(\qtcut)^4$ contributions to the $(\qtcut)^2$ ones.  Higgs boson production
suffers from a larger sensitivity on the cutoff, and the dependence goes from
1\% to 0.2\textperthousand{}, when all the power corrections we computed are
added.  By performing the same numerical comparisons for just the universal
part of the power corrections, we showed that the non-universal contributions
play a crucial role for $Z$ production, while their role is minor in Higgs
boson production.

The knowledge of the power terms is crucial for understanding both the
non-trivial behaviour of cross sections at the boundaries of the phase space,
and the resummation structure at subleading orders.  Within the
$\qt$-subtraction method, the knowledge of the power terms helps in reducing
the cutoff dependence of the cross sections.
While the application of the $\qt$-subtraction method in NLO calculations is
superseded by well-known local subtraction methods, at NNLO it still plays a
major role, also in view of the fact that, as shown in
refs.~\cite{Grazzini:2017mhc, Cieri:2018oms}, the sensitivity to the
numerical value of the cutoff increases at higher orders.

\section*{Acknowledgments}

We thank S.~Catani, D.~de Florian and M.~Grazzini for useful discussions and
comments on the paper.
We thank M.~Ebert, I.~Moult, I.~Stewart, F.~Tackmann, G.~Vita
and H.~X.~Zhu for useful comments on the paper.

\appendix

\section{Partonic phase space and partonic cross sections at NLO}
\label{app:partonic_PS}

In this section we collect some basic formulae used to compute the cross
section of the partonic process
\begin{equation} a(p_1) + b(p_2) \to F(q) + c(k)\,,
\end{equation}
where $a$, $b$ and $c$ are quarks or gluons, in a combination compatible with
the production process of the colourless system $F$.  In parentheses, the
four momenta of the particles are given.

\subsection{Partonic phase space}
The standard Mandelstam relativistic
invariants are given by
\begin{equation}
\label{eq:stu}
  s = \(p_1 + p_2\)^2, \quad t = \(p_1-k\)^2 = -2 p_1\cdot k , \quad u =
\(p_2-k\)^2 = -2 p_2\cdot k , \quad q^2 = Q^2\,,
\end{equation}
and we also define the threshold variable $z$
\begin{equation}
\label{eq:z_def}
z = \frac{Q^2}{s}\,.
\end{equation}
The phase-space volume with the appropriate flux factor  is given by
\begin{eqnarray}
\label{eq:2body_PS}
d\Phi_2 &=& \frac{1}{2s} \frac{d^3q}{(2\pi)^3 2 q^0} \frac{d^3k}{(2\pi)^3 2 k^0}\,
(2\pi)^4\delta^4\!\(p_1 + p_2 - q - k\)
\nonumber\\
 &=& \frac{1}{2s} \frac{1}{(2\pi)^2}  \frac{d^3k}{ 2 k^0} \, \delta\!\((p_1 + p_2
-k)^2 - Q^2\) .
\end{eqnarray}
Since the colourless system recoils against the final coloured parton, their
transverse momenta are equal. Calling $\theta$ the angle between $p_1$ and
$k$, we can write
\begin{eqnarray}
\qt &=& k_{\sss T} = k^0 \sin\theta \,,
\\
t &=& -\sqrt{s} k^0 (1-\cos\theta) \,.
\end{eqnarray}
Inverting the system, we find the relations
\begin{eqnarray}
k^0 &=& -\frac{s \qt^2 + t^2}{2\sqrt{s} t} \,,
\\
\cos\theta &=& \frac{s\qt^2 - t^2}{s \qt^2 + t^2 } \,,
\end{eqnarray}
which lead to an expression of the phase-space volume in terms of $\qt$ and $t$
\begin{equation}
\label{eq:PS_qt_t}
\frac{1}{(2\pi)^2}  \frac{d^3k}{2 k^0} = \frac{1}{4\pi} k^0 dk^0 d\cos\theta
= - \frac{1}{4\pi} \frac{s \qt^2 + t^2}{2\sqrt{s} t} \frac{\sqrt{s}}{s \qt^2 +
  t^2} d\qt^2\,dt = - \frac{1}{8\pi} \frac{dt}{t}  \, d\qt^2 \,.
\end{equation}
On the other hand, using the identity
\begin{equation}
\label{eq:tu_ident}
t\, u = s\, \qt^2\,,
\end{equation}
we can write the argument of the $\delta$ function in eq.~(\ref{eq:2body_PS}) as
%%%% In the following lines, the previous version with the z variable
%% \begin{eqnarray}
%% (p_1+p_2 -k)^2 - Q^2 &=& s + t + u -Q^2
%% = \frac{Q^2}{z} + t + \frac{\qt^2 Q^2}{z\,t}-Q^2
%% \nonumber\\
%% &=&
%% \frac{1}{t} \lq  t^2 + Q^2\( \frac{1}{z} -1\) t + \frac{\qt^2 Q^2}{z} \rq
%% = \frac{1}{t} \(t-t_{+}\)\(t-t_{-}\)\phantom{aaa}
%% \end{eqnarray}
%% where
%% \begin{equation}
%% t_\pm = \frac{Q^{2}}{2z}\,\lq z-1 \pm \sqrt{(1-z)^{2}-4z\,\qtoQsq} \rq \,.
%% \end{equation}
%% \begin{eqnarray}
%% \delta\( \( p_1+p_2 -k \)^2 - Q^2\) \frac{dt}{t} &=&  \frac{t}{t^2 -t_+ t_-} 
%% \lq \delta\(t-t_+\) + \delta\(t-t_-\) \rq dt
%% \nonumber\\
%% &=& \frac{z}{Q^2 \sqrt{(1-z)^{2}-4z\,\qtoQsq}}  \lq \delta\(t-t_+\) +
%% \delta\(t-t_-\) \rq dt
%% \nonumber\\
%% \end{eqnarray}
%% Using eqs.~(****), we can write eq.~(\label{eq:2body_PS}) as
%% \begin{equation}
%% d\Phi_2 = \frac{1}{16\pi} \frac{z^2}{Q^4}
%% \frac{1}{\sqrt{(1-z)^{2}-4z\,\qtoQsq}}  \lq \delta\(t-t_+\) + \delta\(t-t_-\)\rq
%% dt \, d\qt^2
%% \end{equation}
\begin{eqnarray}
(p_1+p_2 -k)^2 - Q^2 &=& s + t + u -Q^2
= s + t + \frac{s \qt^2 }{t}-Q^2
\nonumber\\
&=&
\frac{1}{t} \lq  t^2 + \(s - Q^2\) t + s \qt^2 \rq
= \frac{1}{t} \(t-t_{+}\)\(t-t_{-}\) ,\phantom{aaa} 
\end{eqnarray}
where
\begin{equation}
t_\pm = \frac{1}{2}\,\lq Q^2 -s \pm \sqrt{\(Q^2-s \)^{2} - 4 \,s\, \qt^2} \rq .
\end{equation}
As a consequence, it is possible to write  
\begin{eqnarray}
\label{eq:deltaT}
\delta\( \( p_1+p_2 -k \)^2 - Q^2\) \frac{dt}{t} &=&  \frac{t}{t^2 -t_+ t_-} 
\lq \delta\(t-t_+\) + \delta\(t-t_-\) \rq dt
\nonumber\\
&=& \frac{1}{\sqrt{\(Q^2-s \)^{2} - 4 \,s\, \qt^2}} \lq \delta\(t-t_+\) +
\delta\(t-t_-\) \rq dt \,,
\nonumber\\
\end{eqnarray}
and using eqs.~(\ref{eq:PS_qt_t})
and~(\ref{eq:deltaT}), we can write eq.~(\ref{eq:2body_PS}) as
\begin{equation}
d\Phi_2 = \frac{1}{16\pi} \frac{1}{s}
\frac{1}{\sqrt{\(Q^2-s \)^{2} - 4 \,s\, \qt^2}}
\lq \delta\(t-t_+\) + \delta\(t-t_-\)\rq dt \, d\qt^2 \,.
\end{equation}
We then add a dummy integration over the $z$ variable,
\begin{equation}
d\Phi_2 = \frac{1}{16\pi} \frac{1}{s}
\frac{1}{\sqrt{\(Q^2-s \)^{2} - 4 \,s\, \qt^2}}
\lq \delta\(t-t_+\) + \delta\(t-t_-\)\rq
\delta\!\(z- \frac{Q^2}{s}\) 
dt \, d\qt^2\, dz \,,
\end{equation}
that allows us to rewrite the phase-space volume as
\begin{equation}
\label{eq:PS2_t_qt_z}
d\Phi_2 = \frac{1}{16\pi} \frac{z^2}{Q^4}
\frac{1}{\sqrt{(1-z)^{2}-4z\,\qtoQsq}}  \lq \delta\!\(t-t_+\) + \delta\!\(t-t_-\)\rq
\delta\!\(z- \frac{Q^2}{s}\)
dt \, d\qt^2\, dz \,,
\end{equation}
where
\begin{equation}
\label{eq:tpm_z}
t_\pm = \frac{Q^{2}}{2z}\,\lq z-1 \pm \sqrt{(1-z)^{2}-4z\,\qtoQsq} \rq .
\end{equation}

\subsection{Partonic cross sections}
\label{app:partonic_XS}
We can write the partonic cross sections for a $2 \to 2$ process as
\begin{equation}
d\hat\sigma = \left| {\cal M}(s,t,u) \right|^2 d\Phi_2 \,,
\end{equation}
where ${\cal M}$ is the amplitude for the partonic process, that in general
can be written as a function of the Mandelstam variables $s$, $t$ and $u$.
From eqs.~(\ref{eq:z_def}) and~(\ref{eq:tu_ident}) we can express $s$ and
$u$ as functions of $z$, $\qt$ and $t$, and using eq.~(\ref{eq:PS2_t_qt_z})
we can write
\begin{eqnarray}
\label{eq:disgma_hat}
d\hat\sigma &=& \frac{1}{16\pi} \frac{z^2}{Q^4}
\frac{1}{\sqrt{(1-z)^{2}-4z\,\qtoQsq}}  \lq \delta\!\(t-t_+\) + \delta\!\(t-t_-\)\rq
\nonumber\\
&& \hspace{4cm} {} \times \delta\!\(z- \frac{Q^2}{s}\)
\left| {\cal M}\(z,t,\qt\) \right|^2 dt \, d\qt^2\, dz
\nonumber\\
&=& \frac{1}{16\pi} \frac{z^2}{Q^4\,dz}
\frac{1}{\sqrt{(1-z)^{2}-4z\,\qtoQsq}}
\lq \left| {\cal M}\(z,t_+,\qt\)\right|^2  + \left| {\cal
  M}\(z,t_-,\qt\)\right|^2\rq
\nonumber\\
&& \hspace{4cm} {} \times \delta\!\(\! z- \frac{Q^2}{s}\) d\qt^2\, dz
\nonumber\\
&=&  \dsigdqtz\, \delta\!\(\! z- \frac{Q^2}{s}\) d\qt^2\, dz \,,
\end{eqnarray}
where we have defined
\begin{equation}
\label{eq:diffXS_M}
\dsigdqtz  \equiv \frac{1}{16\pi} \frac{z^2}{Q^4}
\frac{1}{\sqrt{(1-z)^{2}-4z\,\qtoQsq}}
\lq \left| {\cal M}\(z,t_+,\qt\)\right|^2  + \left| {\cal
  M}\(z,t_-,\qt\)\right|^2\rq .
\end{equation}

%% \section{Partonic differential cross sections}

%% \subsection{$\boldsymbol{V}$ production}
%% \subsubsection{$gq \to V q$}
%% \begin{equation}
%% g(p_1) + q(p_2) \to V(q) + q(k)
%% \end{equation}
%% \begin{equation}
%% d\hat\sigma_{qg} =  d\Phi_2 \left| {\cal M} \right|^2
%% \end{equation}

%% \begin{equation}
%% \left| {\cal M} \right|^2 = 8\pi \frac{g^2 \(g_v^2 + g_a^2\)} {c_W^2} \frac{1}{\NC}
%% \TR \as  {\cal A}
%% \end{equation}

%% \begin{equation}
%% {\cal A} = - \lg \frac{t}{s} + \frac{s}{t} + \frac{2 u Q^2}{st} \rg
%% \end{equation}

%% \subsubsection{$q\bar{q} \to V g$}

%% \subsection{$\boldsymbol{H}$ production}
%% \subsubsection{$qg \to H q$}
%% \subsubsection{$gg \to H g$}

\subsection{Squared amplitudes and their soft limit}
\label{app:squared_ampl}
In this section, for completeness, we collect the squared amplitudes $\left|
{\cal M}(s,t,u) \right|^2$ for $V$+jet and $H$+jet, at the lowest order in
$\as$, stripped off of trivial coupling, colour and spin factors.  The
normalization of the following amplitudes is such that
\begin{equation}
\frac{z}{2}\, \frac{\qt^2}{Q^2} \lq \left| {\cal M}\(z,t_+,\qt\)\right|^2  + \left| {\cal
  M}\(z,t_-,\qt\)\right|^2\rq
\end{equation}
is exactly the numerator of eqs.~(\ref{eq:dsigmahat_V_qg})--(\ref{eq:dsigmahat_H_gg}).

Together with the exact squared matrix elements, we give also the soft
behaviour of the amplitudes, using the energy $k^0$ of the final-state parton
as expansion parameter.
We have computed the soft expansion adopting the following procedure: we
first got rid of $s$ in favour of $Q^2$, $t$ and $u$ using the identity
\begin{equation}
s=Q^2 -t -u\,.
\end{equation}
In this way, the only dependence on the energy of the final-state parton is
through $t$ and $u$, that are linearly dependent on  $k^0$
(see eq.~(\ref{eq:stu})).  Then we perform a Laurent expansion in $k^0$, and
define leading soft~(LS) the term proportional to the highest negative power
of $k^0$, next-to-leading soft~(N$^1$LS) the subsequent term, and so on.
Finally we re-express all the soft-expansion contributions in terms of $t$
and $u$.  As a result, at each order of the expansion, all the terms
proportional to a given power of $k^0$ are included, and only them. This is
an unambiguous way to define the softness order of the expansion.

%Our results are summarized in the following.

\subsubsection*{$\boldsymbol{V}$ production}
\begin{itemize}
\item  $q(\bar{q}) +  g \to V + q(\bar{q})$
  
The exact squared amplitude is given by  
\begin{equation}
 \label{eq:M_qg_Vq}
 \left| {\cal M}(s,t,u) \right|^2 = - 2\lq \frac{t}{s}
 + \frac{s}{t} + 2\,\frac{Q^2 u}{st} \rq
 \end{equation}
with soft-expansion terms 
\begin{eqnarray}
\label{eq:M_qg_Vq_LS}
\MLS  &=& -2\,\frac{Q^2}{t}
\\
\label{eq:M_qg_Vq_NLS}
\MNLS{1} &=&  -2\,\frac{u}{t} + 2\
\\
\label{eq:M_qg_Vq_N2LS}
\MNLS{2} &=&  -\frac{2}{Q^2} \lq 2 \frac{u^2}{t} + 2 u+ t \rq
\\
\label{eq:M_qg_Vq_N3LS}
\MNLS{3} &=&  -\frac{2}{Q^4} \lq 2\,
     {\frac{u^3}{t}}+4\,u^2+3\,tu+ t^2 \rq
\\
\MNLS{4} &=& \ldots
\end{eqnarray}

\item  $q + \bar{q} \to V + g$

The exact squared amplitude is given by    
\begin{equation}
 \label{eq:M_qq_Vg}  
\left| {\cal M}(s,t,u) \right|^2 = \frac{4\,Q^4}{tu}
- 4 \lq \frac{ Q^2}{u} + \frac{ Q^2}{t} \rq
+ 2 \lq \frac{u}{t} + \frac{t}{u} \rq
\end{equation}
with soft-expansion terms
\begin{eqnarray}
 \label{eq:M_qq_Vg_LS}  
\MLS &=& 4\,\frac{Q^4}{t\,u}
\\
\label{eq:M_qq_Vg_NLS}  
\MNLS{1} &=& - 4 \lq \frac{ Q^2}{t} + \frac{ Q^2}{u} \rq
\\
\label{eq:M_qq_Vg_N2LS}  
\MNLS{2} &=&  2 \lq \frac{u}{t} + \frac{t}{u} \rq
\\
\label{eq:M_qq_Vg_NnLS}  
\MNLS{{\it n}} &=& 0\,,  \qquad\qquad n\ge 3
\end{eqnarray}
\end{itemize}
We notice that, for $q(\bar{q}) + g \to V + q(\bar{q})$ production, the LS
term of eq.~(\ref{eq:M_qg_Vq_LS}) has only one negative power of $k^0$, while
in $q + \bar{q} \to V + g$ the LS term of eq.~(\ref{eq:M_qq_Vg_LS}) has two
negative powers of $k^0$, in agreement with the eikonal approximation for
soft-gluon emission.

\subsubsection*{$\boldsymbol{H}$ production}

\begin{itemize}
\item $g + q(\bar q) \to H + q(\bar q)$

The exact squared amplitude is given by  
\begin{equation}
 \label{eq:M_gq_Hq}     
\left| {\cal M}(s,t,u) \right|^2  =  -\frac{2}{Q^2}\frac{s^2+u^2}{t}
\end{equation}
with soft-expansion terms
\begin{eqnarray}
 \label{eq:M_gq_Hq_LS}     
\MLS &=& -2\frac{Q^2}{t}
\\
\label{eq:M_gq_Hq_NLS}     
\MNLS{1} &=& 4\frac{u}{t} +4 
\\
\label{eq:M_gq_Hq_N2LS}   
\MNLS{2} &=& -\frac{2}{Q^2} \lq 2\frac{u^2}{t} +2\, u + t \rq
\\
\label{eq:M_gq_Hq_NnLS}   
\MNLS{{\it n}} &=&0\,,    \qquad\qquad n\ge 3
\end{eqnarray}

\item  $g + g \to H + g$

The exact squared amplitude is given by  
\begin{equation}
\label{eq:M_gg_Hg}
\left| {\cal M}(s,t,u) \right|^2 = \frac{2}{Q^2}  \frac{\(Q^2\)^4 + s^4 + t^4 + u^4
}{s\, t\, u}
\end{equation}
with soft-expansion terms 
\begin{eqnarray}
\label{eq:M_gg_Hg_LS}
\MLS &=&  4\,\frac{Q^4}{t\, u}
\\
\label{eq:M_gg_Hg_NLS}
\MNLS{1} &=& -4 \lq \frac{Q^2}{t} + \frac{Q^2}{u} \rq %=\frac{2}{Q^2} \(-2 Q^4 \) \frac{u+t}{t u}
\\
\label{eq:M_gg_Hg_N2LS}
\MNLS{2} &=& 8 \lq \frac{u}{t} +\frac{t}{u} +2\rq % = \frac{2}{Q^2}   4 Q^2 \frac{(u+t)^2}{t \,u} 
\\
\label{eq:M_gg_Hg_N3LS}
\MNLS{3} &=& 0
\\
\label{eq:M_gg_Hg_N4LS}
\MNLS{4} &=& %\frac{2}{Q^2}
     \frac{4}{Q^4} \frac{ \(u^2 + t\,u + t^2 \)^2}{ t\, u}
\\
 \MNLS{5} &=& \ldots    
\end{eqnarray}

\end{itemize}
Similar conclusions to $V$ production can be drawn for $H$ production: one
negative power of $k^0$ in  $g + q(\bar q) \to H + q(\bar q)$, see eq.~(\ref{eq:M_gq_Hq_LS}),
and two negative powers for  $g + g \to H + g$, see eq.~(\ref{eq:M_gg_Hg_LS}).

\section{Process-independent procedure for extending the $\boldsymbol{z}$ integration}
\label{app:integration_recipe}
In this section we describe the procedure followed to extend the integration
range of the $z$ variable up to 1, as displayed in
eq.~(\ref{eq:Ghat_one_def}), performing an expansion in $a$.  We consider an
integral of the following form
\begin{equation}
  \label{eq:I_def}
I = \int_\tau^{1-f(a)} dz\, l(z) \, g(z)\,,
\end{equation}
with $g(z)$ not defined for $z>1-f(a)$, $l(z)$ well behaved for $ \tau \le z
\le 1$ and $l(z)=0$ for $z<\tau$.  We also assume that $l(z)$ is ${\cal
  C}^\infty$, so that we can derive it as many times as necessary.  We then
suppose that $g(z)$ can be written as an expansion in (negative) powers of
$(1-z)$
\begin{equation}
\label{eq:gz_expansion}
g(z) = g_0(z,a) + \frac{g_1(z,a)}{1-z} + \frac{ g_2(z,a)}{(1-z)^2} + \ldots =
\sum_{n=0}^\infty\frac{ g_n(z,a)}{(1-z)^n}\,,
\end{equation}
where, in $z=1$, the $g_i(z,a)$ are not singular or have an integrable
singularity. If not identically zero everywhere, the $g_i(z,a)$ are different
from 0 for $z=1$ and $i\ge 1$, and, in general, the $g_i(z,a)$ functions
contain growing powers of $a$ as $i$ increases.

The point we would like to make here is that the right-hand side of
eq.~(\ref{eq:gz_expansion}) is convergent only for $z\le 1-f(a)$, and it
converges to $g(z)$. For $z > 1-f(a)$ the series does not converge to $g(z)$,
otherwise $g(z)$ would be defined in this region too.

We also assume that we can exchange the order of integration and summation of
the series, and we write eq.~(\ref{eq:I_def}) as
\begin{equation}
I = \sum_{n=0}^\infty I_n\,,
\end{equation}
where
\begin{equation}
  I_n \equiv  \int_\tau^{1-f(a)} dz\, l(z) \,\frac{ g_n(z,a)}{(1-z)^n} =
  \int_0^{1-f(a)} dz\, l(z) \,\frac{ g_n(z,a)}{(1-z)^n} \,,
\end{equation}
where we have extended the $z$-integration down to 0, since $l(z)=0$ for
$z<\tau$.  Each term of the series can now be manipulated as shown in the following.

\subsection*{$\boldsymbol{I_0}$}
\begin{eqnarray}  
I_0 &=&   \int_0^{1-f(a)} dz\, l(z) \,g_0(z,a)
=\int_0^1 dz\, l(z) \,g_0(z,a)  - \int_{1-f(a)}^1 dz\, l(z) \,g_0(z,a)
\end{eqnarray}
is finite and poses no problems.

\subsection*{$\boldsymbol{I_1}$}
\begin{eqnarray}
I_1 &=& {}+ \int_0^{1-f(a)} dz \big[ l(z) - l(1) \big]   \frac{g_1(z,a)}{1-z}
+ \int_0^{1-f(a)} dz \, l(1)  \,   \frac{g_1(z,a)}{1-z}
\nonumber \\
&=&{}  + \int_0^1 dz \big[ l(z) - l(1) \big] \frac{ g_1(z,a)}{1-z}
-\int_{1-f(a)}^1 dz \big[ l(z) - l(1) \big]  \frac{g_1(z,a)}{1-z}
\nonumber\\
&& {} + \int_0^{1-f(a)} dz \, l(1)  \,  \frac{g_1(z,a)}{1-z}\,,
\end{eqnarray}
where we have added and subtracted the first term of the Taylor expansion of
$l(z)$ around the point $z=1$, and performed straightforward manipulations of
the integration limits.  The first and second integrands in the above
equation are well behaved when $z\to 1$, since the numerator goes to zero at
least as fast as $(1-z)$, cancelling the divergence of the denominator.

\subsection*{$\boldsymbol{I_2}$}
In a similar way, we can manipulate $I_2$ to have
\begin{eqnarray}
I_2 &=&
{}+ \int_0^1 dz\big[ l(z) - l(1) - (z-1)\,l^{(1)}(1) \big]   \frac{g_2(z,a)}{(1-z)^2}
\nonumber\\
&& {} -\int_{1-f(a)}^1 dz \big[ l(z) - l(1) - l^{(1)}(1) \, (z-1) \big]  \frac{g_2(z,a)}{(1-z)^2}
\nonumber\\
&& {} + \int_0^{1-f(a)} dz \big[ l(1) + l^{(1)}(1) \, (z-1)\big]   \frac{ g_2(z,a)}{(1-z)^2}\,,
\end{eqnarray}
where we have added and subtracted the first two terms of the Taylor
expansion of $l(z)$ around $z=1$.  Again the first two integrands are finite
when $z \to 1$, since the numerator is $\ord{(1-z)^2}$.

\subsection*{Final expression}
The same procedure can be applied to all the integrals $I_n$ and leads
to the final result
\begin{equation}
\label{eq:I_final}
I = \tilde{I}_1 +\tilde{I}_2 + \tilde{I}_3 \,,
\end{equation}
where
\begin{eqnarray}
%%%%%%%%%%%%%%%%%%%%%%%%%%%%%%%%%%%%%%%%%%%%%%%%%%%%%%%%%%%%%%%%%%%%%%%
%%%                       \tilde{I}_1
%%%%%%%%%%%%%%%%%%%%%%%%%%%%%%%%%%%%%%%%%%%%%%%%%%%%%%%%%%%%%%%%%%%%%%%
\label{eq:Itilde1}
\tilde{I}_1 &=& {} + \int_0^1 dz\, l(z) \,g_0(z,a)
\nonumber \\
&& {}  +  \int_0^1 dz \lq l(z) - l(1) \rq \frac{ g_1(z,a)}{1-z}
\nonumber \\
&& {} + \int_0^1 dz\lq l(z) - l(1) - l^{(1)}(1) \, (z-1) \rq
\frac{g_2(z,a)}{(1-z)^2} +\ldots
\\[5mm]
%%%%%%%%%%%%%%%%%%%%%%%%%%%%%%%%%%%%%%%%%%%%%%%%%%%%%%%%%%%%%%%%%%%%%%%
%%%                       \tilde{I}_2
%%%%%%%%%%%%%%%%%%%%%%%%%%%%%%%%%%%%%%%%%%%%%%%%%%%%%%%%%%%%%%%%%%%%%%%
\tilde{I}_2 &=& {} + \int_0^{1-f(a)} dz \, l(1)  \big[ g(z) - g_0(z,a) \big]
\nonumber\\
 && {} + \int_0^{1-f(a)} dz \, l^{(1)}(1) \, (z-1)  \lq g(z) - g_0(z,a) -
  \frac{g_1(z,a)}{1-z}\rq
\nonumber\\
&& {} + \int_0^{1-f(a)} dz \, \frac{1}{2!}\, l^{(2)}(1) \, (z-1)^2  \lq g(z) - g_0(z,a) -
\frac{g_1(z,a)}{1-z} - \frac{ g_2(z,a)}{(1-z)^2} \rq + \ldots\phantom{aaaa}
\label{eq:Itilde2}
\\[5mm]
%%%%%%%%%%%%%%%%%%%%%%%%%%%%%%%%%%%%%%%%%%%%%%%%%%%%%%%%%%%%%%%%%%%%%%%
%%%                       \tilde{I}_3
%%%%%%%%%%%%%%%%%%%%%%%%%%%%%%%%%%%%%%%%%%%%%%%%%%%%%%%%%%%%%%%%%%%%%%%
\tilde{I}_3 &=& {} - \int_{1-f(a)}^1 dz\, l(z) \,g_0(z,a)
\nonumber\\
&& {} -\int_{1-f(a)}^1 dz \big[ l(z) - l(1) \big]  \frac{g_1(z,a)}{1-z}
\nonumber\\
&& {} -\int_{1-f(a)}^1 dz \lq l(z) - l(1) - l^{(1)}(1) \, (z-1) \rq
\frac{g_2(z,a)}{(1-z)^2} + \ldots
\label{eq:Itilde3}
\end{eqnarray}
Notice that in $\tilde{I}_2$ the sum of the terms of the series add up
to give back $g(z)$, since the upper integration limit is $1-f(a)$, so that
we are within the region of convergence of the series.
The integrals in $\tilde{I}_2$ have to be evaluated exactly analytically, and
this is the harsh part of the calculation.

The integrals in $\tilde{I}_3$ can instead be computed by performing an
expansion in $a$, and this part of the calculation poses no
problems. Examples of resolution of these integrals are given in
appendix~\ref{app:details_derivation}.

Finally, by using of the plus distributions defined in
appendix~\ref{app:plus_distribs}, we can write $\tilde{I}_1$ in a more
compact form
\begin{eqnarray}
\tilde{I}_1 =\! \int_0^1 dz\, l(z) \,g_0(z,a)
  +  \int_0^1 dz \, l(z) \!\lq \frac{ g_1(z,a)}{1-z} \rq_+
  + \int_0^1 dz\, l(z) \!\lq \frac{g_2(z,a)}{(1-z)^2}\rq_{2+} +\ldots\phantom{aaa}
\end{eqnarray}
This completes our process-independent procedure for the manipulation of the
integral in eq.~(\ref{eq:I_def}).

\section{Detailed derivation of the results for $\boldsymbol{V}$ and
  $\boldsymbol{H}$ production}
\label{app:details_derivation}
In this section we present in detail how we applied the method described in
appendix~\ref{app:integration_recipe} to perform the series expansion in $a$
for every production channel of the processes at stake. In particular, we
specify for each channel which functions are assumed to be the $l(z)$ and
$g(z)$ functions of eq.~(\ref{eq:I_def}).

For ease of notation, in the following sections, the subscripts of the parton
luminosities are suppressed, since any misunderstanding is prevented by the
title of the section itself.

Also, in the summary of each of the following sections, a distinction is made
while separating the final result in a universal and a non-universal part.
As detailed in secs.~\ref{sec:part_diff_XS} and~\ref{sec:qt_integration}, the
contributions proportional to the Altarelli--Parisi splitting functions
constitute what we call the universal part of the results. The remaining ones
constitute the non-universal one.

\subsection[${V}$ production: ${qg}$ channel]{$\boldsymbol{V}$ production:
  $\boldsymbol{qg}$ channel} 
The relevant integral, corresponding to that in eq.~(\ref{eq:I_def}), is
\begin{eqnarray}
\label{eq:I_V_qg}
I &=& \int_{\tau}^{1-f(a)} dz \,\Lum\(\frac{\tau}{z}\)\left\{ \frac{1}{2z}\, 
(1+3z) (1-z)\,\sqrt{1-\frac{4az}{(1-z)^{2}}} \right.
  \nonumber \\
  && \left. {} +  \APreg_{qg}(z)\frac{1}{z}
  \left[- \log\frac{az}{(1-z)^{2}} + 2
    \log \frac{1}{2} \left(  \sqrt{1-\frac{4az}{(1-z)^2}}+1 \right)
    \right] \right\} , \phantom{aaaaa}
\end{eqnarray}
where
\begin{equation}
 \APreg_{qg}(z) = 2z^2-2z+1 \,.
\end{equation}
We can express $I$ as the sum of three integrals
\begin{equation}
I = I^a +  I^b +  I^c \,,
\end{equation}
where
\begin{eqnarray}
I^a &=& \int_{\tau}^{1-f(a)} dz \,   \frac{1}{2z}\, 
(1+3z) (1-z) \,\Lum\(\frac{\tau}{z}\) \sqrt{1-\frac{4az}{(1-z)^{2}}}  \,,
\\
I^b &=& - \int_{\tau}^{1-f(a)} dz \, \APreg_{qg}(z) \, \frac{1}{z}\,\Lum\(\frac{\tau}{z}\)
\log\frac{az}{(1-z)^{2}} \,,
\\
I^c &=&  \int_{\tau}^{1-f(a)} dz \, \APreg_{qg}(z)\,  \frac{2}{z}\,
\Lum\(\frac{\tau}{z}\) \log \frac{1}{2} \left( \sqrt{1-\frac{4az}{(1-z)^2} }
+1 \right) ,
\end{eqnarray}
and for each of the three integrals we apply the procedure detailed in
appendix~\ref{app:integration_recipe}.

\subsubsection*{Integral $\boldsymbol{I^a}$}
We define
\begin{eqnarray}
l(z) &=&   \frac{1}{2z}\,  (1+3z) \,\Lum\(\frac{\tau}{z}\) ,
\\[2mm]
g(z) &=& (1-z)\, \sqrt{1-\frac{4az}{(1-z)^{2}}} \,,
\end{eqnarray}
and expanding $g(z)$ according to eq.~(\ref{eq:gz_expansion}), we have
\begin{eqnarray}
g(z) &=& (1-z) - \frac{2az}{1-z}   - \frac{2 a^2 z^2}{(1-z)^3} + \ord{a^3}
\end{eqnarray}
so that
\begin{equation}
g_0(z,a) = 1-z\,, \qquad    g_1(z,a) = - 2az \,, \qquad   g_2(z,a)=0\,,
\qquad   g_3(z,a) = - 2 a^2 z^2\,.
\end{equation}
With this assignment of the different terms of the expansion of $g(z)$, we
perform the integrations in eqs.~(\ref{eq:Itilde1})--(\ref{eq:Itilde3}).

\subsubsection*{Integral $\boldsymbol{I^b}$}
The integrand of $I^b$ is defined up to $z=1$. For this reason, the
computation of this contribution is easier than the previous one. In
particular we can write
\begin{equation}
I^b \equiv I^b_{1} + I^b_{2} \,,
\end{equation}
where
\begin{eqnarray}
I^b_{1} &=& - \int_{0}^1 dz \, \APreg_{qg}(z) \,  \frac{1}{z}
\log\frac{az}{(1-z)^{2}} \,\Lum\(\frac{\tau}{z}\) ,
\\
 I^b_{2} &=& + \int_{1-f(a)}^1 dz \, \APreg_{qg}(z) \, \frac{1}{z}
\log\frac{az}{(1-z)^{2}} \,\Lum\(\frac{\tau}{z}\) .
\end{eqnarray}
Defining
\begin{equation}
l(z) = \left(2z^2-2z+1\right) \frac{1}{z} \,\Lum\(\frac{\tau}{z}\) ,
\end{equation}
we expand it as a power series in $(z-1)$, so that
\begin{equation}
 I^b_{2} =  \sum_{n=0}^\infty \frac{1}{n!} \, l^{(n)}(1) \int_{1-f(a)}^1 dz  \,(z-1)^n 
 \log\frac{az}{(1-z)^{2}}
 \end{equation}
and the integration becomes straightforward.

\subsubsection*{Integral $\boldsymbol{I^c}$}
We define
\begin{eqnarray}
l(z) &=&   \frac{2}{z} \(2z^2-2z+1\) \,\Lum\(\frac{\tau}{z}\) ,
\\
g(z) &=& \log \frac{1}{2} \left( \sqrt{1-\frac{4az}{(1-z)^2} } +1 \right) ,
\end{eqnarray}
and  expanding $g(z)$ according to eq.~(\ref{eq:gz_expansion}), we have
\begin{equation}
g(z) =  -\frac{a z}{(1 - z)^2} - \frac{3}{2}\frac{a^2 z^2}{(1-z)^4} +
\ord{a^3} ,
\end{equation}
so that
\begin{equation}
g_0(z,a) = g_1(z,a)=  g_3(z,a) = 0\,, \qquad      g_2(z,a)= -a z\,,
\qquad    g_4(z,a) =  - \frac{3}{2} a^2 z^2 \,.
\end{equation}
We then perform the integrations in
eqs.~(\ref{eq:Itilde1})--(\ref{eq:Itilde3}).

%\subsubsection{Summary for $\boldsymbol{V}$ production in the  $\boldsymbol{qg}$ channel}
\subsubsection{Summary}
Summarising our results, and writing $I$ in
eq.~(\ref{eq:I_V_qg}) as a sum of the universal and the non-universal part, we
have
\begin{equation}
  I = \Iu + \Id\,,
\end{equation}
where
\begin{eqnarray}
 \Iu &=& {} - \log(a) \int_{0}^1 dz \, \frac{1}{z} \, \APreg_{qg}(z)
  \,\Lum\(\frac{\tau}{z}\) 
 - \int_{0}^1 dz \, \frac{1}{z} \, \APreg_{qg}(z)\,  
\log\frac{z}{(1-z)^{2}} \,\Lum\(\frac{\tau}{z}\) 
\nonumber\\
 && {}
-2 \,a \int_0^1 dz  \,  \APreg_{qg}(z)\,\Lum\(\frac{\tau}{z}\) \lq
\frac{1}{(1-z)^2}\rq_{2+}
 - 3 \, a^2\!\int_0^1 \! dz \,z \,  \APreg_{qg}(z)\, \Lum\(\frac{\tau}{z}\) \!
\lq \frac{1}{(1-z)^4} \rq_{4+}
\nonumber \\[2mm]
&&{} + \lg  - 2 \, \lum + \, \tau \, \deronelum \rg a \log(a)
 +  \lum \, a
\nonumber \\[2mm]
&&{} + \lg  - 3 \, \lum + 3 \, \tau \, \deronelum - \frac{3}{4} \, \tau^2 \,
\dertwolum + \frac{1}{4} \, \tau^3 \, \derthreelum \rg a^2 \log(a) 
\nonumber \\[2mm]
&&{} + \lg  \frac{9}{4} \, \lum - \frac{1}{4} \, \tau \, \deronelum +
\frac{1}{4} \, \tau^2 \, \dertwolum  + \frac{1}{6} \, \tau^3 \, \derthreelum
\rg a^2  + \ord{a^\frac{5}{2}\log(a)},\phantom{aaaa}
\end{eqnarray}
\begin{eqnarray}
\Id  &=&{} + \int_{0}^{1} dz \,  \frac{1}{2z}\, 
  (1+3z) (1-z)\,\Lum\(\frac{\tau}{z}\)
  - a  \int_0^1 dz\,    (1+3z)
  \,\Lum\(\frac{\tau}{z}\)  \lq \frac{1}{1-z} \rq_+
\nonumber\\
&&{}
-  a^2  \int_0^1 \!dz \, z \,  (1+3z) \,\Lum\(\frac{\tau}{z}\) \! \lq
\frac{1}{(1-z)^3}\rq_{3+} 
\nonumber \\[2mm]
&&{} +  2 \, \lum \, a \log(a) - 2 \, \lum \, a
\nonumber \\[2mm]
&&{} + \lg  \frac{3}{2} \, \lum - \frac{3}{2} \, \tau \, \deronelum + \tau^2
\, \dertwolum  \rg a^2 \log(a) 
\nonumber \\[2mm]
&&{} + \lg - \frac{7}{4} \, \lum + \frac{5}{4} \, \tau \, \deronelum +
\frac{1}{2} \, \tau^2 \, \dertwolum \rg a^2
+ \ord{a^\frac{5}{2}\log(a)}.
\end{eqnarray}
%
%% \begin{eqnarray}
%%   I &=& - \log(a) \int_{0}^1 dz \, \frac{1}{z} \, \APreg_{qg}(z)
%%   \,\Lum\(\frac{\tau}{z}\) 
%% \nonumber\\
%%  && {}  - \int_{0}^1 dz \, \frac{1}{z} \, \APreg_{qg}(z)\,  
%% \log\frac{z}{(1-z)^{2}} \,\Lum\(\frac{\tau}{z}\) + \int_{0}^{1} dz \,  \frac{1}{2z}\, 
%% (1+3z) (1-z)\,\Lum\(\frac{\tau}{z}\)
%% \nonumber\\
%%  && {} - a  \int_0^1 dz\,    (1+3z)
%% \,\Lum\(\frac{\tau}{z}\)  \lq \frac{1}{1-z} \rq_+
%% -2 \,a \int_0^1 dz  \,  \APreg_{qg}(z)\,\Lum\(\frac{\tau}{z}\) \lq
%% \frac{1}{(1-z)^2}\rq_{2+}
%% \nonumber\\
%% &&{} -  a^2  \int_0^1 \!dz \, z \,  (1+3z) \,\Lum\(\frac{\tau}{z}\) \! \lq \frac{1}{(1-z)^3}\rq_{3+} 
%%  - 3 \, a^2\!\int_0^1 \! dz \,z \,  \APreg_{qg}(z)\, \Lum\(\frac{\tau}{z}\) \!
%% \lq \frac{1}{(1-z)^4} \rq_{4+}
%% \nonumber\\[2mm]
%% &&{} +  \tau \, \deronelum \, a \log(a)  -  \lum \,  a
%% \nonumber \\[2mm]
%% &&{} + \lq - \frac{3}{2} \, \lum + \frac{3}{2} \, \tau \, \deronelum
%% + \frac{1}{4} \, \tau^2 \, \dertwolum + \frac{1}{4} \, \tau^3 \,
%% \derthreelum   \rq a^2 \log(a) 
%% \nonumber \\[2mm]
%% &&{} + \lq \frac{1}{2} \, \lum +  \tau \, \deronelum
%% + \frac{3}{4} \, \tau^2 \, \dertwolum
%% + \frac{1}{6} \, \tau^3 \, \derthreelum   \rq a^2 + \ord{a^\frac{5}{2}}\,. \phantom{aaa}
%% \end{eqnarray}
%
%
Then, writing  $\Iu$ and $\Id$  in the form
\begin{equation}
 \Iu = \int_0^1 \frac{dz}{z} \,\Lum\(\frac{\tau}{z}\)\gu_{qg}(z)
 \,,\qquad\qquad \Id = \int_0^1 \frac{dz}{z} \,\Lum\(\frac{\tau}{z}\)
 \gd_{qg}(z) \,,
\end{equation}
we get the expression of $\gu_{qg}(z)$ and $\gd_{qg}(z)$ in
eqs.~(\ref{eq:ghat_un_V_qg}) and~(\ref{eq:ghat_nun_V_qg}), respectively.

\subsection[${V}$ production: ${q\bar{q}}$ channel]{$\boldsymbol{V}$
  production: $\boldsymbol{q\bar{q}}$ channel} 
The relevant integral, corresponding to that in eq.~(\ref{eq:I_def}), is
\begin{eqnarray}
\label{eq:I_V_qq}
I &=& \int_{\tau}^{1-f(a)} dz \,\Lum\(\frac{\tau}{z}\)\lg -\frac{2}{z}\, 
(1-z)\,\sqrt{1-\frac{4az}{(1-z)^{2}}} \right.
\nonumber \\
&& \left. +\,\frac{2}{z} \,\APnoreg_{qq}(z)
\lq- \log\frac{az}{(1-z)^{2}} + 2
\log \frac{1}{2} \(  \sqrt{1-\frac{4az}{(1-z)^2}}+1 \)
\rq \rg , \phantom{aa}
\end{eqnarray}
where
\begin{equation}
\APnoreg_{qq}(z) = \frac{1+z^2}{1-z} \,.
\end{equation}
We can express $I$ as the sum of three integrals
\begin{equation}
I = I^a +  I^b +  I^c\,,
\end{equation}
where
\begin{eqnarray}
I^a &=&  -\int_{\tau}^{1-f(a)} dz \,\Lum\(\frac{\tau}{z}\) \frac{2}{z}
(1-z)\,\sqrt{1-\frac{4az}{(1-z)^{2}}} \,,
\\
I^b &=& - \int_{\tau}^{1-f(a)} dz \, \frac{2}{z} \, \APnoreg_{qq}(z) 
 \, \Lum\(\frac{\tau}{z}\)\log\frac{az}{(1-z)^{2}} \,,
\\
I^c &=& \int_{\tau}^{1-f(a)} dz \, \frac{4}{z} \, \APnoreg_{qq}(z) 
 \, \Lum\(\frac{\tau}{z}\)  
    \log \frac{1}{2} \( \sqrt{1-\frac{4az}{(1-z)^2}}+1 \) ,
\end{eqnarray}
and for each of the three integrals we apply the procedure detailed in
appendix~\ref{app:integration_recipe}.

\subsubsection*{Integral $\boldsymbol{I^a}$}
We define
\begin{eqnarray}
l(z) &=&  - \frac{2}{z} \,\Lum\(\frac{\tau}{z}\) ,
\\[2mm]
g(z) &=& (1-z)\, \sqrt{1-\frac{4az}{(1-z)^{2}}} \,,
\end{eqnarray}
and expanding $g(z)$ according to eq.~(\ref{eq:gz_expansion}), we have
\begin{eqnarray}
g(z) &=& (1-z) - \frac{2az}{1-z}   - \frac{2 a^2 z^2}{(1-z)^3} + \ord{a^3} ,
\end{eqnarray}
so that
\begin{equation}
g_0(z,a) = 1-z\,, \qquad    g_1(z,a) = - 2az \,, \qquad   g_2(z,a)=0\,,
\qquad   g_3(z,a) = - 2 a^2 z^2\,.
\end{equation}
We then perform the integrations in
eqs.~(\ref{eq:Itilde1})--(\ref{eq:Itilde3}).

\subsubsection*{Integral $\boldsymbol{I^b}$}
We start by separating $I^b$ into two further integrals, writing
\begin{equation}
I^b = I^{b1} + I^{b2} \,,
\end{equation}
where
\begin{eqnarray}
I^{b1} &=& - \int_{\tau}^{1-f(a)} dz \, \frac{2}{z} \, \APnoreg_{qq}(z) 
\, \Lum\(\frac{\tau}{z}\)\log(az) \,,
\\
I^{b2} &=& + \int_{\tau}^{1-f(a)} dz \, \frac{4}{z} \, \APnoreg_{qq}(z) 
\, \Lum\(\frac{\tau}{z}\)\log(1-z) \,,
\end{eqnarray}
and for each of them we follow our integration and expansion procedure.
\begin{itemize}

\item{\bf Integral $\boldsymbol{I^{b1}}$}
  
We define
\begin{eqnarray}
l(z) &=&  -  \frac{2}{z} \(1+z^2 \) \log(az)\, \Lum\(\frac{\tau}{z}\) ,
\\
g(z) &=& \frac{1}{1-z} \,,
\end{eqnarray}
and we deal with this case as with a case with $g_0=0$, $g_1(z,a)=1$ and all
the other $g_i$ functions equal to 0. Then, we perform the
integrations in eqs.~(\ref{eq:Itilde1})--(\ref{eq:Itilde3}).

\item{\bf Integral $\boldsymbol{I^{b2}}$}
  
We define
\begin{eqnarray}
l(z) &=&   \frac{4}{z}  \(1+z^2 \) \Lum\(\frac{\tau}{z}\) ,
\\
g(z) &=& \frac{\log(1-z)}{1-z} \,,
\end{eqnarray}
and we deal with this case as with a case with $g_0=0$, $g_1(z,a)=\log(1-z)$
and all the other $g_i$ functions equal to 0. Then, we perform the
integrations in eqs.~(\ref{eq:Itilde1})--(\ref{eq:Itilde3}).

\end{itemize}

\subsubsection*{Integral $\boldsymbol{I^c}$}
We define
\begin{eqnarray}
l(z) &=&   \frac{4}{z} \(1+z^2\) \Lum\(\frac{\tau}{z}\) ,
\\
g(z) &=& \frac{1}{1-z}\log \frac{1}{2} \left( \sqrt{1-\frac{4az}{(1-z)^2} } +1 \right) ,
\end{eqnarray}
and  expanding $g(z)$ according to eq.~(\ref{eq:gz_expansion}), we have
\begin{equation}
g(z) =  -\frac{a z}{(1 - z)^3} - \frac{3}{2}\frac{a^2 z^2}{(1-z)^5} +
\ord{a^3} ,
\end{equation}
so that
\begin{equation}
g_0(z,a) = g_1(z,a)=  g_2(z,a) =  g_4(z,a) = 0\,, \quad      g_3(z,a)= -a z\,,
\quad    g_5(z,a) =  - \frac{3}{2} a^2 z^2 \,.
\end{equation}
We then perform the integrations in
eqs.~(\ref{eq:Itilde1})--(\ref{eq:Itilde3}).

%\subsubsection{Summary for $\boldsymbol{V}$ production in the $\boldsymbol{q\bar{q}}$ channel}
\subsubsection{Summary}

%% It can be shown that
%% \begin{equation}
%% \(1+z^2\) \lq \frac{\log(1-z)}{1-z} \rq_+ =  \lq \frac{1+z^2}{1-z}\log(1-z)
%% \rq_+ + \frac{7}{4}\, \delta(1-z)
%% \end{equation}

Summarising our results, and writing $I$ in eq.~(\ref{eq:I_V_qq}) as a sum
of a universal and non-universal part, we have
\begin{equation}
  I = \Iu + \Id\,,
\end{equation}
where
\begin{eqnarray}
\Iu &=& {} -2\log(a) \int_0^1 dz  \, \frac{1}{z} \,\Lum\(\frac{\tau}{z}\)
\APreg_{qq}(z)
\nonumber\\
&& {}
- 2  \int_0^1 dz  \, \frac{1}{z}\, \APnoreg_{qq}(z) \log(z)\, \Lum\(\frac{\tau}{z}\)
+ \int_0^1 dz  \,   \frac{4}{z} \,  (1-z) \, \hat{p}_{qq}(z)
\Lum\(\frac{\tau}{z}\)  \lq \frac{\log(1-z)}{1-z} \rq_+ 
\nonumber\\
&&{} 
-4 a \! \int_0^1 \!dz \, (1-z) \, \hat{p}_{qq}(z) \Lum\(\frac{\tau}{z}\) \lq
\frac{1}{(1-z)^3}\rq_{3+}\!\!
\nonumber\\
&& {} 
- 6 a^2 \! \int_0^1 \!dz \, z \,(1-z) \, \hat{p}_{qq}(z)
\Lum\(\frac{\tau}{z}\) \lq \frac{1}{(1-z)^5}\rq_{5+} 
\nonumber \\[2mm]
&&{} +  \lum \log(a)^2
%+  3 \, \lum  \log(a)
- \frac{\pi^2}{3} \, \lum  
\nonumber \\[2mm]
&&{} + \lg  \lum + 2 \, \tau^2 \, \dertwolum  \rg a \log(a)
 + \lg  - 2 \, \lum + 4 \, \tau \, \deronelum \rg a
\nonumber \\[2mm]
&&{} + \lg  \frac{3}{2} \, \tau^2 \, \dertwolum + \tau^3 \, \derthreelum +
\frac{1}{4} \, \tau^4 \, \derfourlum 
\rg a^2 \log(a) 
\nonumber \\[2mm]
&&{} + \lg - \lum + 2 \, \tau \, \deronelum  + \frac{5}{2} \, \tau^2 \,
\dertwolum + \frac{5}{3} \, \tau^3 \, \derthreelum + \frac{1}{6} \, \tau^4 \,
\derfourlum \rg a^2  
\nonumber\\
&& {} +\ord{a^{\frac{5}{2}}\log(a)},
\end{eqnarray}
\begin{eqnarray}
  \Id &=& {} -\int_{0}^{1} dz \,  \frac{2}{z}\,  (1-z)\,\Lum\(\frac{\tau}{z}\)
+ 4 a  \int_0^1 dz\,  \,\Lum\(\frac{\tau}{z}\)  \lq \frac{1}{1-z} \rq_+
\nonumber\\
&& {}
+ 4 a^2  \int_0^1 dz \, z \,\Lum\(\frac{\tau}{z}\)
\lq\frac{1}{(1-z)^3}\rq_{3+}
\nonumber \\[2mm]
&&{} - 2 \, \lum \, a \log(a) +  2 \, \lum \, a - \, \tau^2 \, \dertwolum \, a^2 \log(a)
\nonumber \\[2mm]
&&{} + \lg   \lum - 2 \, \tau \, \deronelum  - \frac{1}{2} \, \tau^2 \,
\dertwolum  \rg a^2
+\ord{a^{\frac{5}{2}}\log(a)},
\end{eqnarray}
%
%% \begin{eqnarray}
%% I &=& -2\log(a) \int_0^1 dz  \, \frac{1}{z} \,\Lum\(\frac{\tau}{z}\)
%% \APreg_{qq}(z)
%% \nonumber\\
%% && {}-\int_{0}^{1} dz \,  \frac{2}{z}\,  (1-z)\,\Lum\(\frac{\tau}{z}\)
%% - 2  \int_0^1 dz  \, \frac{1}{z}\, \APnoreg_{qq}(z) \log(z)\, \Lum\(\frac{\tau}{z}\)
%% \nonumber\\
%% &&{}+ \int_0^1 dz  \,   \frac{4}{z} \,  (1-z) \, \hat{p}_{qq}(z) \Lum\(\frac{\tau}{z}\)  \lq \frac{\log(1-z)}{1-z} \rq_+
%% \nonumber\\
%% &&{}+ 4 a  \int_0^1 dz\,  \,\Lum\(\frac{\tau}{z}\)  \lq \frac{1}{1-z} \rq_+  
%% -4 a \! \int_0^1 \!dz \, (1-z) \, \hat{p}_{qq}(z) \Lum\(\frac{\tau}{z}\) \lq
%% \frac{1}{(1-z)^3}\rq_{3+}\!\!
%% \nonumber\\
%% && {} + 4 a^2  \int_0^1 dz \, z \,\Lum\(\frac{\tau}{z}\)
%% \lq\frac{1}{(1-z)^3}\rq_{3+}
%% - 6 a^2 \! \int_0^1 \!dz \,z (1-z) \, \hat{p}_{qq}(z) \Lum\(\frac{\tau}{z}\) \lq \frac{1}{(1-z)^5}\rq_{5+}
%% %
%% %
%% \nonumber \\[2mm]
%% &&{} +  \lum \log^2(a)
%% %+   3 \, \lum  \,\log(a)
%% - \frac{\pi^2}{3} \, \lum  
%%  +  2 \, \tau^2 \, \dertwolum \, a \log(a)
%%  +  4 \, \tau \, \deronelum \, a
%% \nonumber \\[2mm]
%% &&{} + \lq  + \frac{1}{2} \, \tau^2 \, \dertwolum + \, \tau^3 \, \derthreelum
%% + \frac{1}{4} \, \tau^4 \, \derfourlum \rq a^2 \log(a) 
%% \nonumber \\[2mm]
%% &&{} + \lq + 2 \, \tau^2 \, \dertwolum + \frac{5}{3} \, \tau^3 \,
%% \derthreelum +\frac{1}{6} \, \tau^4 \, \derfourlum  \rq a^2 
%% \nonumber\\
%% && {} +\ord{a^{\frac{5}{2}}\log(a)}\,,
%% \end{eqnarray}
where we have written the $(1+z^2)$ terms coming from the numerator of the
$q\bar{q}$ splitting function as
\begin{equation}
1+z^2 = (1-z) \, \hat{p}_{qq}(z) \,.
\end{equation}
Then, writing  $\Iu$ and $\Id$  in the form
\begin{equation}
 \Iu = \int_0^1 \frac{dz}{z} \,\Lum\(\frac{\tau}{z}\)\gu_{q\bar{q}}(z)
 \,,\qquad\qquad \Id = \int_0^1 \frac{dz}{z} \,\Lum\(\frac{\tau}{z}\)
 \gd_{q\bar{q}}(z) \,,
\end{equation}
we get the expression of $\gu_{q\bar{q}}(z)$ and $\gd_{q\bar{q}}(z)$ in
eqs.~(\ref{eq:ghat_un_V_qq}) and~(\ref{eq:ghat_nun_V_qq}), respectively.
%% Then, writing $I$ in the form
%% \begin{equation}
%%  I = \int_0^1  \frac{dz}{z} \,\Lum\(\frac{\tau}{z}\)  \hat{g}^{\sss (1)}_{q\bar{q}}(z)\,,
%% \end{equation}
%% we get the expression of $\hat{g}^{\sss (1)}_{q\bar{q}}(z)$ in
%% eq.~(\ref{eq:Ghat_V_qq}).

%\section{Detailed derivation of the results for $\boldsymbol{H}$ production}

\subsection[${H}$ production: ${gq}$ channel]{$\boldsymbol{H}$ production:
  $\boldsymbol{gq}$ channel} 
The relevant integral, corresponding to that in eq.~(\ref{eq:I_def}), is
\begin{eqnarray}
\label{eq:I_H_gq}  
  I &=& \int_{\tau}^{1-f(a)} d z  \,\mathcal{L}\(\frac{\tau}{z}\)\lg
  -\frac{3(1-z)^2}{2z^2}\,\sqrt{1-\frac{4az}{(1-z)^2}} \right. 
  \nonumber \\
  && \left. +\,\frac{1}{z} \,\APreg_{gq}(z) 
  \lq- \log\frac{az}{(1-z)^2} + 2
    \log \frac{1}{2} \(  \sqrt{1-\frac{4az}{(1-z)^2}}+1 \)  
    \rq \rg ,
\end{eqnarray}
where
\begin{equation}
\APreg_{gq}(z) = \frac{z^2-2z+2}{z} \,.
\end{equation}
We can express $I$ as the sum of three integrals
\begin{equation}
I = I^a +  I^b +  I^c \,,
\end{equation}
where
\begin{eqnarray}
  I^a &=&  \int_{\tau}^{1-f(a)} dz \,
\lq - \frac{3}{2}\frac{(1-z)^{2}}{z^2} \rq
\Lum\(\frac{\tau}{z}\) \sqrt{1-\frac{4az}{(1-z)^2}} \,,
\\
I^b &=&  \int_{\tau}^{1-f(a)} dz \,
\( -\frac{1}{z} \) \APreg_{gq}(z)\, \Lum\(\frac{\tau}{z}\) 
\log\frac{az}{(1-z)^{2}} \,,
\\
I^c &=&  \int_{\tau}^{1-f(a)} dz \,
 \frac{2}{z}  \, \APreg_{gq}(z)
\,\Lum\(\frac{\tau}{z}\) \log \frac{1}{2} \left( \sqrt{1-\frac{4az}{(1-z)^2}
} +1 \right) ,
\end{eqnarray}
and for each of the three integrals we apply the procedure detailed in
appendix~\ref{app:integration_recipe}.

\subsubsection*{Integral $\boldsymbol{I^a}$}
We define
\begin{eqnarray}
l(z) &=&   -\frac{3}{2}\,\frac{1}{z^2} \,\Lum\(\frac{\tau}{z}\) ,
\\[2mm]
g(z) &=& (1-z)^2\, \sqrt{1-\frac{4az}{(1-z)^2}} \,,
\end{eqnarray}
and expanding $g(z)$ according to eq.~(\ref{eq:gz_expansion}), we have
\begin{eqnarray}
g(z) &=& (1-z)^2 - 2az   - \frac{2 a^2 z^2}{(1-z)^2} + \ord{a^3} ,
\end{eqnarray}
so that
\begin{equation}
g_0(z,a) = (1-z)^2 - 2az\,, \qquad    g_1(z,a) = 0 \,, \qquad
g_2(z,a) = - 2 a^2 z^2\,, \qquad   g_3(z,a) = 0\,.
\end{equation}
We then perform the integrations in
eqs.~(\ref{eq:Itilde1})--(\ref{eq:Itilde3}).

\subsubsection*{Integral $\boldsymbol{I^b}$}
The integrand of $I^b$ is defined up to $z=1$. Thus, the
computation of this contribution is easier than the previous one. In
particular, we can proceed by separating it into two further integrals
\begin{equation}
I^b \equiv I^b_{1} + I^b_{2} \,,
\end{equation}
where
\begin{eqnarray}
I^b_{1} &=&  \int_{0}^1 dz \, \( -\frac{1}{z} \) \APreg_{gq}(z) \Lum\(\frac{\tau}{z}\)
\log\frac{az}{(1-z)^{2}} \,,
\\
 I^b_{2} &=&  \int_{1-f(a)}^1 dz \, \frac{1}{z} \, \APreg_{gq}(z) \, \Lum\(\frac{\tau}{z}\)
 \log\frac{az}{(1-z)^{2}} \,.
 \end{eqnarray}
Then, after defining
\begin{equation}
l(z) =  \lq \frac{z^2-2z+2}{z^2} \rq \Lum\(\frac{\tau}{z}\) ,
\end{equation}
we expand $I^b_2$ as a power series in $(z-1)$, so that
\begin{eqnarray}
 I^b_{2} &=&  \sum_{n=0}^\infty \frac{1}{n!} \, l^{(n)}(1) \int_{1-f(a)}^1 dz  \,(z-1)^n 
 \log\frac{az}{(1-z)^{2}}\,,
\end{eqnarray}
and this integration is straightforward to be performed.

\subsubsection*{Integral $\boldsymbol{I^c}$}
We define
\begin{eqnarray}
l(z) &=&   2\, \frac{z^2-2z+2}{z^2} \,\Lum\(\frac{\tau}{z}\) ,
\\
g(z) &=& \log \frac{1}{2} \left( \sqrt{1-\frac{4az}{(1-z)^2} } +1 \right) ,
\end{eqnarray}
and  expanding $g(z)$ according to eq.~(\ref{eq:gz_expansion}), we have
\begin{equation}
g(z) =  -\frac{a z}{(1 - z)^2} - \frac{3}{2}\frac{a^2 z^2}{(1-z)^4} +
\ord{a^3} ,
\end{equation}
so that
\begin{equation}
g_0(z,a) = g_1(z,a)=  g_3(z,a) = 0\,, \qquad      g_2(z,a)= -a z\,,
\qquad    g_4(z,a) =  - \frac{3}{2} a^2 z^2 \,.
\end{equation}
We then perform the integrations in
eqs.~(\ref{eq:Itilde1})--(\ref{eq:Itilde3}).

%\subsubsection{Summary for $\boldsymbol{H}$ production in the $\boldsymbol{gq}$ channel}
\subsubsection{Summary}
Summarising our results, and writing $I$ in eq.~(\ref{eq:I_H_gq}) as a sum of
a universal and non-universal part, we have
\begin{equation}
  I = \Iu + \Id\,,
\end{equation}
where
\begin{eqnarray}
\Iu &=&  -  \log(a) \int_{0}^1 dz \,\frac{1}{z} \, \APreg_{gq}(z) \, \Lum\(\frac{\tau}{z}\)
-  \int_{0}^1 dz \, \frac{1}{z} \, \APreg_{gq}(z) \, \Lum\(\frac{\tau}{z}\)
  \log\frac{z}{(1-z)^{2}}
  \nonumber \\
&&{}  - 2\,a \int_0^1 dz \,   \, \APreg_{gq}(z) \,\Lum\(\frac{\tau}{z}\)  \lq \frac{1}{(1-z)^2}\rq_{2+}
  \nonumber \\
&&{} - 3 \, a^2\!\int_0^1 dz \,  z \, \APreg_{gq}(z)  \,\Lum\(\frac{\tau}{z}\) 
\lq \frac{1}{(1-z)^4} \rq_{4+}
\nonumber \\[2mm]
&&{} + \lg \, \lum + \, \tau \, \deronelum \rg a \log(a) + \lum \, a
\nonumber \\[2mm]
&&{} + \lg 3 \, \tau \, \deronelum + \frac{3}{2} \, \tau^2 \, \dertwolum +
\frac{1}{4} \, \tau^3 \, \derthreelum \rg a^2 \log(a) 
\nonumber \\[2mm]
&&{} + \lg 2 \, \lum + \frac{17}{4} \, \tau \,\deronelum
+ \frac{7}{4} \, \tau^2 \, \dertwolum  + \frac{1}{6} \, \tau^3 \, \derthreelum  \rg a^2 
+ \ord{a^\frac{5}{2}\log(a)} ,\phantom{aaaaa}
\end{eqnarray}
\begin{eqnarray}
\Id &=&
 - \frac{3}{2}  \int_{0}^{1} dz \,  \frac{1}{z^2} (1-z)^2 
 \,\Lum\(\frac{\tau}{z}\)
+ 3\,a \int_{0}^{1} dz \, \frac{1}{z}  \,\Lum\(\frac{\tau}{z}\)
 \nonumber \\
 &&{} +  3 \, a^2  \int_0^1 dz \, \Lum\(\frac{\tau}{z}\)  \lq \frac{1}{(1-z)^2}\rq_{2+}
  \nonumber \\
  &&{}
\nonumber \\[2mm]
&&{}  - \frac{3}{2} \, \tau \, \deronelum \, a^2 \log(a)
 + \lg - \frac{3}{2} \, \lum  - \frac{3}{4} \, \tau \, \deronelum \rg a^2
  + \ord{a^\frac{5}{2}\log(a)} .
\end{eqnarray}
%
%% \begin{eqnarray}
%%   I &=&  -  \log(a) \int_{0}^1 dz \, \frac{1}{z}\, \APreg_{gq}(z) \, \Lum\(\frac{\tau}{z}\)  
%%   \nonumber \\
%% &&{}   -  \int_{0}^1 dz \, \frac{1}{z} \, \APreg_{gq}(z) \, \Lum\(\frac{\tau}{z}\)
%%   \log\frac{z}{(1-z)^{2}} 
%%   - \frac{3}{2} \int_{0}^{1} dz \, \frac{1}{z^2}\,(1-z)^2 \,\Lum\(\frac{\tau}{z}\)
%%  \nonumber \\
%%  &&{} +3\, a \int_{0}^{1} dz \, \frac{1}{z} \,\Lum\(\frac{\tau}{z}\)
%%   - 2\,a \int_0^1 dz \,   \, \APreg_{gq}(z) \,\Lum\(\frac{\tau}{z}\)  \lq \frac{1}{(1-z)^2}\rq_{2+}
%%  \nonumber \\
%%  &&{} +  3 \, a^2  \int_0^1 dz \, \Lum\(\frac{\tau}{z}\)  \lq \frac{1}{(1-z)^2}\rq_{2+}
%%   - 3 \, a^2\!\int_0^1 dz \,  z \, \APreg_{gq}(z)  \,\Lum\(\frac{\tau}{z}\) 
%% \lq \frac{1}{(1-z)^4} \rq_{4+}
%% \nonumber \\[2mm]
%% &&{} + \lg \, \lum + \, \tau \, \deronelum \rq a \log(a) +  \lum \, a
%% \nonumber \\[2mm]
%% &&{} + \lq \frac{3}{2} \, \tau \, \deronelum + \frac{3}{2} \, \tau^2 \, \dertwolum +
%% \frac{1}{4} \, \tau^3 \, \derthreelum \rg a^2 \log(a) 
%% \nonumber \\[2mm]
%% &&{} + \lg \frac{1}{2} \, \lum + \frac{7}{2} \, \tau \, \deronelum + \frac{7}{4} \, \tau^2 \,
%% \dertwolum + \frac{1}{6} \, \tau^3 \, \derthreelum \rg a^2 
%% \nonumber \\[2mm]
%% &&{} + \ord{a^\frac{5}{2}} . \phantom{aaaaa} 
%% \end{eqnarray}
%
%
Then, writing  $\Iu$ and $\Id$  in the form
\begin{equation}
 \Iu = \int_0^1 \frac{dz}{z} \,\Lum\(\frac{\tau}{z}\)\gu_{gq}(z)
 \,,\qquad\qquad \Id = \int_0^1 \frac{dz}{z} \,\Lum\(\frac{\tau}{z}\)
 \gd_{gq}(z) \,,
\end{equation}
we get the expression of $\gu_{gq}(z)$ and $\gd_{gq}(z)$ in
eqs.~(\ref{eq:ghat_un_H_gq}) and~(\ref{eq:ghat_nun_H_gq}), respectively.
%% Then, writing  $I$  in the form
%% \begin{equation}
%%  I = \int_0^1   \frac{dz}{z} \,\Lum\(\frac{\tau}{z}\) \hat{g}^{\sss (1)}_{gq}(z)\,,
%% \end{equation}
%% we get the expression of $\hat{g}^{\sss (1)}_{gq}(z)$ in eq.~(\ref{eq:Ghat_H_gq}).

\subsection[${H}$ production: ${gg}$ channel]{$\boldsymbol{H}$ production: $\boldsymbol{gg}$ channel}
The relevant integral, corresponding to that in eq.~(\ref{eq:I_def}), is
\begin{eqnarray}
\label{eq:I_H_gg}  
  I &=& \int_\tau^{1-f(a)} dz \,\mathcal{L}\(\frac{\tau}{z}\)\lg
  -\frac{11}{3} \frac{(1-z)^3}{z^2}\,\sqrt{1-\frac{4az}{(1-z)^2}} \right.
 % \nonumber \\
 % &&
  +\, \frac{2}{3}\, a \,\frac{1-z}{z} \sqrt{1-\frac{4az}{(1-z)^2}}
  \nonumber \\
  &&\left.  {} + \frac{2}{z}\, \APnoreg_{gg}(z) 
  \lq - \log\frac{az}{(1-z)^2} + 2 \log \frac{1}{2} \(
  \sqrt{1-\frac{4az}{(1-z)^2}}+1 \)  \rq \rg ,  \phantom{aaa}
\end{eqnarray}
where
\begin{equation}
\APnoreg_{gg}(z)= \frac{2 (z^2-z+1)^2}{z(1-z)} \,.
\end{equation}
We can express $I$ as the sum of four integrals
\begin{equation}
I = I^{a1} + I^{a2} +  I^b +  I^c \,,
\end{equation}
where
\begin{eqnarray}
  I^{a1} &=&  \int_{\tau}^{1-f(a)} dz
  \( -\frac{11}{3} \) \frac{(1-z)^3}{z^2} \, \Lum\(\frac{\tau}{z}\)
  \sqrt{1-\frac{4az}{(1-z)^2}} \,, 
\\
  I^{a2} &=&  \int_{\tau}^{1-f(a)} dz \,
 \frac{2}{3}\, a \,\frac{1-z}{z} \, \Lum\(\frac{\tau}{z}\) \sqrt{1-\frac{4az}{(1-z)^2}} \,,
  \\
  I^b &=&  \int_{\tau}^{1-f(a)} dz \, \( -\frac{2}{z}\)  \APnoreg_{gg}(z) 
  \Lum\(\frac{\tau}{z}\)\log\frac{az}{(1-z)^{2}}  \,,
\\
I^c &=& \int_{\tau}^{1-f(a)} dz \,\frac{4}{z} \,\APnoreg_{gg}(z)
 \, \Lum\(\frac{\tau}{z}\)  
    \log \frac{1}{2} \( \sqrt{1-\frac{4az}{(1-z)^2}}+1 \) ,
\end{eqnarray}
and for each of the four integrals we apply the procedure detailed in
appendix~\ref{app:integration_recipe}.

\subsubsection*{Integral $\boldsymbol{I^{a1}}$}
We define
\begin{eqnarray}
l(z) &=&  -\frac{11}{3} \,\frac{1}{z^2} \,\Lum\(\frac{\tau}{z}\) ,
\\[2mm]
g(z) &=& (1-z)^3\, \sqrt{1-\frac{4az}{(1-z)^{2}}} \,,
\end{eqnarray}
and expanding $g(z)$ according to eq.~(\ref{eq:gz_expansion}), we have
\begin{eqnarray}
g(z) &=& (1-z)^3 - {2az}(1-z)   - \frac{2 a^2 z^2}{1-z} + \ord{a^3} ,
\end{eqnarray}
so that
\begin{eqnarray}
&&g_0(z,a) =  (1-z)^3 - {2az}(1-z)\,, \qquad    g_1(z,a) = - 2 a^2 z^2\,,
\\[2mm]
&&g_2(z,a) = 0\,, \qquad    g_3(z,a) = \ord{a^3}\,.
\end{eqnarray}
We then perform the integrations in
eqs.~(\ref{eq:Itilde1})--(\ref{eq:Itilde3}).

\subsubsection*{Integral $\boldsymbol{I^{a2}}$}
We define
\begin{eqnarray}
l(z) &=&  \frac{2}{3} \, a \,\frac{1}{z} \,\Lum\(\frac{\tau}{z}\) ,
\\[2mm]
g(z) &=& (1-z)\, \sqrt{1-\frac{4az}{(1-z)^{2}}} \,,
\end{eqnarray}
and expanding $g(z)$ according to eq.~(\ref{eq:gz_expansion}), we have
\begin{eqnarray}
g(z) &=& (1-z) - \frac{2az}{1-z}   - \frac{2 a^2 z^2}{(1-z)^3} + \ord{a^3} ,
\end{eqnarray}
so that
\begin{equation}
g_0(z,a) =  (1-z)\,, \qquad    g_1(z,a) = - 2 a z\,, \qquad    g_2(z,a) =
0\,, \qquad    g_3(z,a) = - 2 a^2 z^2\,. 
\end{equation}
We then perform the integrations in
eqs.~(\ref{eq:Itilde1})--(\ref{eq:Itilde3}).

\subsubsection*{Integral $\boldsymbol{I^b}$}
We start by separating $I^b$ into two further integrals
\begin{equation}
I^b = I^{b1} + I^{b2} \,,
\end{equation}
where
\begin{eqnarray}
I^{b1} &=&  \int_{\tau}^{1-f(a)} dz   \( -\frac{2}{z}\)  \APnoreg_{gg}(z) \,
 \Lum\(\frac{\tau}{z}\)\log(az) \,,
\\
I^{b2} &=&  \int_{\tau}^{1-f(a)} dz  \, \frac{4}{z}\,  \APnoreg_{gg}(z) 
\, \Lum\(\frac{\tau}{z}\)\log(1-z) \,,
\end{eqnarray}
and for each of them we follow our integration and expansion procedure.
\begin{itemize}
  
\item{\bf Integral $\boldsymbol{I^{b1}}$}
  
We define
\begin{eqnarray}
l(z) &=&  -\frac{4\(z^2-z+1\)^2}{z^2} \log(az)\, \Lum\(\frac{\tau}{z}\) ,
\\
g(z) &=& \frac{1}{1-z} \,,
\end{eqnarray}
We deal with this case as with a case with $g_0=0$, $g_1(z,a)=1$ and all the
other $g_i$ functions equal to 0. Then, we perform the integrations in
eqs.~(\ref{eq:Itilde1})--(\ref{eq:Itilde3}).

\item{\bf Integral $\boldsymbol{I^{b2}}$}
  
We define
\begin{eqnarray}
l(z) &=&   \frac{8\(z^2-z+1\)^2}{z^2} \Lum\(\frac{\tau}{z}\)
\\
g(z) &=& \frac{\log(1-z)}{1-z} 
\end{eqnarray}
We deal with this case as with a case with $g_0=0$, $g_1(z,a)=\log(1-z)$ and all
the other $g_i$ functions equal to 0. Then, we perform the integrations in
eqs.~(\ref{eq:Itilde1})--(\ref{eq:Itilde3}).

\end{itemize}

\subsubsection*{Integral $\boldsymbol{I^c}$}
We define
\begin{eqnarray}
l(z) &=&   \frac{8\(z^2-z+1\)^2}{z^2 } \, \Lum\(\frac{\tau}{z}\) ,
\\
g(z) &=& \frac{1}{1-z}\log \frac{1}{2} \left( \sqrt{1-\frac{4az}{(1-z)^2} } +1 \right) ,
\end{eqnarray}
and  expanding $g(z)$ according to eq.~(\ref{eq:gz_expansion}), we have
\begin{equation}
g(z) =  -\frac{a z}{(1 - z)^3} - \frac{3}{2}\frac{a^2 z^2}{(1-z)^5} +
\ord{a^3} ,
\end{equation}
so that
\begin{equation}
g_0(z,a) = g_1(z,a)=  g_2(z,a) =  g_4(z,a) = 0\,, \quad      g_3(z,a)= -a z\,,
\quad    g_5(z,a) =  - \frac{3}{2} a^2 z^2 \,.
\end{equation}
We then perform the integrations in
eqs.~(\ref{eq:Itilde1})--(\ref{eq:Itilde3}).

%\subsubsection{Summary for $\boldsymbol{H}$ production in the $\boldsymbol{gg}$ channel}
\subsubsection{Summary}
Summarising our results, and writing $I$ in eq.~(\ref{eq:I_H_gg}) as a sum of a
universal and non-universal part, we have
\begin{equation}
  I = \Iu + \Id\,,
\end{equation}
where
\begin{eqnarray}
\Iu &=&
-2\log(a) \int_0^1 dz  \, \frac{1-z}{z} \, \APnoreg_{gg}(z)
 \Lum\(\frac{\tau}{z}\)   \lq \frac{1}{1-z} \rq_+
\nonumber\\
&& {}-2 \int_0^1 dz \,  \frac{1}{z} \, \APnoreg_{gg}(z) \log(z)
\Lum\(\frac{\tau}{z}\)  
\nonumber\\
&& {}  + 4 \int_0^1 dz  \,  \frac{1-z}{z} \, \APnoreg_{gg}(z)
\Lum\(\frac{\tau}{z}\)  \lq \frac{\log(1-z)}{1-z} \rq_+ 
\nonumber\\
&& {}- 4\,a  \int_0^1 dz \,(1-z) \,\APnoreg_{gg}(z)  \,
\Lum\(\frac{\tau}{z}\) \lq \frac{1}{(1-z)^3}\rq_{3+} 
\nonumber\\
&& {}- 6 \, a^2  \int_0^1 dz
\, z(1-z)  \,\APnoreg_{gg}(z)  \Lum\(\frac{\tau}{z}\) \lq \frac{1}{(1-z)^5}\rq_{5+}
\nonumber \\[2mm]
&&{} +  \lum \, \log^2(a) - \frac{\pi^2}{3} \, \lum
\nonumber \\[2mm]
&&{}
 + \lg 8 \, \lum + 2 \, \tau^2 \, \dertwolum \rg a \log(a)
 + \lg  - 2 \, \lum  + 4 \, \tau \, \deronelum \rg a
\nonumber \\[2mm]
&&{} + \lg  6 \, \lum + 6 \, \tau^2 \,
\dertwolum + \, \tau^3 \, \derthreelum + \frac{1}{4} \, \tau^4 \, \derfourlum\rg a^2 \log(a) 
\nonumber \\[2mm]
&&{} + \lg  - \frac{3}{2} \, \lum + 11 \, \tau \, \deronelum  + \frac{11}{2}
\, \tau^2 \, \dertwolum + \frac{5}{3} \, \tau^3 \, \derthreelum + \frac{1}{6}
\, \tau^4 \, \derfourlum \rg a^2 
\nonumber \\[2mm]
&& {} + \ord{a^\frac{5}{2}\log(a)},
\end{eqnarray}
\begin{eqnarray}
\Id &=& -\frac{11}{3} \int_{0}^{1} dz \, \frac{1}{z^2} \,(1-z)^3 
\,\Lum\(\frac{\tau}{z}\)
 + 8  a\int_{0}^{1} dz \,   \frac{1}{z} (1-z)
\,\Lum\(\frac{\tau}{z}\)
\nonumber\\
&&{}  +  6 \,a^2 \!\int_0^1 dz\, \Lum\(\frac{\tau}{z}\)  \lq \frac{1}{1-z} \rq_+
\nonumber \\
&&{}   - 3 \, \lum \, a^2 \log(a)  - \frac{5}{2} \, \lum \, a^2 +
\ord{a^\frac{5}{2}\log(a)} .
\end{eqnarray}
%
%% \begin{eqnarray}
%% I &=& -2\log(a) \int_0^1 dz  \, \frac{1-z}{z} \, \APnoreg_{gg}(z)
%%  \Lum\(\frac{\tau}{z}\)   \lq \frac{1}{1-z} \rq_+
%% \nonumber\\
%% && {} - \frac{11}{3} \int_{0}^{1} dz \,  \frac{1}{z^2} (1-z)^3 
%% \,\Lum\(\frac{\tau}{z}\)
%% -2 \int_0^1 dz \,  \frac{1}{z} \, \APnoreg_{gg}(z) \log(z) \Lum\(\frac{\tau}{z}\)  
%% \nonumber\\
%% && {}  + 4 \int_0^1 dz  \,  \frac{1-z}{z} \, \APnoreg_{gg}(z)
%% \Lum\(\frac{\tau}{z}\)  \lq \frac{\log(1-z)}{1-z} \rq_+  
%% \nonumber\\
%% && {} + 8\, a \int_0^1 dz \, \frac{1}{z} (1-z)
%% \,\Lum\(\frac{\tau}{z}\)
%% - 4\,a  \int_0^1 dz \,(1-z) \,\APnoreg_{gg}(z)  \, \Lum\(\frac{\tau}{z}\) \lq
%% \frac{1}{(1-z)^3}\rq_{3+}
%% \nonumber\\
%% && {}  + 6 \,a^2 \!\int_0^1 dz\, \Lum\(\frac{\tau}{z}\)  \lq \frac{1}{1-z} \rq_+
%% - 6 \, a^2  \int_0^1 dz
%% \, z(1-z)  \,\APnoreg_{gg}(z)  \Lum\(\frac{\tau}{z}\) \lq \frac{1}{(1-z)^5}\rq_{5+}
%% \nonumber \\[2mm]
%% &&{} +  \lum \log^2(a)  -  \frac{\pi^2}{3} \, \lum  + \lq  8 \, \lum  +2 \, \tau^2 \, \dertwolum \rq a \log(a)
%% \nonumber \\[2mm]
%% &&{}
%% + \lq - 2 \, \lum + 4 \, \tau \, \deronelum  \rq a
%% \nonumber \\[2mm]
%% &&{} + \lq + 3 \, \lum  + 6 \, \tau^2 \, \dertwolum + \tau^3 \, \derthreelum
%% + \frac{1}{4} \, \tau^4 \, \derfourlum \rq a^2 \log(a) 
%% \nonumber \\[2mm]
%% &&{} + \lq - 4 \, \lum  + 11 \, \tau \, \deronelum + \frac{11}{2} \, \tau^2 \, \dertwolum +
%% \frac{5}{3} \, \tau^3 \, \derthreelum + \frac{1}{6} \, \tau^4 \,
%% \derfourlum \rq a^2 
%% \nonumber\\
%% && {} + \ord{a^\frac{5}{2}\log(a)} \,.
%% \end{eqnarray}
%
%
Then, writing  $\Iu$ and $\Id$  in the form
\begin{equation}
 \Iu = \int_0^1 \frac{dz}{z} \,\Lum\(\frac{\tau}{z}\)\gu_{gg}(z)
 \,,\qquad\qquad \Id = \int_0^1 \frac{dz}{z} \,\Lum\(\frac{\tau}{z}\)
 \gd_{gg}(z) \,,
\end{equation}
we get the expression of $\gu_{gg}(z)$ and $\gd_{gg}(z)$ in
eqs.~(\ref{eq:ghat_un_H_gg}) and~(\ref{eq:ghat_nun_H_gg}), respectively.
%% Then, writing  $I$  in the form
%% \begin{equation}
%%  I = \int_0^1   \frac{dz}{z} \,\Lum\(\frac{\tau}{z}\) \hat{g}^{\sss (1)}_{gg}(z)\,,
%% \end{equation}
%% we get the expression of $\hat{g}^{\sss (1)}_{gg}(z)$ in eq.~(\ref{eq:Ghat_H_gg}).

\subsection[Study of a universal term of the form ${1/(1-z)}$]{Study of a
  universal term of the form $\boldsymbol{1/(1-z)}$} 
\label{app:soft_coll_univ}
In this section we apply the procedure described in
appendix~\ref{app:integration_recipe} to study the universal part of the
Altarelli--Parisi splitting functions that accounts for soft radiation,
i.e.~the $z\to 1$ limit. In this approximation, the Altarelli--Parisi
splitting functions $\APnoreg_{qq}(z)$ and $\APnoreg_{gg}(z)$ behave like
$1/(1-z)$. The relevant integral, corresponding to that in
eq.~(\ref{eq:I_def}), is given by
\begin{eqnarray}
  \Iu &=& \int_\tau^{1-f(a)} dz \,\mathcal{L}\(\frac{\tau}{z}\)
\frac{1}{z} \, p(z)
  \lq - \log\frac{az}{(1-z)^2} + 2 \log \frac{1}{2} \(
  \sqrt{1-\frac{4az}{(1-z)^2}}+1 \)  \rq\!, \phantom{aaaa}
\end{eqnarray}
where
\begin{equation}
p(z)= \frac{1}{1-z}\,.
\end{equation}
We write $I$ as the sum of two integrals
\begin{equation}
I =   I^b +  I^c\,,
\end{equation}
where
\begin{eqnarray}
  I^b &=&  \int_{\tau}^{1-f(a)} dz \, \( -\frac{1}{z}\)  p(z) 
  \Lum\(\frac{\tau}{z}\)\log\frac{az}{(1-z)^{2}}  \,,
\\
I^c &=& \int_{\tau}^{1-f(a)} dz \,\frac{2}{z} \,p(z)
 \, \Lum\(\frac{\tau}{z}\)  
    \log \frac{1}{2} \( \sqrt{1-\frac{4az}{(1-z)^2}}+1 \),
\end{eqnarray}
and for each of the two integrals we apply the procedure detailed in
appendix~\ref{app:integration_recipe}.

\subsubsection*{Integral $\boldsymbol{I^b}$}
We write $I^b$ as sum of two further integrals
\begin{equation}
I^b = I^{b1} + I^{b2}\,,
\end{equation}
where
\begin{eqnarray}
I^{b1} &=&  \int_{\tau}^{1-f(a)} dz   \( -\frac{1}{z}\) p(z) \,
 \Lum\(\frac{\tau}{z}\)\log(az)\,,
\\
I^{b2} &=&  \int_{\tau}^{1-f(a)} dz  \, \frac{2}{z}\,  p(z) 
\, \Lum\(\frac{\tau}{z}\)\log(1-z)\,,
\end{eqnarray}

\subsubsection*{Integral $\boldsymbol{I^{b1}}$}
We define
\begin{eqnarray}
l(z) &=&  -\frac{1}{z} \log(az)\, \Lum\(\frac{\tau}{z}\),
\\
g(z) &=& \frac{1}{1-z} \,,
\end{eqnarray}
and we treat this case as the case with $g_0(z,a)=0$ and $g_1(z,a)=1$ and all the
other $g_i$ functions equal to 0. We then perform the integrations in
eqs.~(\ref{eq:Itilde1})--(\ref{eq:Itilde3}).

\subsubsection*{Integral $\boldsymbol{I^{b2}}$}
We define
\begin{eqnarray}
l(z) &=&   \frac{2}{z} \, \Lum\(\frac{\tau}{z}\),
\\[2mm]
g(z) &=& \frac{\log(1-z)}{1-z} \,,
\end{eqnarray}
and we treat this case as the case with $g_0(z,a)=0$ and $g_1(z,a)=\log(1-z)$ and
all the other $g_i$ functions equal to 0. We then perform the integrations in
eqs.~(\ref{eq:Itilde1})--(\ref{eq:Itilde3}).

\subsubsection*{Integral $\boldsymbol{I^c}$}
We define
\begin{eqnarray}
l(z) &=&   \frac{2}{z } \, \Lum\(\frac{\tau}{z}\) ,
\\
g(z) &=& \frac{1}{1-z}\log \frac{1}{2} \left(\sqrt{1-\frac{4az}{(1-z)^2}} +1 \right), 
\end{eqnarray}
and  expanding $g(z)$ according to eq.~(\ref{eq:gz_expansion}), we have
\begin{equation}
g(z) =  -\frac{a z}{(1 - z)^3} - \frac{3}{2}\frac{a^2 z^2}{(1-z)^5} + \ord{a^3},
\end{equation}
so that
\begin{equation}
g_0(z,a) = g_1(z,a)=  g_2(z,a) =  g_4(z,a) = 0\,, \quad      g_3(z,a)= -a z\,,
\quad    g_5(z,a) =  - \frac{3}{2} a^2 z^2 \,.
\end{equation}
We then perform the integrations in
eqs.~(\ref{eq:Itilde1})--(\ref{eq:Itilde3}).

%\subsubsection{Summary for the $\boldsymbol{1/(1-z)}$ term}
\subsubsection{Summary}

Summarising our results, we have
\begin{eqnarray}
  \Iu &=& {} -\log(a) \int_0^1 dz  \, \frac{1}{z } 
  \Lum\(\frac{\tau}{z}\)   \lq \frac{1}{1-z} \rq_+
\nonumber \\[2mm]
&&{}   - \int_0^1 dz  \, \frac{1}{z (1-z)} \log(z)\,
\Lum\(\frac{\tau}{z}\)
+  \int_0^1 dz  \,   \frac{2}{z} \Lum\(\frac{\tau}{z}\) \lq
  \frac{\log(1-z)}{1-z} \rq_+
\nonumber \\[2mm]
&&{}
-2 a  \int_0^1 dz \, \Lum\(\frac{\tau}{z}\) \lq \frac{1}{(1-z)^3}\rq_{3+}
- 3 a^2  \int_0^1 dz \,
z \, \Lum\(\frac{\tau}{z}\) \lq
\frac{1}{(1-z)^5}\rq_{5+}
\nonumber \\[2mm]
&&{} +  \frac{1}{4} \, \lum  \log^2(a)  - \frac{\pi^2}{12} \, \lum  
 + \lg \tau \, \deronelum  + \frac{1}{2} \, \tau^2 \, \dertwolum   \rg a \log(a)
\nonumber \\[2mm]
&&{} + \lg \frac{1}{2} \, \lum +  \tau \, \deronelum  \rg a
\nonumber \\[2mm]
&&{} + \lg \frac{3}{4} \, \tau^2 \, \dertwolum + \frac{1}{2} \, \tau^3 \,
\derthreelum + \frac{1}{16} \, \tau^4 \, \derfourlum \rg a^2 \log(a)
\nonumber \\[2mm]
&&{} + \lg  - \frac{1}{8} \, \lum + \frac{1}{2} \, \tau \, \deronelum + \frac{13}{8}
\, \tau^2 \, \dertwolum
+ \frac{7}{12} \, \tau^3 \, \derthreelum  + \frac{1}{24} \,
\tau^4 \, \derfourlum  \rg a^2
\nonumber\\
&& {} + \ord{a^\frac{5}{2}\log(a)}.
\end{eqnarray}
Then, writing $\Iu$ in the form 
\begin{equation}
 \Iu = \int_0^1 dz \, \frac{1}{z} \,\Lum\(\frac{\tau}{z}\) \gu(z)\,,
\end{equation}
we get
\begin{eqnarray}
  \gu(z) &=&  {} +  \frac{1}{4} \, \delta(1 - z)  \log^2(a)
  -  \lq \frac{1}{1-z} \rq_+ \log(a)
 - \frac{\pi^2}{12} \, \delta(1 - z)  
\nonumber \\[2mm]
&&{} 
  -   \, \frac{1}{(1-z)} \log(z)\,
+     2 \lq   \frac{\log(1-z)}{1-z} \rq_+
\nonumber \\[2mm]
&&{} + \lg \frac{1}{2} \delta^{(2)}(1 - z) - \delta^{(1)}(1 - z) \rg a
\log(a)
\nonumber \\[2mm]
&&{}
+ \lg  - \frac{1}{2}\, \delta(1 - z) + \delta^{(1)}(1 - z) -2  z  \lq
\frac{1}{(1-z)^3}\rq_{3+}\rg a 
\nonumber \\[2mm]
&&{} + \lg \frac{3}{4} \, \delta^{(2)}(1 - z) - \frac{1}{2} \,\delta^{(3)}(1 -
z) + \frac{1}{16} \,\delta^{(4)}(1 - z) \rg a^2 \log(a) 
\nonumber \\[2mm]
&&{} + \lg \frac{1}{8}\, \delta(1 - z) + \frac{1}{2}\, \delta^{(1)}(1 - z) -
\frac{5}{8} \,\delta^{(2)}(1 - z)   - \frac{1}{12} \,\delta^{(3)}(1 - z)
\right.
\nonumber \\[2mm]
&&{} \hspace{1cm} \left. +  \frac{1}{24} \,\delta^{(4)}(1 - z)
-3 z^2  \lq \frac{1}{(1-z)^5}\rq_{5+}  \rg a^2
\nonumber\\[2mm]
&& {} + \ord{a^\frac{5}{2}\log(a)}.
\end{eqnarray}

\section{Altarelli--Parisi splitting functions}
\label{app:AP}
The zero-order Altarelli--Parisi splitting functions are defined as
\begin{eqnarray}
P_{qq}(z) &=& P_{\bar{q}\bar{q}}(z) = \CF \lq \frac{1+z^2}{(1-z)_+} + \frac{3}{2}
\delta(1-z)\rq = \CF \lq \frac{1+z^2}{1-z}\rq_+
\nonumber
\\[2mm]
&\equiv& \CF \, \APreg_{qq}(z)  + \frac{3}{2} \CF\, \delta(1-z)\,,
\\[2mm]
P_{qg}(z) &=& P_{\bar{q}g}(z) = \TR \lq  z^2 + (1-z)^2 \rq
= \TR \lq  2 z^2 -2 z+ 1 \rq \equiv \TR \, \APreg_{qg}(z) \,,
\\[2mm]
P_{gq}(z) &=& P_{g\bar{q}}(z) = \CF \lq \frac{1+(1-z)^2}{z} \rq =
\CF \lq \frac{z^2 -2z + 2}{z}\rq \equiv \CF \, \APreg_{gq}(z) \,,
\\[2mm]
P_{gg}(z) &=&  2 \CA \lq \frac{z}{(1-z)_+} + \frac{1-z}{z} + z(1-z)\rq +
\frac{1}{6}\lq 11\CA -4 \nf\TR \rq \delta(1-z) 
\nonumber\\
&\equiv& \CA \, \APreg_{gg}(z) + \frac{1}{6}\lq 11\CA -4 \nf\TR \rq
\delta(1-z) \,.
\end{eqnarray}
The unregularised Altarelli--Parisi splitting functions are given by
\begin{eqnarray}
\hat{P}_{qq}(z) &=& \hat{P}_{\bar{q}\bar{q}}(z)  = \CF  \frac{1+z^2}{1-z}
\equiv \CF \, \APnoreg_{qq}(z) \,,
%% \\[2mm]
%% \hat{P}_{qg}(z) &=& \hat{P}_{\bar{q}g}(z) = \TR \lq  z^2 + (1-z)^2 \rq
%% = \TR \lq  2 z^2 -2 z+ 1 \rq \equiv \TR  \,\APnoreg_{qg}(z) 
%% \\[2mm]
%% \hat{P}_{gq}(z) &=& \hat{P}_{g\bar{q}}(z) = \CF \lq \frac{1+(1-z)^2}{z} \rq =
%% \CF \lq \frac{z^2 -2z + 2}{z}\rq \equiv \CF  \,\APnoreg_{gq}(z)
\\[2mm]
\hat{P}_{gg}(z) &=&  2 \,\CA \lq \frac{z}{1-z} + \frac{1-z}{z} + z(1-z)\rq
= \CA \frac{2 \(z^2-z+1\)^2}{z(1-z)} \equiv  \CA   \,\APnoreg_{gg}(z) \,. \phantom{aaa}
\end{eqnarray}

\section{Plus distributions}
\label{app:plus_distribs}
We define a plus distribution of order $n$ as
\begin{equation}
\int_0^1 dz\, l(z) \lq g(z) \rq_{n+} \equiv \int_0^1 dz \lg
l(z) - \sum_{i=0}^{n-1}  \frac{1}{i!} \, l^{(i)}(1)\, (z-1)^i \rg  g(z) \,,
\end{equation}
where $g(z)$ has a pole of order $n$ for $z=1$, and $l(z)$ is a continuous
function around $z=1$, together with all its derivatives up to order $(n-1)$.
For example, the first three plus distributions read
\begin{eqnarray}
\int_0^1 dz\, l(z) \lq g(z)\rq_+ &\equiv& \int_0^1 dz\lg l(z) - l(1) \rg g(z)\,,
\\
\int_0^1 dz\, l(z) \lq g(z) \rq_{2+} &\equiv& \int_0^1 dz  \lg
l(z)-l(1)- l^{(1)}(1)\,(z-1) \rg  g(z) \,,
\\
\int_0^1 dz\, l(z) \lq g(z) \rq_{3+} &\equiv& \int_0^1 dz \lg
l(z)-l(1)- l^{(1)}(1)\,(z-1) -\frac{1}{2!}\,l^{(2)}(1)\,(z-1)^2 \rg  g(z)
\,. \phantom{aaaa}
\end{eqnarray}
With simple manipulations, some useful identities follow
\begin{eqnarray}
\label{eq:ratio+}
\int_0^1 dz\, l(z) \lq \frac{n(z)}{d(z)}\rq_+ &=& 
\int_0^1 dz \lg l(z) \, n(z) \lq \frac{1}{d(z)}\rq_+ - l(1)\, n(z)
\lq \frac{1}{d(z)} \rq_+ \rg ,
\\
\label{eq:ratio2+}
\int_0^1 dz\, l(z) \lq \frac{n(z)}{d(z)}\rq_{2+}\!  &=& \int_0^1 dz \lg
l(z)\, n(z) \lq \frac{1}{d(z)}\rq_{2+} - l(1) \, n(z) \lq \frac{1}{d(z)}\rq_{2+} \right.
\nonumber\\
&&\left. {} - l^{(1)}(1)\, n(z)  \lq \frac{z-1}{d(z)}\rq_{+} \rg ,
\\
\label{eq:ratio3+}
\int_0^1 dz\, l(z) \lq \frac{n(z)}{d(z)}\rq_{3+} \!\!&=& \int_0^1 dz \lg
l(z)\, n(z) \lq \frac{1}{d(z)}\rq_{3+} - l(1) \, n(z) \lq \frac{1}{d(z)}\rq_{3+} \right.
\nonumber\\
&& \left.  {} - l^{(1)}(1)\, n(z)  \lq \frac{z-1}{d(z)}\rq_{2+} -
\frac{1}{2!}\,l^{(2)}(1)\, n(z)  \lq \frac{(z-1)^2}{d(z)}\rq_{+} \rg , \phantom{aaaa}
\end{eqnarray}
and, in general, 
\begin{equation}
\label{eq:ratio_p+}
\int_0^1 dz\, l(z) \!\lq \frac{n(z)}{d(z)}\rq_{p+} \!\!= \!\int_0^1\!\! dz  \lg
l(z)\, n(z)\! \lq \frac{1}{d(z)}\rq_{p+}\!\! -
\sum_{i=0}^{p-1} \frac{1}{i!} \, l^{(i)}(1) \, n(z)\! \lq
\frac{(z-1)^i}{d(z)}\rq_{(p-i)+} \rg . \phantom{aaaa}
\end{equation}

%\bibliography{paper}

\providecommand{\href}[2]{#2}\begingroup\raggedright\endgroup

\end{document}